\documentclass[12pt]{article}
\input{ckt-simplify.sty}
\input{MyQcircuit.sty}
\usepackage{epsfig}
\usepackage{amsmath}
\usepackage{amssymb}

\begin{document}

\title{Simplifying Quantum Circuits\\
via\\
Circuit Invariants and Dressed CNOTs
 }

\author{Robert R. Tucci\\
        P.O. Box 226\\
        Bedford,  MA   01730\\
        tucci@ar-tiste.com}

\date{ \today}

\maketitle

\vskip2cm
\section*{Abstract}
Quantum Compiling Algorithms
 decompose (exactly,
without approximations) an
arbitrary $2^\nb$ unitary matrix
acting on $\nb$ qubits, into
a sequence of elementary operations (SEO).
There are many possible
ways of decomposing
a unitary matrix into a SEO,
and some of these decompositions
have shorter length (are more efficient)
than others.
Finding an optimum
(shortest) decomposition
is a very hard task, and is not
our intention here. A less ambitious,
more doable task is to find
methods for optimizing
small segments of a SEO.
Call these methods piecewise optimizations.
Piecewise optimizations
involve replacing a
small quantum circuit by
an equivalent one with fewer
CNOTs. Two circuits
are said to be equivalent
if one of them multiplied by some
external
local operations
equals the other.
This equivalence relation
between circuits has
its own class functions,
which we call circuit invariants.
Dressed CNOTs are
a simple yet very useful generalization
of standard CNOTs.
After discussing circuit invariants
and dressed CNOTs, we
give some methods for simplifying
2-qubit and 3-qubit circuits.
We include with this paper
software (written in Octave/Matlab)
that checks many
of the algorithms proposed in
the paper.

\tableofcontents
\section{Introduction}
\label{sec-introduction}

Quantum Compiling Algorithms
 decompose (exactly,
without approximations) an
arbitrary $2^\nb$ unitary matrix
acting on $\nb$ qubits, into
a sequence of elementary operations (SEO).
By elementary operations
we mean operations that act on only
a few (usually 1 or 2) qubits (for example,
all single-qubit rotations and CNOTs.)
The most efficient quantum
compiling algorithms to date
are based on a recursive
application of the Cosine-Sine Decomposition (CSD),
a technique first proposed
 in Ref.\cite{Tuc-rud}.
An implementation of the algorithm
of Ref.\cite{Tuc-rud} may be
found in the computer program
called Qubiter (patented,
C++ source code publicly available at
www.ar-tiste.com).
Long after Ref.\cite{Tuc-rud} and Qubiter
came out, many papers
on quantum compiling
via recursive CSD have appeared. These can be
easily tracked down
by making a keyword search in ArXiv or Google
for something like
(``Cosine-Sine" and  ``Decomposition"
and ``quantum") .

We will call the number of CNOTs in a SEO its length.
(Single-qubit rotations are not counted
because these can be performed
much faster than CNOTs.)
Of course, there are many possible
ways of decomposing
a unitary matrix into a SEO,
and some of these decompositions
have shorter length (are more efficient)
than others. The algorithm of
Ref.\cite{Tuc-rud} per se does not
yield the shortest SEO.
Finding an optimum
(shortest) decomposition
is a very hard task, and is not
our intention here. A less ambitious,
more doable task is to find
methods for optimizing
small segments of a SEO.
Call these methods piecewise optimizations.
The hope is that
given any SEO, one
can apply piecewise optimization
methods to reduce the original
SEO into an equivalent SEO
whose length is much less, and
might even be close to the
shortest possible length.
An analogy to our piecewise optimization
strategy is the following. Think of
a SEO as being
like a path between two points in a
manifold. If this path
is initially unnecessary long,
one might hope to make it a little
less so by breaking it into
pieces and optimizing the length
of each piece. Breaking it into pieces again,
and optimizing each piece again. And so on.

Piecewise optimizations
involve replacing a
small quantum circuit by
an equivalent one with fewer
CNOTs. Two circuits
are said to be equivalent
if one of them multiplied by some
external
local operations
equals the other.
By external local
operations, we mean
single-qubit rotations
applied at the beginning
or end of the circuit.
This equivalence relation
between circuits has
its own class functions,
which we call circuit invariants.
Many excellent papers
already exist on the use of
such  invariants in quantum
computing.
See, for example, Refs.
\cite{Rai},
\cite{Gra},
\cite{Mak},
and
\cite{She}.
Such invariants are a crucial ingredient of
this paper. (However, the paper
does not assume that the reader possesses
any prior knowledge
about these invariants. The paper
is self-contained in this regard.)

Besides circuit invariants,
another important ingredient
of this paper is what we call
dressed CNOTs (DC-NOTs). DC-NOTs are
a simple yet very useful generalization
of standard CNOTs. To my knowledge,
this paper
is the first one to consider
DC-CNOTs.
DC-NOTs are
convenient
because they lump together
a  CNOT and some
single-qubit rotations.
Modulo
external local operations,
one
can express
any circuit solely in terms
of a single type of circuit element
(DC-CNOTs), rather than having to
express it with two different types of circuit
elements (CNOTs and single-qubit
rotations).

After discussing circuit invariants
and DC-NOTs, this paper
gives some methods for simplifying
2-qubit and 3-qubit circuits.

Much is
already known about
simplifying 2-qubit circuits.
Ref.\cite{VD} shows,
via Cartan's KAK decomposition\cite{Tuc-KAK},
that a 2-qubit circuit with any number of
CNOTs
can always be reduced to a circuit
with 3 CNOTs.
Refs.\cite{VD} and \cite{She}
give necessary and
sufficient conditions
for when a 2-qubit circuit
with 3 CNOTs reduces to fewer
than 3 CNOTs.
In this paper, we spend
some time
  re-proving
  these already
known 2-qubit results
using the new language
of circuit invariants
and DC-NOTs. This
exercise yields
new techniques and new
geometrical insights
that
were lacking in previous proofs.

In this paper, we  also present some interesting
new ways of simplifying
 3-qubit circuits.
 Our results for
 3-qubit circuits rely
heavily on our results for
2-qubit circuits.

We include with this paper
software (written in Octave/Matlab)
that checks many
of the algorithms proposed in
the paper. In the header of each section,
and in the Table of Contents,
each section name is followed by
a list in square brackets of the names
of the Octave m-files relevant to that section.
Our software is not intended to be
very efficient, or to be free of
all conceivable loopholes. It is
only intended to be a proof of principle
of our algorithms.
\section{Notation
\\{\footnotesize\tt[
global\_declarations.m, global\_defs.m, simul\_real\_svd.m, Gamma\_rep.m,\\
sig.m, check\_dcnots.m, factor\_SU2pow2\_matrix.m, factor\_SU2pow3\_matrix.m,\\
test\_factor\_su2pow.m, get\_normal\_unit\_vec.m, get\_unit\_vec.m
]}}
\label{sec-notation}

In this section, we discuss notation,
linguistic idiosyncrasies and abbreviations
that will be used in subsequent sections.
If any notation in this paper remains
unclear to the reader after reading
this section, he should consult
Ref.\cite{Paulinesia},
a review article, written by
the author of this paper, that
uses the same notation as this paper.

We will often use the symbol
$\nb=0,1,2,\ldots$ for
number of bits,
and $\ns = 2^\nb$
for the corresponding number
of states.

We will often abbreviate
$\cos(\alpha)$ and
$\sin(\alpha)$ by $c_\alpha$ and
$s_\alpha$, respectively.
We will often use a subscript of $f$
to denote the final value of
quantity that changes (e.g., $\hata$ changes to
$\hata_f$).
When we say $b=\pm a$,
we mean $b\in \{a,-a\}$.
When we write $X_{\alpha\rarrow\beta}$,
we mean, the quantity
obtained by replacing $\alpha$
by $\beta$ everywhere in $X$.
Likewise, by
$X_{\alpha\darrow\beta}$
we mean, the quantity
obtained by swapping $\alpha$
and $\beta$ everywhere in $X$.
When we say ``$A (ditto, A')$ is $B(ditto, B')$"
we mean ``$A$ is $B$ and $A'$ is $B'$".
LHS and RHS will stand for
left-hand side and right-hand side.
``RHON basis" will stand for "right-handed
orthonormal basis".

Let $Bool=\{0,1\}$.
Let $\RR$ denote the real numbers,
$\CC$ the complex numbers, $\ZZ$
all integers (positive and negative).
For integers $a$ and $b$,
 $\ZZ_{a,b}$ will denote all integers from
$a$ to $b$, including $a$ and $b$. If $\Omega$
is anyone  of the symbols $>, \geq, <, \leq$,
and $S$ is any set, define
$S^{\Omega\;0}= \{x\in S: x\;\Omega\;0\}$
if the right hand side is defined.
For example, $\ZZ^{>0}$
are the positive integers.
As usual, for any set $S$
and $r, p,q\in \ZZ^{>0}$,
$S^r$ will denote the set of
r-tuples of $S$, and $S^{p\times q}$,
the set of $p\times q$ matrices with
entries in $S$.

As usual, $U(\ns)$
will denote the
$\ns\times\ns$ unitary matrices, and $SU(\ns)$
the special (i.e., with determinant=1)
elements of $U(\ns)$.
Given any $A\in U(\ns)$,
we define $\hat{A}$ by
$\hat{A} = A/[\det(A)]^{\frac{1}{\ns}}$,
where we choose the principal branch of
the function $(\cdot)^{\frac{1}{\ns}}$.
We will refer to
$\hat{A}$ as the
``special counterpart" of $A$.
(here the adjective ``special"
again means ``with determinant=1").

$\RR^3$ will denote the set of
all 3 dimensional
real vectors, and
$\unitvecs=\{x\in\RR^3:|x|=1\}$.
As is common in the Physics literature,
a letter with an arrow (ditto, caret)
over it,
as in $\veca$ (ditto, $\hata$)
will denote an element of $\RR^3$
(ditto, $\unitvecs$).
$\veca$ and $\hata$
will be treated as column vectors
when they appear in matrix expressions.

Let $\veca_j\in \RR^3$ for $j\in \ZZ_{1,r}$.
We will use the following non-standard notation
for r-fold cross products:
\beq
\manyx{\veca_1\veca_2\veca_3\ldots\veca_r}=
(\cdots((\veca_1\times\veca_2)\times\veca_3)\cdots\times \veca_r)
\;.
\eeq
For example, $\manyx{\veca_1\veca_2\veca_3\veca_4} =
((\veca_1\times\veca_2)\times\veca_3)\times \veca_4$.
Of course, an (r+2)-fold cross-product can be
reduced to an r-fold cross-product using
the well known ``BAC minus CAB" identities:
for $\veca,\vecb,\vecc\in\RR^3$,
$\veca\times(\vecb\times\vecc) =
\vecb(\veca\cdot\vecc)-\vecc(\veca\cdot\vecb)$
and
$(\veca\times\vecb)\times\vecc =
\vecb(\veca\cdot\vecc)-\veca(\vecb\cdot\vecc)$.
For example, if $\hata,\hatb$
are perpendicular unit vectors,
then $\manyx{\hata\hatb\hatb}=-\hata$.

Suppose $\veca, \vecb\in \RR^3$.
$angle(\veca,\vecb)$ will
denote the angle between
$\veca$ and $\vecb$,
defined up to $2\pi$.
We will say $\veca$ is
parallel to $\vecb$ and
write $\veca\parallel\vecb$
iff $\veca\times\vecb=0$;
i.e., iff $\veca=\pm \vecb$,
or $\veca=0$, or $\vecb=0$.
We will say
$\veca$ is
perpendicular to $\vecb$ and
write $\veca\perp\vecb$ iff $\veca\cdot\vecb=0$.
For $\vecb\neq 0$, define $\along{\veca}{\vecb}$,
{\bf the part of $\veca$ along $\vecb$},
by

\beq
\along{\veca}{\vecb}=
\frac{(\veca\cdot\vecb)\vecb}{|\vecb|^2}
\;.
\eeq
For $\vecb\neq 0$, define $\across{\veca}{\vecb}$,
{\bf the part of $\veca$ across $\vecb$},
by

\beq
\across{\veca}{\vecb}
= \veca - \along{\veca}{\vecb}
=
\veca -\frac{(\veca\cdot\vecb)\vecb}{|\vecb|^2}
=
\frac{-\manyx{\veca\vecb\vecb}}{|\vecb|^2}
\;.
\eeq

For any square matrix $A$,
$A^T$ will denote its transpose,
$A^*$, its complex conjugate,
and $A^\dagger = A^{*T}$, its Hermitian
conjugate.
$\delta_{i,j}$ will denote the Kronecker delta
function.(It equals one if $i=j$
and zero otherwise.)

Let $I_2,\sigx, \sigy, \sigz$ be the
2d identity matrix and Pauli matrices.
Sometimes, we set $(X_1,X_2,X_3)=(X,Y,Z)$
and denote the Pauli matrices by
$\sigma_{X_1}, \sigma_{X_2}, \sigma_{X_3}$.
Suppose $W \in\{X,Y,Z\}$.
Define the number operators:
$n_W = \frac{1-\sigma_W}{2}$
and $\nbar_W = \frac{1+\sigma_W}{2}$.
Note that $(-1)^{n_W} = \sigma_W$.
Usually,
$n_Z$ is denoted merely  by $n$
and $\nbar_Z$ by $\nbar$.
If $W_j \in\{1, X,Y,Z\}$
for $j\in \ZZ_{1,\nb}$, let
$\sigma_{W_1,W_2, \ldots, W_\nb} =
\sigma_{W_1}\otimes \sigma_{W_2}
\otimes\ldots \sigma_{W_\nb}$,
where any incidence of
$\sigma_1$ on the RHS
is replaced by $I_2$.
For example, $\sigma_{XY1} =
\sigx\otimes\sigy\otimes I_2$.

$H = \frac{1}{\sqrt{2}}
\left[\begin{array}{cc}1&1
\\1&-1\end{array}\right]$
is the one-bit Hadamard matrix
and $H^{\otimes\nb}$ is its
$\nb$-fold tensor product.
$H$ satisfies $H^2=1$, $H\sigx H=\sigz$,
$H\sigz H= \sigx$ and $H\sigy H = -\sigy$.

Suppose $a_0\in \RR$ and $\veca\in\RR^3$.
We will abbreviate $\vec{\sigma}\cdot\veca$
by $\sigma_\veca$. The standard terminology
is to call
$a_0 + i\sigma_\veca$
a {\bf quaternion}, and to call
$\sigma_\veca$
a vector quaternion (divided by $i$).
To shorten
this terminology, we will
refer to $\sigma_\veca$ as a {\bf Paulion},
and call $\veca$ its {\bf defining vector}.
If $|\veca|=1$, we will
call $\sigma_\veca$ a {\bf unit Paulion}.
One can reduce a product of
two Paulions
by using the identity
$\sigma_{\veca}\sigma_{\vecb}=
\veca\cdot\vecb + i \sigma_{\veca\times\vecb}$.
For $\hata\in \unitvecs$, define number operators
$n_\hata = \frac{1-\sigma_\hata}{2}$ and
$\nbar_\hata = \frac{1+\sigma_\hata}{2}$.
Note that $(-1)^{n_\hata} = \sigma_\hata$.
If $W_j \in\unitvecs$ or $W_j=1$
for $j\in \ZZ_{1,\nb}$, let
$\sigma_{W_1,W_2, \ldots, W_\nb} =
\sigma_{W_1}\otimes \sigma_{W_2}
\otimes\ldots \sigma_{W_\nb}$.

Suppose $\calm$
is the set of all matrices
$M\in \CC^{4\times 4}$
that can be expressed in the form
$M=\sum_k \sigma_{\veca_k,\vecb_k}$, where
$\veca_k, \vecb_k\in \RR^3$ for all $k$.
Suppose $\call$ is the set of all
matrices $L\in \RR^{3\times 3}$
that can be expressed in the form
$L=\sum_k \veca_k \vecb_k^T$,
where $\veca_k, \vecb_k\in \RR^3$ for all $k$.
For every $M\in \calm$,
let $\Gamma(M)$ or $M^\Gamma$
represent the $3\times3$
matrix with entries
$\frac{1}{4}\tr(\sigma_{X_i,X_j}M)$,
where $i,j\in \ZZ_{1,3}$.
(The symbol $\Gamma$ was chosen
to evoke the mental picture of
a column vector times a row vector;
such is the output of the function $\Gamma(\cdot)$).
For every  $L\in \call$,
define $\Gamma^{-1}(L) =
\sum_{i,j}\sigma_{X_i,X_j}L_{i,j}$.
It's easy to check that
$\Gamma \Gamma^{-1} = \Gamma^{-1}\Gamma =1$
so the map $\Gamma:\calm\rarrow \call$ is 1-1 onto.
Let $lin(\calm)$ be
the set of linear combinations over $\CC$
of elements of $\calm$,
and $lin(\call)$ of $\call$.
The map $\Gamma$
can be extended to $\overline{\Gamma}:
\CC + lin(\calm)\rarrow
\CC + lin(\call)$,
$\overline{\Gamma}(\lam + \sum_i \alpha_i M_i)
= \lam + \sum_i \alpha_i M_i^\Gamma$.
$\overline{\Gamma}$ is also a 1-1 onto map.
Henceforth, we will use $\Gamma$
to refer to both $\Gamma$
and its extension $\overline{\Gamma}$.
Given a matrix $A\in \CC + lin(\calm)$,
we will call $A^\Gamma$ its Gamma representation.
Often, in contexts where this
 will not lead to
confusion, we will drop the
$\Gamma$ superscript and denote
$A^\Gamma$ simply by $A$.

The next theorem, although almost
trivial,
will be used frequently in this paper.

\begin{theo}\label{th-double-cover}
The map $f:\unitvecs\times\unitvecs\rarrow SU(2)$,
$f(\hata, \hatb) = \sigma_\hata\sigma_\hatb$
is well defined and onto. In other words:
(well-defined)
If $\hata, \hatb\in \unitvecs$, then
$f(\hata, \hatb)\in SU(2)$.
(onto)
If $U\in SU(2)$, then there
exist $\hata,\hatb\in  \unitvecs$
such that $U=f(\hata, \hatb)$.
\end{theo}
\proof

(well defined)
Given $\hata,\hatb\in\unitvecs$,
one can always find an angle $\theta$
such that $\hata\cdot\hatb=c_\theta$
and $|\hata\times\hatb|=s_\theta$.
Let $\hatw = \frac{\hata\times \hatb}
{|\hata\times \hatb|}$.
It follows that
$\sigma_\hata\sigma_\hatb =
\hata\cdot\hatb + i\sigma_{\hata\times\hatb}
=e^{i\theta\sigma_\hatw}\in SU(2)$.

(onto) Given
$U= e^{i\theta\sigma_\hatw}$,
where $\hatw\in\unitvecs$ and
$\theta\in \RR$, one can
always find
a (non-unique) pair of unit vectors
$\hata$ and $\hatb$
in the plane perpendicular
to $\hatw$, such that
$\theta=angle(\hata,\hatb)$,
and
$\hata\times\hatb$
points in the $\hatw$
direction. Hence,
$\hata\cdot\hatb = c_\theta$
and
$\hata\times\hatb = s_\theta\hatw$.
It follows that
$\sigma_\hata\sigma_\hatb =
\hata\cdot\hatb + i\sigma_{\hata\times\hatb}
=e^{i\theta\sigma_\hatw}$.

\qed

\begin{figure}[h]
    \begin{center}
    \epsfig{file=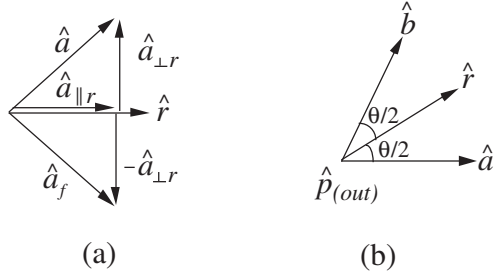, height=1.5in}
    \caption{(a)If
    $\sigma_\hatr \sigma_\hata \sigma_\hatr
    =\sigma_{\hata_f}$, then $\hata_f$
    is obtained
    by reflecting $\hata$ on $\hatr$.(b)
    Suppose  $\hatb$ is the result of rotating
    $\hata$ by an angle $\theta$. Then
    $\hatb$ can be obtained by reflecting $\hata$
    on the bisector $\hatr$
    of the angle between $\hata$ and $\hatb$.}
    \label{fig-paulion-rot}
    \end{center}
\end{figure}

One has:

\beq
\sigma_\hatr
\sigma_\hata
\sigma_\hatr
=
\sigma_\hatr
(\sigma_{\along{\hata}{\hatr}} +
\sigma_{\across{\hata}{\hatr}})
\sigma_\hatr
=
\sigma_{\along{\hata}{\hatr}}
-\sigma_{\across{\hata}{\hatr}}
=
\sigma_{
\along{\hata}{\hatr}
-\across{\hata}{\hatr}
}
=
\sigma_{\hata_f}
\;.
\label{eq-paulion-reflection}
\eeq
A geometrical interpretation
of this identity
is shown in
Fig.\ref{fig-paulion-rot}a.
The similarity transformation
$\sigma_\hatr(\cdot)\sigma_\hatr$
takes the Paulion
$\sigma_\hata$ to $\sigma_{\hata_f}$,
where $\hata_f$ is the reflection
of $\hata$ on  $\hatr$.

Suppose $\hata, \hatb\in\unitvecs$,
and we want to find
$U\in SU(2)$ such that
$\sigma_\hatb=
U^\dagger\sigma_\hata U$.
Such a $U$ can be
constructed as a product of
two Paulions
(See Fig.\ref{fig-paulion-rot}b).
Indeed,
let $\theta=angle(\hata,\hatb)$
and $\hatp= \frac{\hata\times\hatb}
{|\hata\times\hatb|}$.
Let $\hatr$ be the vector
that bisects the angle
between $\hata$ and $\hatb$,
and is oriented so that $\hata\times\hatr$
points along $\hatp$. Note that
$\hatb$ can be obtained
by reflecting $\hata$ on the bisector $\hatr$.
Hence

\beq
\sigma_\hata\sigma_\hatr  =
e^{i\frac{\theta}{2}\sigma_\hatp}
\;,\;\;
\sigma_\hatb=
\sigma_\hatr\sigma_\hata\sigma_\hatr
\;.
\eeq
Combining these two results yields

\beq
\sigma_\hatb
=
(\sigma_\hatr\sigma_\hata)
\sigma_\hata
(\sigma_\hata\sigma_\hatr)
=
e^{-i\frac{\theta}{2}\sigma_\hatp}
\sigma_\hata
e^{i\frac{\theta}{2}\sigma_\hatp}
\;.
\eeq

\section{Invariants for Quantum Circuits}
\label{sec-ckt-invariants}

In this section, we will discuss
circuit invariants; i.e.,
functions that map
all equivalent circuits to the same value.
By equivalent circuits we mean circuits that are
equal,
modulo external local operations.

Suppose $A$ and $B$ are
elements of $U(\ns)$ ( i.e., they
are $\nb$-qubit gates).
We will say $A$ and $B$
are {\bf equivalent under  local operations
on the right hand side (LO-RHS)},
and write
$A\sim_R B$, iff there exist $U_j\in U(2)$
for $j\in \ZZ_{0,\nb-1}$
such that

\beq
B = A (U_{\nb-1} \otimes \ldots \otimes U_2 \otimes U_1 \otimes U_0)
\;.
\label{eq-def-lo-rhs}
\eeq
$\sim_R$ is clearly an
equivalence relation as it
is symmetric, reflexive and transitive.

Henceforth, we will say that a function
$\chi$ with domain $U(\ns)$ is
a LO-RHS invariant if for any
$A,B\in U(\ns)$, $A\sim_R B$ implies that
$\chi(A) = e^{i\zeta}\chi(B)$
for some $\zeta\in \RR$ ($\zeta$
may depend on $A,B$).

A frequent goal is
to find a complete set of
scalar invariant functions; that is, a
set of functions
$\chi_j:U(\ns)\rarrow \RR$ such that
for any $A,B \in U(\ns)$,
$A\sim_R B$
if and only if
 $\chi_j(A)=\chi_j(B)$ for all $j$.
An extensive literature already
exist on such invariants. They
were first studied by Group Theorists,
and, in more recent times, they
have been used by Quantum Computerists
\cite{Rai},
\cite{Gra},
\cite{Mak},
\cite{She}.

One can define an analogous
equivalence relation $\sim_L$ for local operations
on the left hand side (LO-LHS),
and an equivalence
relation $\sim_{LR}$
for local operations on both sides (LO-2S).
Of course, the equivalence classes (e-classes)
of $\sim_R$
are a disjoint partition of
$U(\ns)$. Ditto for the e-classes of $\sim_{L}$
and $\sim_{LR}$.
It's also clear that any e-class for $\sim_R$
is contained in an e-class for $\sim_{LR}$,
and that some e-classes of $\sim_{LR}$
contain more than one e-class of
$\sim_R$. (In fact, the e-classes
of $\sim_R$ contained within a single e-class
of $\sim_{LR}$,
can be labeled by the elements of $U(2)^{\otimes \nb}$).

Note that for any $\vec{a}\in \RR^3$,

\beq
\sigy \sigma^T_{\veca}\sigy
=
-\sigma_{\veca}
\;.\label{eq-neg-sig}
\eeq
Hence, for $\theta\in\RR$ and $\vec{a}\in \RR^3$,

\beq
\sigy [e^{i(\theta+\sigma_\veca)}]^T \sigy=
e^{i(\theta-\sigma_\veca)}
\;.
\eeq
Thus, when $U\in SU(2)$ (but not
when $U\in U(2)$),
$\sigy U^T \sigy =U^{-1}=U^\dagger$.

For any $A\in U(\ns)$, define
a quadratic (second order in $A$) invariant

\beq
A^{(2)} =
A \sigy^{\otimes \nb} A^T \sigy^{\otimes \nb}
\;.
\eeq
For example, for $A\in U(4)$,
$A^{(2)} = A\sigyy A^T \sigyy$.

\begin{theo}\label{th-quad-invar}

\begin{enumerate}
\item[]
\item[(a)]For $A,B\in SU(4)$,
$A\sim_R B$ if and only if
$A^{(2)}=(-1)^n B^{(2)}$ for some $n\in \ZZ$.

\item[(b)] For $A,B\in U(4)$,
$A\sim_R B$ if and only if
$A^{(2)}=e^{i\zeta} B^{(2)}$ for some $\zeta\in \RR$.
\end{enumerate}
\end{theo}
\proof

\begin{enumerate}
\item[(a)]

Assume $A, B\in SU(4)$.
$A$ can always be represented in the form

\beq
A = i^{n(A)}
\exp(i a_{jk}\sigma_{X_j X_k})
\exp(i a'_j\sigma_{X_j 1})
\exp(i a_k \sigma_{1 X_k})
\;,
\label{eq-A-su4}
\eeq
where $n(A)\in\ZZ$
and $a_{jk},a'_j, a_k\in \RR$. (Note that
$\det(i I_4)=1$ so $\det(A)=1$.)
We are using Einstein's implicit summation convention,
and $j,k$ range over $\{1,2,3\}$.
By Eqs.(\ref{eq-neg-sig})
and (\ref{eq-A-su4}),

\beq
\sigyy A^T \sigyy= i^{n(A)}
\exp(-i a_k \sigma_{1 X_k})
\exp(-i a'_j\sigma_{X_j 1})
\exp(i a_{jk}\sigma_{X_j X_k})
\;.
\eeq
Thus

\beq
A^{(2)} = (-1)^{n(A)}
\exp(i 2a_{jk}\sigma_{X_j X_k})
\;.
\label{eq-A-su4-invar}
\eeq
Likewise, $B$ can be represented in the form

\beq
B = i^{n(B)}
\exp(i b_{jk}\sigma_{X_j X_k})
\exp(i b'_j\sigma_{X_j 1})
\exp(i b_k \sigma_{1 X_k})
\;,
\label{eq-B-su4}
\eeq
where $n(B)\in \ZZ$
and $b_{jk},b'_j, b_k\in \RR$.
Then

\beq
B^{(2)} = (-1)^{n(B)}
\exp(i 2b_{jk}\sigma_{X_j X_k})
\;.
\label{eq-B-su4-invar}
\eeq

\rproof Suppose $A\sim_R B$.
Looking at Eqs.(\ref{eq-def-lo-rhs}),
(\ref{eq-A-su4})
and (\ref{eq-B-su4}), we see that
for every $j,k$, there exists
an integer $n_{jk}$ such that
$a_{jk} = b_{jk} + \pi n_{jk}$.
Therefore,

\beq
\exp(i 2a_{jk}\sigma_{X_j X_k})=
\exp(i 2b_{jk}\sigma_{X_j X_k})
\;.
\label{eq-gist-rproof}
\eeq
Therefore, looking at
Eqs.(\ref{eq-A-su4-invar})
and (\ref{eq-B-su4-invar}), we see that
there exists an integer $n$
such that $A^{(2)}=(-1)^n B^{(2)}$.

\lproof Suppose
 $A^{(2)}=(-1)^n B^{(2)}$.
Then, looking at
Eqs.(\ref{eq-A-su4-invar})
and (\ref{eq-B-su4-invar}), we see that
for every $j,k$, there exists
an integer $n_{jk}$ such that
$2a_{jk} = 2b_{jk} + \pi n_{jk}$.
Therefore,

\beq
\exp(i a_{jk}\sigma_{X_j X_k})=
\exp(i b_{jk}\sigma_{X_j X_k})
\prod_{j,k}[i \sigma_{X_j X_k}]^{n_{jk}}
\;.
\label{eq-gist-lproof}
\eeq
Therefore,
from Eqs.(\ref{eq-def-lo-rhs}),
(\ref{eq-A-su4})
and (\ref{eq-B-su4}), we see that
$A\sim_R B$.

\item[(b)]

Assume $A, B\in U(4)$.
Eqs.(\ref{eq-A-su4}) and (\ref{eq-A-su4-invar})
still apply except that we must replace in them
$i^{n(A)}$ by $e^{i\zeta(A)}$ and
$(-1)^{n(A)}$ by $e^{i 2\zeta(A)}$
for some $\zeta(A)\in\RR$.
Eqs.(\ref{eq-B-su4}) and (\ref{eq-B-su4-invar})
still apply except that we must replace in them
$i^{n(B)}$ by $e^{i\zeta(B)}$ and
$(-1)^{n(B)}$ by $e^{i 2\zeta(B)}$
for some $\zeta(B)\in\RR$.

\rproof Suppose $A\sim_R B$.
Eq.(\ref{eq-gist-rproof})
still applies so
there exists $\zeta\in \RR$
such that $A^{(2)}=e^{i\zeta} B^{(2)}$.

\lproof Suppose
 $A^{(2)}=e^{i\zeta} B^{(2)}$.
Eq.(\ref{eq-gist-lproof})
still applies so $A\sim_R B$.
\end{enumerate}

\qed

By virtue of Theorem \ref{th-quad-invar},
the absolute value of the
 entries of the matrix $A^{(2)}$
are a complete set of LO-RHS scalar
invariants for
$\nb=2$. Theorem \ref{th-quad-invar}(a)
reflects the fact that when $A,B\in SU(4)$,
since $A$ and $B$ must both
have unit determinant,
the only local operations
connecting $A$ and $B$ are either
elements of $SU(2)$ or $i$ or products of these.
Applying an $SU(2)$ gate
 to the RHS of $A$
does not change $A^{(2)}$,
whereas applying $i$ changes
$A^{(2)}$ to its negative.

Now suppose $\nb=3$.
One can
represent any $A\in SU(8)$ as

\begin{eqnarray}\label{eq-su8-repres}
A &=& e^{i\frac{\pi}{4}n(A)}
\exp(i a_{jkr}\sigma_{X_j X_k X_r})\nonumber\\
&&\;\;\;\exp(i a''_{jk}\sigma_{1X_j X_k})
\exp(i a'_{jk}\sigma_{X_j 1 X_k})
\exp(i a_{jk}\sigma_{X_j X_k 1})\nonumber\\
&&\;\;\;\;\;\;\exp(i a''_{j}\sigma_{X_j 11})
\exp(i a'_{j}\sigma_{1X_j 1})
\exp(i a_{j}\sigma_{11X_j})
\;.
\end{eqnarray}
When the continuous parameters of $A$ are small,
\beq
A^{(2)} \approx e^{i\frac{\pi}{2}n(A)}[1 +
2i (a''_{jk}\sigma_{1X_j X_k}+
a'_{jk}\sigma_{X_j 1 X_k}+
a_{jk}\sigma_{X_j X_k 1})]
\;.
\eeq
This $A^{(2)}$ is independent of
the $a_{jkr}$ parameters.
So, for $A,B\in SU(8)$,
$A^{(2)}=\pm B^{(2)}$ or $A^{(2)}=\pm i B^{(2)}$ is
a necessary but not a sufficient
condition for $A\sim_R B$.
More invariants than
just $A^{(2)}$ are needed for $\nb>2$.

\begin{figure}[h]
    \begin{center}
    \epsfig{file=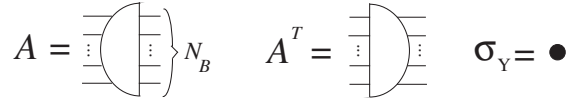, height=.5in}
    \caption{Key to symbols used in
    Figs.\ref{fig-3bit-invar} and \ref{fig-show-are-invars}.
    $A\in SU(\ns)$.}
    \label{fig-invars-symb}
    \end{center}
\end{figure}

\begin{figure}[h]
    \begin{center}
    \epsfig{file=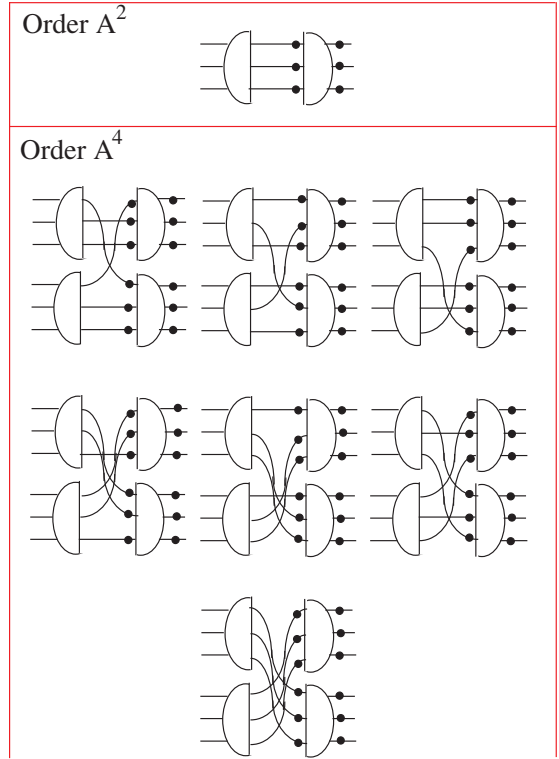, height=4in}
    \caption{Second and fourth order
    invariants under LO-RHS for a circuit
    with 3 bits. $A\in SU(8)$.}
    \label{fig-3bit-invar}
    \end{center}
\end{figure}

\begin{figure}[h]
    \begin{center}
    \epsfig{file=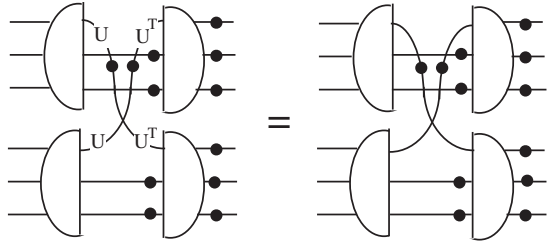, height=1.25in}
    \caption{Why the diagrams shown in
    Fig.\ref{fig-3bit-invar} are
    invariant under LO-RHS.
    $A\in SU(8)$ and $U\in SU(2)$.}
    \label{fig-show-are-invars}
    \end{center}
\end{figure}

Higher order invariants can be generated
as follows. We will represent them diagrammatically
using the symbols defined in Fig.\ref{fig-invars-symb}.
Fig.\ref{fig-3bit-invar} shows
second and fourth order
invariants under LO-RHS for a circuit
with 3 bits. The same idea can
be used to generate invariants of order
equal to any even number,
for any number of qubits.
Fig.\ref{fig-show-are-invars}
explains why the circuits portrayed in
Fig.\ref{fig-3bit-invar} are invariant under LO-RHS.
Roughly speaking, if we apply a
$U\in SU(2)$ to the RHS of $A\in SU(8)$,
then, in the diagram of a fourth order invariant,
a copy of $U$ must be inserted
next to each of the 4 copies of $A$.
And these 4 copies of $U$ annihilate
each other. This paper will only
use the second order invariant $A^{(2)}$. We
will not even use Group Theory in this paper.
For information on the group
theoretic underpinnings of quantum circuit invariants,
see, for example, Ref.\cite{Rai}.

\section{Dressed CNOTs
\\{\footnotesize\tt[
dr11.m,
dr110.m,
dr011.m,
dr101.m,
]}}
\label{sec-dc-nots}

In this section, we define
dressed CNOTs, a simple yet
powerful generalization of the
standard CNOT. We also discuss some simple
properties of dressed CNOTs
that will be used in subsequent
sections.

The {\bf controlled NOT (CNOT)}
with control bit 1 and target bit 0, is defined by

\beq
\begin{array}{c}
\Qcircuit @C=1em @R=1em @!R{
&\timesgate\qwx[1]
&\qw
\\
&\dotgate
&\qw
}
\end{array}
= (-1)^{n_X(0)n(1)}
=\sigma_X(0)^{n(1)}
\;.
\eeq
Now suppose $U$ and $V$
are arbitrary elements of $SU(2)$.
Define $\hata$ and $\hata'$ by
$U\sigx U^\dagger = \sigma_\hata$
and
$V\sigz V^\dagger = \sigma_{\hata'}$.
Then a {\bf dressed CNOT (DC-NOT)}
connecting bits 0 and 1, is defined by

\beq
\begin{array}{c}
\Qcircuit @C=1em @R=1em @!R{
&\ovalgate{\hata}\qwx[1]
&\qw
\\
&\ovalgate{\hata'}
&\qw
}
\end{array}
=
\begin{array}{c}
\Qcircuit @C=1em @R=1em @!R{
&\gate{U}
&\timesgate\qwx[1]
&\gate{U^\dagger}
&\qw
\\
&\gate{V}
&\dotgate
&\gate{V^\dagger}
&\qw
}
\end{array}
=(-1)^{n_\hata(0) n_{\hata'}(1)}
=\sigma_\hata(0)^{n_{\hata'}(1)}
=\sigma_{\hata'}(1)^{n_{\hata}(0)}
\;.
\eeq
We will refer to the vectors
$\hata'$ and $\hata$ as the
{\bf defining vectors} of the DC-NOT.

Sometimes in this paper, we will draw a circuit
containing one or more DC-NOTs whose oval nodes
are empty.
By this we will mean  that the omitted defining vectors
are arbitrary and their precise value
is unimportant in that context.

Consider the wire corresponding to
bit $\mu$ in a quantum circuit.
Within the bit-$\mu$ wire,
consider  two adjacent
oval nodes belonging to
two different DC-NOTs:
$
\begin{array}{c}
\Qcircuit @C=1em @R=.5em @!R{
&\ovalgate{\hata}
&\ovalgate{\hatb}
&\qw
}
\end{array}
\;.
$
If $\hatb\parallel \hata$,
we will say there is a {\bf breach}
at that position in the bit-$\mu$ wire.
If $\hatb\perp \hata$,
we will say there is a {\bf foil}
at that position in the bit-$\mu$ wire.

\begin{theo}

\beq
\begin{array}{c}
\Qcircuit @C=1em @R=1em @!R{
&\ovalgate{\hata}\qwx[1]
&\qw
\\
&\ovalgate{\hata'}
&\qw
}
\end{array}
=
\frac{1}{2}
( 1 + \sigma_{1,\hata} +
\sigma_{\hata',1} - \sigma_{\hata',\hata})
\;.
\eeq
\end{theo}
\proof

\beq
\sigma_{\hata'}(1)^{n_{\hata}(0)}
=
\sigma_{\hata'}(1)n_\hata(0) + \nbar_\hata(0)
=
\frac{1}{2}
( 1 + \sigma_{1,\hata} +
\sigma_{\hata',1} - \sigma_{\hata',\hata})
\;.
\eeq

\qed

\begin{theo}
\beq
\left(
\begin{array}{c}
\Qcircuit @C=1em @R=1em @!R{
&\ovalgate{\hata}\qwx[1]
&\qw
\\
&\ovalgate{\hata'}
&\qw
}
\end{array}
\right)^2
=1
\;.
\eeq
\end{theo}
\proof

$\sigma_\hata(0)^{2n_{\hata'}(1)}=1$.

\qed

\begin{theo}
\beq
\begin{array}{c}
\Qcircuit @C=1em @R=1em @!R{
&\ovalgate{-\hata}\qwx[1]
&\qw
\\
&\ovalgate{\hata'}
&\qw
}
\end{array}
=
\begin{array}{c}
\Qcircuit @C=1em @R=1em @!R{
&\qw
&\ovalgate{\hata}\qwx[1]
&\qw
\\
&\gate{\sigma_{\hata'}}
&\ovalgate{\hata'}
&\qw
}
\end{array}
\;.
\label{eq-dcnot-with-neg}
\eeq
\end{theo}
\proof
\beq
[-\sigma_\hata(0)]^{n_{\hata'}(1)}=
(-1)^{n_{\hata'}(1)}\sigma_\hata(0)^{n_{\hata'}(1)}=
\sigma_{\hata'}(1)\sigma_\hata(0)^{n_{\hata'}(1)}
\;.
\eeq

\qed

In subsequent sections, we
will often need to calculate
the effect of a similarity transformation
produced by pre and post multiplying an operator
by the same DC-NOT. The next
theorem will be useful
for performing such calculations.

\begin{theo}
\beq
\begin{array}{c}
\Qcircuit @C=1em @R=1em @!R{
&\ovalgate{\hatb}\qwx[1]
&\gate{\sigma_{\veca}}
&\ovalgate{\hatb}\qwx[1]
&\qw
\\
&\ovalgate{\hatb'}
&\qw
&\ovalgate{\hatb'}
&\qw
}
\end{array}
=
\sigma_{1,\along{\veca}{\hatb}}+
\sigma_{\hatb',\across{\veca}{\hatb}}
\;.
\label{eq-sim-trans-of-sig-veca}
\eeq
\end{theo}
\proof

Clearly,
\beq
\begin{array}{c}
\Qcircuit @C=1em @R=1em @!R{
&\ovalgate{\hatb}\qwx[1]
&\gate{\sigma_{\along{\veca}{\hatb}}}
&\ovalgate{\hatb}\qwx[1]
&\qw
\\
&\ovalgate{\hatb'}
&\qw
&\ovalgate{\hatb'}
&\qw
}
\end{array}
=
\sigma_{1,\along{\veca}{\hatb}}
\;.
\label{eq-sim-transf-parallel-part}
\eeq
On the other hand,

\beq
\begin{array}{c}
\Qcircuit @C=1em @R=1em @!R{
&\ovalgate{\hatb}\qwx[1]
&\gate{\sigma_{\across{\veca}{\hatb}}}
&\ovalgate{\hatb}\qwx[1]
&\qw
\\
&\ovalgate{\hatb'}
&\qw
&\ovalgate{\hatb'}
&\qw
}
\end{array}
=
\begin{array}{c}
\Qcircuit @C=1em @R=1em @!R{
&\gate{\sigma_{\across{\veca}{\hatb}}}
&\ovalgate{-\hatb}\qwx[1]
&\ovalgate{\hatb}\qwx[1]
&\qw
\\
&\qw
&\ovalgate{\hatb'}
&\ovalgate{\hatb'}
&\qw
}
\end{array}
=
\sigma_{\hatb',\across{\veca}{\hatb}}
\;.
\label{eq-sim-transf-perp-part}
\eeq

\qed
\section{Wake Identities}
\label{sec-wake-ids}

In this section we prove what we call
 a ``wake identity". We call it thus
because in it, one DC-NOT is pushed
through another, producing
a third DC-NOT as its ``wake".

\begin{theo}
Suppose $\hata'\perp\hatb'$.

\begin{subequations}
\begin{eqnarray}
\begin{array}{c}
\Qcircuit @C=1em @R=1em @!R{
&\ovalgate{\hatb}\qwx[1]
&\qw
&\qw
&\qw
\\
&\ovalgate{\hatb'}
&\foil
&\ovalgate{\hata'}\qwx[1]
&\qw
\\
&\qw
&\qw
&\ovalgate{\hata''}
&\qw
}
\end{array}
&=&
\begin{array}{c}
\Qcircuit @C=1em @R=1em @!R{
&\qw
&\ovalgate{\hatb}\qwx[1]
&\ovalgate{\hatb}\qwx[2]
&\qw
\\
&\ovalgate{\hata'}\qwx[1]
&\ovalgate{\hatb'}
&\qw
&\qw
\\
&\ovalgate{\hata''}
&\qw
&\ovalgate{\hata''}
&\qw
}
\end{array}
\label{eq-wake-on-right}
\\
&=&
\begin{array}{c}
\Qcircuit @C=1em @R=1em @!R{
&\ovalgate{\hatb}\qwx[2]
&\qw
&\ovalgate{\hatb}\qwx[1]
&\qw
\\
&\qw
&\ovalgate{\hata'}\qwx[1]
&\ovalgate{\hatb'}
&\qw
\\
&\ovalgate{\hata''}
&\ovalgate{\hata''}
&\qw
&\qw
}
\end{array}
\;.
\end{eqnarray}
\end{subequations}
\end{theo}
\proof

\beqa
\begin{array}{c}
\Qcircuit @C=1em @R=1em @!R{
&\qw
&\qw
&\ovalgate{\hatb}\qwx[1]
&\qw
&\qw
&\qw
\\
&\ovalgate{\hata'}\qwx[1]
&\foil
&\ovalgate{\hatb'}
&\foil
&\ovalgate{\hata'}\qwx[1]
&\qw
\\
&\ovalgate{\hata''}
&\qw
&\qw
&\qw
&\ovalgate{\hata''}
&\qw
}
\end{array}
&=&
\sigma_{\hata'}(1)^{n_{\hata''}(2)}
\sigma_{\hatb'}(1)^{n_{\hatb}(0)}
\sigma_{\hata'}(1)^{n_{\hata''}(2)}
\\
&=&
[(-1)^{n_{\hata''}(2)}
\sigma_{\hatb'}(1)]^{n_{\hatb}(0)}
\\
&=&
(-1)^{n_{\hata''}(2)n_{\hatb}(0)}
\sigma_{\hatb'}(1)^{n_{\hatb}(0)}
\\
&=&
\begin{array}{c}
\Qcircuit @C=1em @R=1em @!R{
&\ovalgate{\hatb}\qwx[2]
&\ovalgate{\hatb}\qwx[1]
&\qw
\\
&\qw
&\ovalgate{\hatb'}
&\qw
\\
&\ovalgate{\hata''}
&\qw
&\qw
}
\end{array}
\;.
\eeqa

\qed
\section{Swapper Identities
\\{\footnotesize\tt[
swap\_t3.m,
test\_swap\_t3.m
]}}
\label{sec-swapper-ids}

In this section, we discuss certain
DC-NOT identities associated with
the qubit Exchange Operator (a.k.a.
Swap Operator or Swapper).

We will represent the Swapper by a double
arrow connecting the two qubits being swapped.
By definition, the Swapper satisfies

\beq
\begin{array}{c}
\Qcircuit @C=1em @R=1em @!R{
&\gate{U}
&\uarrowgate\qwx[1]
&\qw
\\
&\qw
&\darrowgate
&\qw
}
\end{array}
=
\begin{array}{c}
\Qcircuit @C=1em @R=1em @!R{
&\qw
&\uarrowgate\qwx[1]
&\qw
\\
&\qw
&\darrowgate
&\gate{U}
}
\end{array}
\;
\eeq
for any $U\in U(2)$.
As is well known (for a proof,
see, for example, Ref.\cite{Paulinesia}),
 the Swapper can be expressed
as a product of 3 CNOTs:

\beq
\begin{array}{c}
\Qcircuit @C=1em @R=1em @!R{
&\uarrowgate\qwx[1]
&\qw
\\
&\darrowgate
&\qw
}
\end{array}
=
\begin{array}{c}
\Qcircuit @C=1em @R=1em @!R{
&\timesgate\qwx[1]
&\dotgate\qwx[1]
&\timesgate\qwx[1]
&\qw
\\
&\dotgate
&\timesgate
&\dotgate
&\qw
}
\end{array}
\;.
\eeq
The next theorem shows that
the Swapper can also be expressed as a
product of 3 DC-NOTs.

\begin{theo}
Suppose $\hata\perp\hatb$,
$U\in SU(2)$,
$U^\dagger \sigma_{\hata} U = \sigma_{\hata'}$,
and
$U^\dagger \sigma_{\hatb} U = \sigma_{\hatb'}$.

\beqa
\begin{array}{c}
\Qcircuit @C=1em @R=1em @!R{
&\uarrowgate\qwx[1]
&\qw
\\
&\darrowgate
&\qw
}
\end{array}
&=&
\begin{array}{c}
\Qcircuit @C=1em @R=1em @!R{
&\ovalgate{\hata}\qwx[1]
&\foil
&\ovalgate{\hatb}\qwx[1]
&\foil
&\ovalgate{\hata}\qwx[1]
&\qw
\label{eq-swapper-two-vecs}
\\
&\ovalgate{\hatb}
&\foil
&\ovalgate{\hata}
&\foil
&\ovalgate{\hatb}
&\qw
}
\end{array}
\\
&=&
\begin{array}{c}
\Qcircuit @C=1em @R=1em @!R{
&\ovalgate{\hata}\qwx[1]
&\foil
&\ovalgate{\hatb}\qwx[1]
&\foil
&\ovalgate{\hata}\qwx[1]
&\gate{U}
&\qw
\\
&\ovalgate{\hatb'}
&\foil
&\ovalgate{\hata'}
&\foil
&\ovalgate{\hatb'}
&\gate{U^\dagger}
&\qw
}
\end{array}
\label{eq-swapper-four-vecs}
\;.
\eeqa
\end{theo}
\proof

Since $\hata\perp\hatb$,
there exists $V\in SU(2)$
such that
$V^\dagger \sigx V = \sigma_{\hata}$
and $V^\dagger \sigz V = \sigma_{\hatb}$.
Then

\beq
\begin{array}{c}
\Qcircuit @C=1em @R=1em @!R{
&\uarrowgate\qwx[1]
&\qw
\\
&\darrowgate
&\qw
}
\end{array}
=
\begin{array}{c}
\Qcircuit @C=1em @R=1em @!R{
&\gate{V^\dagger}
&\uarrowgate\qwx[1]
&\gate{V}
&\qw
\\
&\gate{V^\dagger}
&\darrowgate
&\gate{V}
&\qw
}
\end{array}
=
\begin{array}{c}
\Qcircuit @C=1em @R=1em @!R{
&\gate{V^\dagger}
&\timesgate\qwx[1]
&\dotgate\qwx[1]
&\timesgate\qwx[1]
&\gate{V}
&\qw
\\
&\gate{V^\dagger}
&\dotgate
&\timesgate
&\dotgate
&\gate{V}
&\qw
}
\end{array}
=
\begin{array}{c}
\Qcircuit @C=1em @R=1em @!R{
&\ovalgate{\hata}\qwx[1]
&\ovalgate{\hatb}\qwx[1]
&\ovalgate{\hata}\qwx[1]
&\qw
\\
&\ovalgate{\hatb}
&\ovalgate{\hata}
&\ovalgate{\hatb}
&\qw
}
\end{array}
\;.
\eeq
This proves Eq.(\ref{eq-swapper-two-vecs}).
Eq.(\ref{eq-swapper-four-vecs}) follows from

\beq
\begin{array}{c}
\Qcircuit @C=1em @R=1em @!R{
&\uarrowgate\qwx[1]
&\qw
\\
&\darrowgate
&\qw
}
\end{array}
=
\begin{array}{c}
\Qcircuit @C=1em @R=1em @!R{
&\qw
&\uarrowgate\qwx[1]
&\qw
&\gate{U}
&\qw
\\
&\gate{U^\dagger}
&\darrowgate
&\gate{U}
&\gate{U^\dagger}
&\qw
}
\end{array}
\;.
\eeq

\qed

We will refer to the next
identity, Eq.(\ref{eq-2thirds}),
 as the 2/3-Swapper identity,
because its LHS contains  2/3 of a
Swapper.

\begin{figure}[h]
    \begin{center}
    \epsfig{file=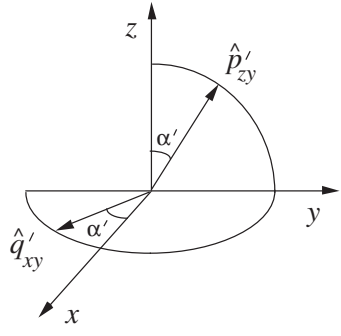, height=1.75in}
    \caption{Orientation of vectors $\hatqxy'$
    and $\hatpzy'$. Note that
    $H\sigma_{\hatpzy'}H = \sigma_{\hatqxy'}$.
    The same picture, but
    omitting all primes, describes $\hatqxy$
    and $\hatpzy$.}
    \label{fig-qxy-pzy}
    \end{center}
\end{figure}

\begin{theo}\label{th-2thirds}

For any $\alpha\in\RR$,
\begin{subequations}
\label{eq-2thirds}
\begin{eqnarray}
\begin{array}{c}
\Qcircuit @C=1em @R=1em @!R{
&\ovalgate{\hatx}\qwx[1]
&\ovalgate{\hatz}\qwx[1]
&\qw
\\
&\ovalgate{\hatx}
&\ovalgate{\hatz}
&\qw
}
\end{array}
&=&
\begin{array}{c}
\Qcircuit @C=1em @R=1em @!R{
&\ovalgate{\hatqxy}\qwx[1]
&\ovalgate{\hatz}\qwx[1]
&\gate{U}
&\qw
\\
&\ovalgate{\hatqxy'}
&\ovalgate{\hatz}
&\gate{U'}
&\qw
}
\end{array}
\label{eq-2thirds-a}
\\
&=&
\begin{array}{c}
\Qcircuit @C=1em @R=1em @!R{
&\gate{U}
&\ovalgate{\hatx}\qwx[1]
&\ovalgate{\hatpzy'}\qwx[1]
&\qw
\\
&\gate{U'}
&\ovalgate{\hatx}
&\ovalgate{\hatpzy}
&\qw
}
\end{array}
\label{eq-2thirds-b}
\;,
\end{eqnarray}
\end{subequations}
where (see Fig.\ref{fig-qxy-pzy})

\beq
\hatpzy = c_\alpha \hatz + s_\alpha \haty
\;,\;\;
\hatpzy' = (\hatpzy)_{\alpha\rarrow \alpha'}
\;,
\eeq

\beq
\hatqxy = c_\alpha \hatx - s_\alpha \haty
\;,\;\;
\hatqxy' = (\hatqxy)_{\alpha\rarrow \alpha'}
\;,
\eeq
($p$ vector has a positive sign
in front of $s_\alpha$,
$q$ vector has a negative one)
and

\beq
U = e^{i \frac{\alpha}{2} \sigz}
e^{-i \frac{\alpha'}{2} \sigx}
\;,\;\;
U' = (U)_{\alpha\darrow \alpha'}
\;.
\label{eq-2thirds-u-def}
\eeq
Note that the left-hand sides
of Eqs.(\ref{eq-2thirds-a})
and (\ref{eq-2thirds-b}) are independent
of the two angles $\alpha$ and $\alpha'$;
only their
right-hand sides depend on these angles.
\end{theo}
\proof

From the expression of Swapper as
a product of 3 CNOTs, we get
\beq
\begin{array}{c}
\Qcircuit @C=1em @R=1em @!R{
&\ovalgate{\hatz}\qwx[1]
&\ovalgate{\hatx}\qwx[1]
&\ovalgate{\hatz}\qwx[1]
&\qw
\\
&\ovalgate{\hatz}
&\ovalgate{\hatx}
&\ovalgate{\hatz}
&\qw
}
\end{array}
=
\begin{array}{c}
\Qcircuit @C=1em @R=1em @!R{
&\uarrowgate\qwx[1]
&\gate{H}
&\qw
\\
&\darrowgate
&\gate{H}
&\qw
}
\end{array}
\;.
\eeq
From Fig.\ref{fig-qxy-pzy},
it follows that
\beq
\sigma_{\hatqxy} =
e^{i \frac{\alpha}{2} \sigz}
\sigma_\hatx
e^{-i \frac{\alpha}{2} \sigz}
\;.
\eeq
Thus

\beqa
\lefteqn{
\begin{array}{c}
\Qcircuit @C=1em @R=1em @!R{
&\ovalgate{\hatz}\qwx[1]
&\ovalgate{\hatqxy}\qwx[1]
&\ovalgate{\hatx}\qwx[1]
&\ovalgate{\hatz}\qwx[1]
&\qw
\\
&\ovalgate{\hatz}
&\ovalgate{\hatqxy'}
&\ovalgate{\hatx}
&\ovalgate{\hatz}
&\qw
}
\end{array}
=
\begin{array}{c}
\Qcircuit @C=1em @R=1em @!R{
&\ovalgate{\hatz}\qwx[1]
&\ovalgate{\hatqxy}\qwx[1]
&\ovalgate{\hatz}\qwx[1]
&\uarrowgate\qwx[1]
&\gate{H}
&\qw
\\
&\ovalgate{\hatz}
&\ovalgate{\hatqxy'}
&\ovalgate{\hatz}
&\darrowgate
&\gate{H}
&\qw
}
\end{array}
}
\\
&=&
\begin{array}{c}
\Qcircuit @C=1em @R=1em @!R{
&\gate{e^{i \frac{\alpha}{2} \sigz}}
&\ovalgate{\hatz}\qwx[1]
&\ovalgate{\hatx}\qwx[1]
&\ovalgate{\hatz}\qwx[1]
&\gate{e^{-i \frac{\alpha}{2} \sigz}}
&\uarrowgate\qwx[1]
&\gate{H}
&\qw
\\
&\gate{e^{i \frac{\alpha'}{2} \sigz}}
&\ovalgate{\hatz}
&\ovalgate{\hatx}
&\ovalgate{\hatz}
&\gate{e^{-i \frac{\alpha'}{2} \sigz}}
&\darrowgate
&\gate{H}
&\qw
}
\end{array}
\\
&=&
\begin{array}{c}
\Qcircuit @C=1em @R=1em @!R{
&\gate{e^{i \frac{\alpha}{2} \sigz}}
&\uarrowgate\qwx[1]
&\gate{H}
&\gate{e^{-i \frac{\alpha}{2} \sigz}}
&\uarrowgate\qwx[1]
&\gate{H}
&\qw
\\
&\gate{e^{i \frac{\alpha'}{2} \sigz}}
&\darrowgate
&\gate{H}
&\gate{e^{-i \frac{\alpha'}{2} \sigz}}
&\darrowgate
&\gate{H}
&\qw
}
\end{array}
\\
&=&
\begin{array}{c}
\Qcircuit @C=1em @R=1em @!R{
&\freegate{
e^{i \frac{\alpha}{2} \sigz}
e^{-i \frac{\alpha'}{2} \sigx}
}
\\
&\freegate{
e^{i \frac{\alpha'}{2} \sigz}
e^{-i \frac{\alpha}{2} \sigx}
}
}
\end{array}
\;.
\eeqa
\qed

The next theorem follows immediately
from the previous one, by a change of basis.

\begin{theo}\label{th-2thirds-ab}
Suppose $\alpha\in \RR$,
$\hata\perp\hatb$, and $\hata'\perp\hatb'$. Then

\beqa
\begin{array}{c}
\Qcircuit @C=1em @R=1em @!R{
&\ovalgate{\hatb}\qwx[1]
&\foil
&\ovalgate{\hata}\qwx[1]
&\qw
\\
&\ovalgate{\hatb'}
&\foil
&\ovalgate{\hata'}
&\qw
}
\end{array}
&=&
\begin{array}{c}
\Qcircuit @C=1em @R=1em @!R{
&\ovalgate{\hatb_f}\qwx[1]
&\foil
&\ovalgate{\hata}\qwx[1]
&\gate{U}
&\qw
\\
&\ovalgate{\hatb'_f}
&\foil
&\ovalgate{\hata'}
&\gate{U'}
&\qw
}
\end{array}
\\
&=&
\begin{array}{c}
\Qcircuit @C=1em @R=1em @!R{
&\gate{U}
&\ovalgate{\hatb}\qwx[1]
&\foil
&\ovalgate{\hata'_f}\qwx[1]
&\qw
\\
&\gate{U'}
&\ovalgate{\hatb'}
&\foil
&\ovalgate{\hata_f}
&\qw
}
\end{array}
\;,
\eeqa
where

\beq
\hata_f = c_\alpha \hata + s_\alpha \manyx{\hata\hatb}
\;,\;\;
\hata'_f = c_{\alpha'} \hata' + s_{\alpha'} \manyx{\hata'\hatb'}
\;,
\eeq

\beq \hatb_f = c_\alpha \hatb - s_\alpha \manyx{\hata\hatb}
\;,\;\; \hatb'_f = c_{\alpha'} \hatb' - s_{\alpha'}
\manyx{\hata'\hatb'}
\;,
\eeq
and

\beq
U = e^{i \frac{\alpha}{2} \sigma_\hata}
e^{-i \frac{\alpha'}{2} \sigma_\hatb}
\;,\;\;
U' = e^{i \frac{\alpha'}{2} \sigma_{\hata'}}
e^{-i \frac{\alpha}{2} \sigma_{\hatb'}}
\;.
\eeq
\end{theo}
\proof

Just change basis in the space
where bit 0 (ditto, bit 1) lives so that
 $(\hatx,\haty,\hatz)$
is replaced by
$(\hatb,\manyx{\hata\hatb},\hata)$
(ditto, $(\hatb',\manyx{\hata'\hatb'},\hata')$).

\qed

We will refer to the next identity,
Eq.(\ref{eq-2thirds-split}),
as the 1/3 Swapper identity.

\begin{theo}\label{th-2thirds-split}
\beq
\begin{array}{c}
\Qcircuit @C=1em @R=1em @!R{
&\ovalgate{\hatx}\qwx[1]
&\ovalgate{\hatqxy}\qwx[1]
&\qw
\\
&\ovalgate{\hatx}
&\ovalgate{\hatqxy'}
&\qw
}
\end{array}
=
\begin{array}{c}
\Qcircuit @C=1em @R=1em @!R{
&\ovalgate{\hatz}\qwx[1]
&\ovalgate{\hatpzy'}\qwx[1]
&\gate{U^{\dagger}}
&\qw
\\
&\ovalgate{\hatz}
&\ovalgate{\hatpzy}
&\gate{U^{'\dagger}}
&\qw
}
\end{array}
\;,
\label{eq-2thirds-split}
\eeq
where all variables are defined as in
Theorem \ref{th-2thirds}.
\end{theo}
\proof

From the Hermitian conjugate of
Eq.(\ref{eq-2thirds-a}), one gets
\beq
\begin{array}{c}
\Qcircuit @C=1em @R=1em @!R{
&\ovalgate{\hatz}\qwx[1]
&\ovalgate{\hatx}\qwx[1]
&\ovalgate{\hatqxy}\qwx[1]
&\qw
\\
&\ovalgate{\hatz}
&\ovalgate{\hatx}
&\ovalgate{\hatqxy'}
&\qw
}
\end{array}
=
\begin{array}{c}
\Qcircuit @C=1em @R=1em @!R{
&\gate{U^\dagger}\qwx[1]
&\ovalgate{\hatz}\qwx[1]
&\qw
\\
&\gate{U^{'\dagger}}
&\ovalgate{\hatz}
&\qw
}
\end{array}
\;.
\eeq
Let $LHS$ and $RHS$ stand for the
left and right hand sides of
Eq.(\ref{eq-2thirds-split}).
Pre-multiplying
 both sides of the last
equation by
$
\begin{array}{c}
\Qcircuit @C=1em @R=1em @!R{
&\freeovalgate{\hatz}\qwx[1]
\\
&\freeovalgate{\hatz}
}
\end{array}
$
yields

\beq
LHS =
\begin{array}{c}
\Qcircuit @C=1em @R=1em @!R{
&\ovalgate{\hatz}\qwx[1]
&\gate{U^\dagger}
&\ovalgate{\hatz}\qwx[1]
&\qw
\\
&\ovalgate{\hatz}
&\gate{U^{'\dagger}}
&\ovalgate{\hatz}
&\qw
}
\end{array}
=
RHS
\;.
\eeq

\qed
\section{DC-NOT Similarity Transformation Identities
\\{\footnotesize\tt[
sim\_trans\_t4.m,
test\_sim\_trans\_t4.m
]}}
\label{sec-sim-trans-ids}

In this section, we present
some identities which contain
 a similarity transformation
produced by pre and post multiplying
an operator by the same DC-NOT.

We will refer to the next theorem as
the DC-NOT similarity transformation identity.

\begin{theo}\label{th-sim-trans}
For any $\alpha, \lam\in \RR$,

\beq
\begin{array}{c}
\Qcircuit @C=1em @R=1em @!R{
&\ovalgate{\hatx}\qwx[1]
&\gate{
c_\alpha \sigx + s_\alpha \sigz
}
&\ovalgate{\hatx}\qwx[1]
&\qw
\\
&\ovalgate{\hatx}
&\gate{
s_\alpha \sigx + c_\alpha \sigz
}
&\ovalgate{\hatx}
&\qw
}
\end{array}
=
\begin{array}{c}
\Qcircuit @C=1em @R=1em @!R{
&\ovalgate{\hatqxy}\qwx[1]
&\gate{
c_\alpha \sigma_\hatqxy + s_\alpha \sigz
}
&\ovalgate{\hatqxy}\qwx[1]
&\qw
\\
&\ovalgate{\hatqxy}
&\gate{
s_\alpha \sigma_\hatqxy + c_\alpha \sigz
}
&\ovalgate{\hatqxy}
&\qw
}
\end{array}
\;,
\label{eq-sim-trans-id}
\eeq
where $\hatqxy = c_\lam \hatx - s_\lam \haty$.
Note that the LHS of Eq.(\ref{eq-sim-trans-id})
 equals its RHS evaluated at $\lam=0$.
\end{theo}
\proof

 Since

\beq
 \manyx{\hatqxy\hatz} =
 -(c_\lam\haty + s_\lam \hatx)
\;,
\eeq
it follows that

\beqa
\sigma_{\hatqxy\hatqxy}
+ \sigma_{\manyx{\hatqxy\hatz}\manyx{\hatqxy\hatz}}
&=&
\sigma_{
c_\lam\hatx - s_\lam \haty,
c_\lam\hatx - s_\lam \haty
}
+ \sigma_{
c_\lam\haty + s_\lam \hatx,
c_\lam\haty + s_\lam \hatx
}
\\
&=&
\sigma_{\hatx\hatx} + \sigma_{\haty\haty}
\;.
\eeqa
Let LHS and RHS denote the left-hand side
and right-hand side, respectively,
of Eq.(\ref{eq-sim-trans-id}). Then,
using Eq.(\ref{eq-sim-trans-of-sig-veca}),

\beqa
RHS &=&
\begin{array}{c}
\Qcircuit @C=1em @R=1em @!R{
&\freeovalgate{\hatqxy}\qwx[1]
\\
&\freeovalgate{\hatqxy}
}
\end{array}
(s_\alpha \sigma_{\hatq_{xy},1}
+c_\alpha\sigma_{\hatz,1})
\begin{array}{c}
\Qcircuit @C=1em @R=1em @!R{
&\freeovalgate{\hatqxy}\qwx[1]
\\
&\freeovalgate{\hatqxy}
}
\end{array}
\begin{array}{c}
\Qcircuit @C=1em @R=1em @!R{
&\freeovalgate{\hatqxy}\qwx[1]
\\
&\freeovalgate{\hatqxy}
}
\end{array}
(c_\alpha \sigma_{1,\hatq_{xy}}
+s_\alpha\sigma_{1,\hatz})
\begin{array}{c}
\Qcircuit @C=1em @R=1em @!R{
&\freeovalgate{\hatqxy}\qwx[1]
\\
&\freeovalgate{\hatqxy}
}
\end{array}
\\
&=&
(s_\alpha \sigma_{\hatqxy1}
+ c_\alpha \sigma_{\hatz\hatqxy})
(c_\alpha \sigma_{1,\hatqxy}
+ s_\alpha \sigma_{\hatqxy\hatz})
\\
&=&
s_\alpha c_\alpha
(\sigma_{\hatqxy\hatqxy}
+ \sigma_{\manyx{\hatqxy\hatz}\manyx{\hatqxy\hatz}})
+ c^2_\alpha \sigma_{\hatz 1}
+ s^2_\alpha \sigma_{1\hatz}
\\
&=&
s_\alpha c_\alpha
(\sigma_{\hatx\hatx}
+ \sigma_{\haty\haty})
+ c^2_\alpha \sigma_{\hatz 1}
+ s^2_\alpha \sigma_{1\hatz}
\\
&=& LHS
\;.
\eeqa
\qed

It is convenient to define, for any $\xi\in \RR$,

\beq
\hatp_{w_1,w_2}^\xi = c_\xi\hatw_1 + s_\xi\hatw_2
\;,\;\;
\hatq_{w_1,w_2}^\xi = c_\xi\hatw_1 - s_\xi\hatw_2
\;.
\label{eq-general-p-q-def}
\eeq
(The $\hatp$ vectors have a positive sign
in front of the sine function whereas the $\hatq$
vectors have a negative one).

\begin{figure}[h]
    \begin{center}
    \epsfig{file=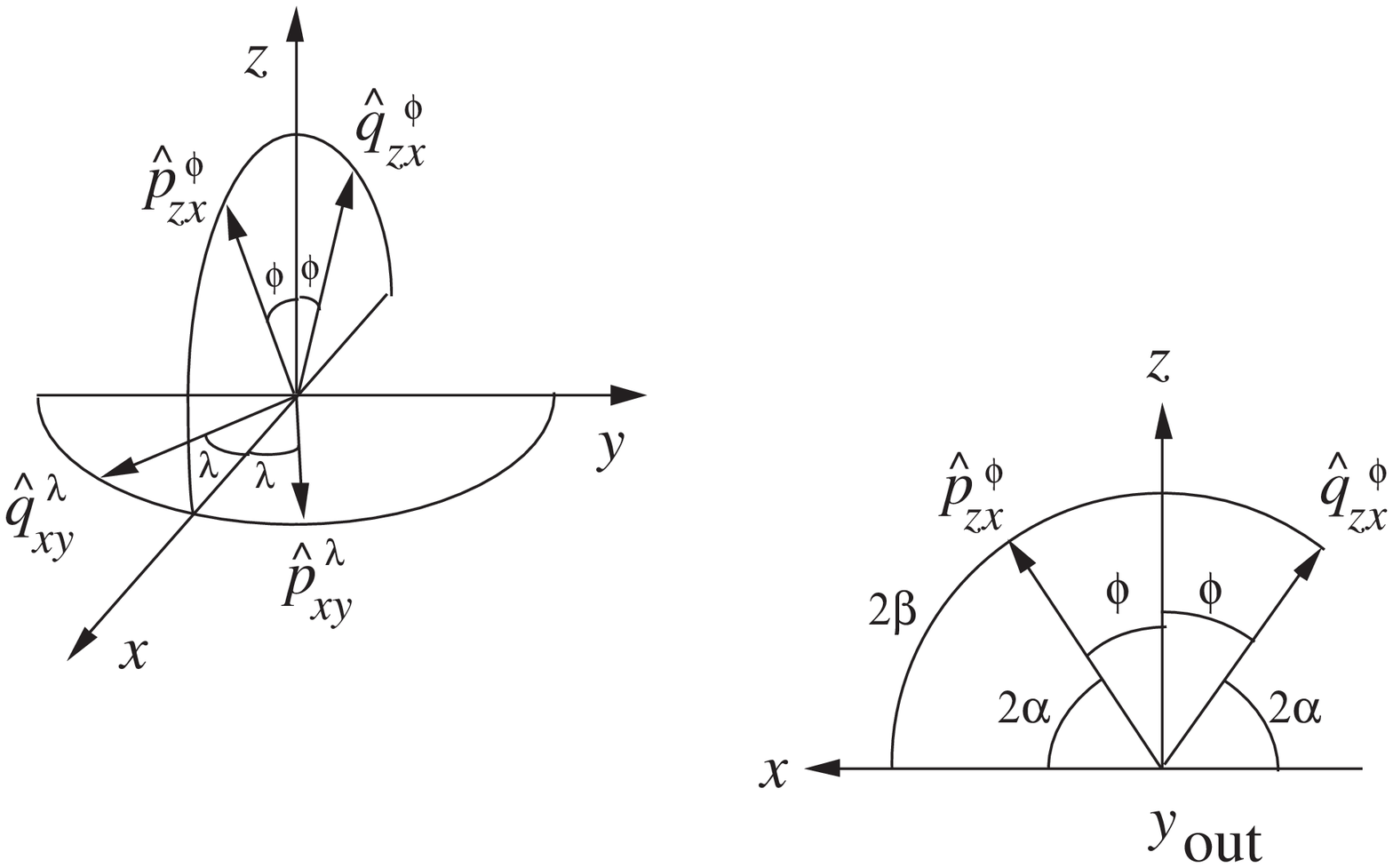, height=2.75in}
    \caption{Variables used in
    Theorem \ref{th-split-sim-trans}.}
    \label{fig-split-sim-trans}
    \end{center}
\end{figure}

The next theorem follows from
the DC-NOT similarity transformation identity.

\begin{theo}\label{th-split-sim-trans}

For any $\phi,\lam\in \RR$,

\beq
\begin{array}{c}
\Qcircuit @C=1em @R=1em @!R{
&\ovalgate{\hatpzx^\phi}\qwx[1]
&\ovalgate{\hatx}\qwx[1]
&\ovalgate{\hatqxy^\lam}\qwx[1]
&\qw
\\
&\ovalgate{\hatqzx^\phi}
&\ovalgate{\hatx}
&\ovalgate{\hatqxy^\lam}
&\qw
}
\end{array}
=
\begin{array}{c}
\Qcircuit @C=1em @R=1em @!R{
&\ovalgate{\hata_f}\qwx[1]
&\gate{U}
&\qw
\\
&\ovalgate{\hata'_f}
&\gate{U'}
&\qw
}
\end{array}
\;,
\label{eq-split-sim-trans-id}
\eeq
where (see Fig.\ref{fig-split-sim-trans})

\beq
\hata_f = c_\lam \hatpzx^\phi + s_\lam \haty
\;,\;\;
\hata'_f = c_\lam \hatqzx^\phi + s_\lam \haty
\;,
\eeq
and

\beq
U = (c_\alpha \sigx + s_\alpha\sigz)
(c_\alpha \sigma_{\hatqxy^\lam} + s_\alpha\sigz)
\;,\;\;
U'=(U)_{\alpha\rarrow \beta}
\;,
\eeq
where

\beq
2\alpha = \frac{\pi}{2} - \phi
\;,\;\;
2\beta = \pi -2\alpha
\;.
\eeq

\end{theo}
\proof

From Fig.\ref{fig-split-sim-trans},
it follows that

\beq
\hatpzx^\phi =
e^{i\alpha\sigy}\sigx e^{-i\alpha\sigy}
\;,
\eeq
and

\beq
\hatqzx^\phi =
e^{i\beta\sigy}\sigx e^{-i\beta\sigy}
\;.
\eeq
Let LHS and RHS denote the left-hand side and
right-hand side, respectively,
of Eq.(\ref{eq-split-sim-trans-id}). Then

\beq
LHS =
\begin{array}{c}
\Qcircuit @C=1em @R=1em @!R{
&\gate{e^{i\alpha\sigy}}
&\ovalgate{\hatx}\qwx[1]
&\gate{e^{-i\alpha\sigy}}
&\ovalgate{\hatx}\qwx[1]
&\ovalgate{\hatqxy^\lam}\qwx[1]
&\qw
\\
&\gate{ e^{i\beta\sigy}}
&\ovalgate{\hatx}
&\gate{e^{-i\beta\sigy}}
&\ovalgate{\hatx}
&\ovalgate{\hatqxy^\lam}
&\qw
}
\end{array}
\;,
\eeq
and

\beq
RHS =
\begin{array}{c}
\Qcircuit @C=1em @R=1em @!R{
&\gate{e^{i\alpha\sigy}}
&\ovalgate{\hatpxy^\lam}\qwx[1]
&\gate{ e^{-i\alpha\sigy}}
&\gate{U}
&\qw
\\
&\gate{e^{i\beta\sigy}}
&\ovalgate{\hatpxy^\lam}
&\gate{ e^{-i\beta\sigy}}
&\gate{U'}
&\qw
}
\end{array}
\;.
\eeq
Therefore, Eq.(\ref{eq-split-sim-trans-id}) is
equivalent to the assertion that

\beq
\begin{array}{c}
\Qcircuit @C=1em @R=1em @!R{
&\ovalgate{\hatx}\qwx[1]
&\gate{e^{-i\alpha\sigy}}
&\ovalgate{\hatx}\qwx[1]
&\qw
\\
&\ovalgate{\hatx}
&\gate{e^{-i\beta\sigy}}
&\ovalgate{\hatx}
&\qw
}
\end{array}
=
\begin{array}{c}
\Qcircuit @C=1em @R=1em @!R{
&\ovalgate{\hatpxy^\lam}\qwx[1]
&\gate{e^{-i\alpha\sigy}U}
&\ovalgate{\hatqxy^\lam}\qwx[1]
&\qw
\\
&\ovalgate{\hatpxy^\lam}
&\gate{e^{-i\beta\sigy} U'}
&\ovalgate{\hatqxy^\lam}
&\qw
}
\end{array}
\;.
\eeq
Now pre-multiply each side of the last
equation by $\sigxx$

\beq
\begin{array}{c}
\Qcircuit @C=1em @R=1em @!R{
&\ovalgate{\hatx}\qwx[1]
&\gate{c_\alpha\sigx + s_\alpha\sigz}
&\ovalgate{\hatx}\qwx[1]
&\qw
\\
&\ovalgate{\hatx}
&\gate{c_\beta\sigx + s_\beta\sigz}
&\ovalgate{\hatx}
&\qw
}
\end{array}
=
\begin{array}{c}
\Qcircuit @C=1em @R=1em @!R{
&\ovalgate{\hatqxy^\lam}\qwx[1]
&\gate{c_\alpha\sigma_{\hatqxy^\lam} + s_\alpha\sigz}
&\ovalgate{\hatqxy^\lam}
&\qw
\\
&\ovalgate{\hatqxy^\lam}
&\gate{c_\beta\sigma_{\hatqxy^\lam} + s_\beta\sigz}
&\ovalgate{\hatqxy^\lam}
&\qw
}
\end{array}
\;.
\eeq
The preceding equation
follows from Theorem \ref{th-sim-trans}
and the fact that $\alpha + \beta = \pi/2$.

\qed

The next theorem is a simple
variation of the previous one.
(The left-hand sides
 of Eqs.(\ref{eq-split-sim-trans-id})
and (\ref{eq-split-sim-trans-id2})
differ only in that one circuit has two
$q$'s in the
bit-1 wire
whereas the other circuit has two
$p$'s.)

\begin{theo}\label{th-split-sim-trans2}

\beq
\begin{array}{c}
\Qcircuit @C=1em @R=1em @!R{
&\ovalgate{\hatpzx^\phi}\qwx[1]
&\ovalgate{\hatx}\qwx[1]
&\ovalgate{\hatqxy^\lam}\qwx[1]
&\qw
\\
&\ovalgate{\hatpzx^\phi}
&\ovalgate{\hatx}
&\ovalgate{\hatpxy^\lam}
&\qw
}
\end{array}
=
\begin{array}{c}
\Qcircuit @C=1em @R=1em @!R{
&\gate{\sigz}
&\ovalgate{\hata'_f}\qwx[1]
&\gate{U'\sigz}
&\qw
\\
&\qw
&\ovalgate{\hata_f}
&\gate{U\sigma_{\hatqxy^\lam}\sigx}
&\qw
}
\end{array}
\;,
\label{eq-split-sim-trans-id2}
\eeq
where all variables are
defined as in Theorem \ref{th-split-sim-trans}.
\end{theo}
\proof

Let $LHS_{\ref{eq-split-sim-trans-id}}$
represent the left-hand side
of Eq.(\ref{eq-split-sim-trans-id}),
and $LHS_{\ref{eq-split-sim-trans-id2}}$, the left-hand
side of Eq.(\ref{eq-split-sim-trans-id2}).
Then

\beqa
\begin{array}{c}
\Qcircuit @C=1em @R=1em @!R{
&\qw
&\qw
\\
&\gate{\sigz}
&\qw
}
\end{array}
LHS_{\ref{eq-split-sim-trans-id}}
\begin{array}{c}
\Qcircuit @C=1em @R=1em @!R{
&\gate{\sigma_{\hatqxy^\lam}\sigx}
&\qw
\\
&\gate{\sigz}
&\qw
}
\end{array}
&=&
\begin{array}{c}
\Qcircuit @C=1em @R=1em @!R{
&\ovalgate{\hatpzx^\phi}\qwx[1]
&\ovalgate{\hatx}\qwx[1]
&\ovalgate{\hatqxy^\lam}\qwx[1]
&\gate{\sigma_{\hatqxy^\lam}\sigx}
&\qw
\\
&\ovalgate{\hatpzx^\phi}
&\ovalgate{-\hatx}
&\ovalgate{-\hatqxy^\lam}
&\qw
}
\end{array}
\\
&=&
\begin{array}{c}
\Qcircuit @C=1em @R=1em @!R{
&\ovalgate{\hatpzx^\phi}\qwx[1]
&\ovalgate{\hatx}\qwx[1]
&\gate{\sigx}
&\ovalgate{\hatqxy^\lam}\qwx[1]
&\gate{(\sigma_{\hatqxy^\lam})^2\sigx}
&\qw
\\
&\ovalgate{\hatpzx^\phi}
&\ovalgate{\hatx}
&\qw
&\ovalgate{\hatqxy^\lam}
&\qw
}
\end{array}
\\
&=&
\begin{array}{c}
\Qcircuit @C=1em @R=1em @!R{
&\uarrowgate\qwx[1]
&\qw
\\
&\darrowgate
&\qw
}
\end{array}
LHS_{\ref{eq-split-sim-trans-id2}}
\begin{array}{c}
\Qcircuit @C=1em @R=1em @!R{
&\uarrowgate\qwx[1]
&\qw
\\
&\darrowgate
&\qw
}
\end{array}
\;.
\eeqa
The right-hand sides of
Eqs.(\ref{eq-split-sim-trans-id})
and
(\ref{eq-split-sim-trans-id2})
must be related
in the same way as their left-hand sides.

\qed
\section{LO-RHS Invariant for Circuits with\\
Two Qubits, and Multiple DC-NOTs}
\label{sec-two-bit-dcnot-rhs-invariants}

In previous sections we defined
the LO-RHS invariant $A^{(2)}$
for  any $A\in U(\ns)$.
We also defined DC-NOTs and discussed
some of their properties. In this section,
we combine these two concepts:
we  calculate $A^{(2)}$
when $A$ is a product of one or more DC-NOTs
acting on the same two qubits.

Henceforth, we will denote
the product of $r$
DC-NOTs (all acting on the same two qubits)
by the symbol
$\calg_r$ followed by a list
(enclosed in parenthesis) of
its arguments. Sometimes,
if this doesn't lead to confusion,
its list
of arguments will be omitted. Thus,

\beq
\calg_r
\left(
\begin{array}{cccc}
\hata_r & \cdots & \hata_2 & \hata_1\\
\hata'_r & \cdots & \hata'_2 & \hata'_1
\end{array}
\right)=
\begin{array}{c}
\Qcircuit @C=1em @R=1em @!R{
&\ovalgate{\hata_r}\qwx[1]
&\qw
\\
&\ovalgate{\hata'_r}
&\qw
}
\end{array}
\cdots
\begin{array}{c}
\Qcircuit @C=1em @R=1em @!R{
&\ovalgate{\hata_2}\qwx[1]
&\ovalgate{\hata_1}\qwx[1]
&\qw
\\
&\ovalgate{\hata'_2}
&\ovalgate{\hata'_1}
&\qw
}
\end{array}
\;.
\eeq

The determinant of $\calg_r$
equals either plus or minus one. Indeed,

\beq
\det
\left(
\begin{array}{c}
\Qcircuit @C=1em @R=1em @!R{
&\ovalgate{\hata}\qwx[1]
&\qw
\\
&\ovalgate{\hata'}
&\qw
}
\end{array}
\right)
=
\det
\left(
\begin{array}{c}
\Qcircuit @C=1em @R=1em @!R{
&\timesgate\qwx[1]
&\qw
\\
&\dotgate
&\qw
}
\end{array}
\right)
=\det
\left[
\begin{array}{cc}
I_2 & 0\\
0   & \sigx
\end{array}
\right]
=
-1
\;.
\eeq
Since $\det(AB)=\det(A)\det(B)$, it follows that
for $r=\ZZ^{>0}$,

\beq
\det(\calg_r)= (-1)^r
\;.
\eeq

It is convenient to define a matrix
 $\hat{\calg}_r$ by

\beq
\hat{\calg}_r = (-1)^{\frac{r}{4}}\calg_r =
i^{\frac{r}{2}}\calg_r
\;.
\eeq
Henceforth, we will refer to
$\hat{\calg}_r$ as the
{\bf special counterpart} of $\calg_r$.
(Here the adjective ``special"
means ``having unit determinant").
$\hat{\calg}_r\in U(4)$ and
$\det(\hat{\calg}_r)=1$, so $\hat{\calg}_r\in SU(4)$.

Since $\sigy \sigma_\hata^T \sigy =
\sigma_{-\hata}$,

\beqa
\calg^{(2)}_r &=&
\calg_r \sigyy \calg_r^T \sigyy
\\
&=&
\begin{array}{c}
\Qcircuit @C=1em @R=1em @!R{
&\ovalgate{\hata_{r}}\qwx[1]
&\qw
\\
&\ovalgate{\hata'_{r}}
&\qw
}
\end{array}
\cdots
\begin{array}{c}
\Qcircuit @C=1em @R=1em @!R{
&\ovalgate{\hata_{2}}\qwx[1]
&\ovalgate{\hata_{1}}\qwx[1]
&\ovalgate{-\hata_{1}}\qwx[1]
&\ovalgate{-\hata_{2}}\qwx[1]
&\qw
\\
&\ovalgate{\hata'_{2}}
&\ovalgate{\hata'_{1}}
&\ovalgate{-\hata'_{1}}
&\ovalgate{-\hata'_{2}}
&\qw
}
\end{array}
\cdots
\begin{array}{c}
\Qcircuit @C=1em @R=1em @!R{
&\ovalgate{-\hata_{r}}\qwx[1]
&\qw
\\
&\ovalgate{-\hata'_{r}}
&\qw
}
\end{array}
\;.
\eeqa
For $r\in \ZZ^{>0}$,
$\calg^{(2)}_r$
obeys the following
recursion relation:

\beq
\calg^{(2)}_{r+1}
=
\begin{array}{c}
\Qcircuit @C=1em @R=1em @!R{
&\ovalgate{\hata_{r+1}}\qwx[1]
&\qw
\\
&\ovalgate{\hata'_{r+1}}
&\qw
}
\end{array}
\calg_r^{(2)}
\begin{array}{c}
\Qcircuit @C=1em @R=1em @!R{
&\ovalgate{-\hata_{r+1}}\qwx[1]
&\qw
\\
&\ovalgate{-\hata'_{r+1}}
&\qw
}
\end{array}
\;.
\eeq

Note that the LO-RHS invariants of
$\calg_r$ and of its
special counterpart $\hat{\calg}_r$
are related by

\beq
\hat{\calg}_r^{(2)} =
i^r \calg_r^{(2)}
\;.
\label{eq-invar-of-hat-graph}
\eeq

The remainder of Section
\ref{sec-two-bit-dcnot-rhs-invariants} consists
of 4 subsections which
give explicit formulas for
$\calg^{(2)}_r$ for $r$
from 1 to 4. These 4 subsections
are very useful, but
make for dry reading when considered in isolation;
they only come alive and prove their mettle as we
start using them in subsequent sections.
Thus, the reader is advised  not to spend
too much time on them during his first reading
of this paper. He should skim the 4 subsections, and
then come back to them as the need
arises.
\subsection{Invariant for Circuits with
1 DC-NOT
\\{\footnotesize\tt[
ckt\_invar123.m
]}}

\label{sec-invariants-1cnot}

This part of our program is dedicated to the letters
$\calg^{(2)}_1$.

\begin{theo}
\beq
\calg^{(2)}_1 =
\begin{array}{c}
\Qcircuit @C=1em @R=1em @!R{
&\ovalgate{\hata}\qwx[1]
&\ovalgate{-\hata}\qwx[1]
&\qw
\\
&\ovalgate{\hata'}
&\ovalgate{-\hata'}
&\qw
}
\end{array}
=
-
\sigma_{\hata',\hata}
\;.
\label{eq-invariant-1bit}
\eeq
\end{theo}
\proof

\beq
\begin{array}{c}
\Qcircuit @C=1em @R=1em @!R{
&\ovalgate{\hata}\qwx[1]
&\ovalgate{-\hata}\qwx[1]
&\qw
\\
&\ovalgate{\hata'}
&\ovalgate{-\hata'}
&\qw
}
\end{array}
=
\begin{array}{c}
\Qcircuit @C=1em @R=1em @!R{
&\ovalgate{\hata}\qwx[1]
&\ovalgate{-\hata}\qwx[1]
&\qw
&\gate{\sigma_{-\hata}}
\\
&\ovalgate{\hata'}
&\ovalgate{\hata'}
&\qw
&\qw
}
\end{array}
=
\begin{array}{c}
\Qcircuit @C=1em @R=1em @!R{
&\freegate{\sigma_{-\hata}}
\\
&\freegate{\sigma_{\hata'}}
}
\end{array}
\;.
\eeq

\qed
\subsection{Invariant for Circuits with
 2 DC-NOTs
 \\{\footnotesize\tt[
ckt\_invar123.m,
diag\_ckt\_invar2.m,
diag\_ckt\_invar2\_aux.m,\\
test\_diag\_invar2.m
]}}

\label{sec-invariants-2cnots}

This part of our program is dedicated to the letters
$\calg^{(2)}_2$.

\begin{theo}\label{th-invar2}
\beqa
\calg^{(2)}_2 &=&
\begin{array}{c}
\Qcircuit @C=1em @R=1em @!R{
&\ovalgate{\hatb}\qwx[1]
&\ovalgate{\hata}\qwx[1]
&\ovalgate{-\hata}\qwx[1]
&\ovalgate{-\hatb}\qwx[1]
&\qw
\\
&\ovalgate{\hatb'}
&\ovalgate{\hata'}
&\ovalgate{-\hata'}
&\ovalgate{-\hatb'}
&\qw
}
\end{array}\\
&=&
\lam_{2r} + i\lam_{2i} + \Lam_{2r} + i\Lam_{2i}
\;,
\label{eq-invariant-2bit}
\eeqa
where

\beq
\lam_{2r}= (\hata\cdot\hatb)(\hata'\cdot\hatb')
\;,
\eeq

\beq
\lam_{2i}= 0
\;,
\eeq

\beq
\Lam_{2r}=
-\sigma_{\manyx{\hata'\hatb'\hatb'},\manyx{\hata\hatb\hatb}}
\;,
\label{eq-def-Lam2r}
\eeq

\beq
\Lam_{2i}=
\hata\cdot\hatb\sigma_{\hata'\times\hatb',\hatb}
+
\hata'\cdot\hatb'\sigma_{\hatb',\hata\times\hatb}
\;.
\label{eq-def-Lam2i}
\eeq
\end{theo}

\proof

An explicit expression
for $\calg^{(2)}_1$ was given
in Section \ref{sec-invariants-1cnot}.
Eq.(\ref{eq-sim-trans-of-sig-veca})
shows how to calculate the effect of DC-NOT
similarity transformations.
Using these two results,
one gets

\beqa
\calg^{(2)}_2 &=&
\begin{array}{c}
\Qcircuit @C=1em @R=1em @!R{
&\freeovalgate{\hatb}\qwx[1]
\\
&\freeovalgate{\hatb'}
}
\end{array}
[-\sigma_{\hata',\hata}]
\begin{array}{c}
\Qcircuit @C=1em @R=1em @!R{
&\freeovalgate{-\hatb}\qwx[1]
\\
&\freeovalgate{-\hatb'}
}
\end{array}
\\
&=&
\begin{array}{c}
\Qcircuit @C=1em @R=1em @!R{
&\freeovalgate{\hatb}\qwx[1]
\\
&\freeovalgate{\hatb'}
}
\end{array}
\sigma_{\hata',1}
\begin{array}{c}
\Qcircuit @C=1em @R=1em @!R{
&\freeovalgate{\hatb}\qwx[1]
\\
&\freeovalgate{\hatb'}
}
\end{array}
\begin{array}{c}
\Qcircuit @C=1em @R=1em @!R{
&\freeovalgate{\hatb}\qwx[1]
\\
&\freeovalgate{\hatb'}
}
\end{array}
\sigma_{1,\hata}
\begin{array}{c}
\Qcircuit @C=1em @R=1em @!R{
&\freeovalgate{\hatb}\qwx[1]
\\
&\freeovalgate{\hatb'}
}
\end{array}
\begin{array}{c}
\Qcircuit @C=1em @R=1em @!R{
&\freeovalgate{\hatb}\qwx[1]
&\freeovalgate{-\hatb}\qwx[1]
\\
&\freeovalgate{\hatb'}
&\freeovalgate{-\hatb'}
}
\end{array}
(-1)
\\
&=&
(\sigma_{\along{\hata'}{\hatb'},1}+
\sigma_{\across{\hata'}{\hatb'},\hatb})
(\sigma_{1,\along{\hata}{\hatb}}+
\sigma_{\hatb,\across{\hata}{\hatb}})
\sigma_{\hatb',\hatb}
\;
\\
&=&
\left\{
\begin{array}{l}
(\hata'\cdot\hatb')(\hata\cdot\hatb)\\
-\sigma_{\manyx{\hata'\hatb'\hatb'},
\manyx{\hata\hatb\hatb}}\\
+i\left[
\hata\cdot\hatb\sigma_{\hata'\times\hatb',\hatb}
+
\hata'\cdot\hatb'\sigma_{\hatb',\hata\times\hatb}
\right]
\end{array}
\right.
\;.
\eeqa
\qed

\begin{theo}
\beq
[\Lam_{2r},\Lam_{2i}]=0
\;,
\eeq

\beq
(\Lam_{2r}^\Gamma)^T \Lam_{2i}^\Gamma=0
\;,
\eeq

\beq
\Lam_{2r}^\Gamma (\Lam_{2i}^\Gamma)^T=0
\;.
\eeq

\end{theo}
\proof

This follows easily from
Eqs.(\ref{eq-def-Lam2r}) and (\ref{eq-def-Lam2i}).

\qed

It is convenient to parameterize
the expression for $\calg^{(2)}_2$
given by Theorem \ref{th-invar2},
using as few parameters as possible.

\begin{figure}[h]
    \begin{center}
    \epsfig{file=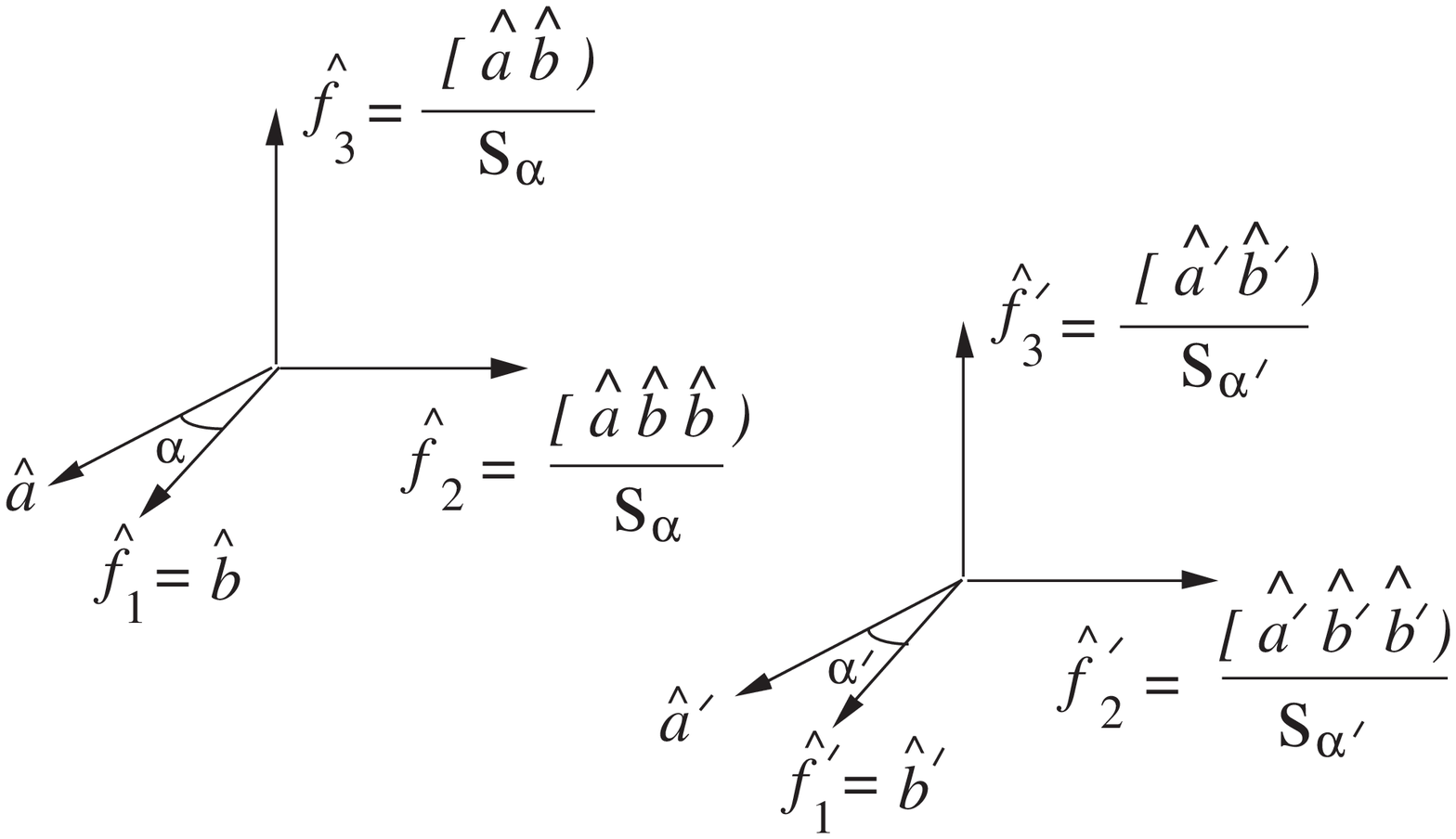, height=2.25in}
    \caption{Principal parameters of $\calg_2^{(2)}$.}
    \label{fig-principal-2cnot}
    \end{center}
\end{figure}

\begin{theo}
$\calg^{(2)}_2$ can be
parameterized with 2 real numbers
$\alpha,\alpha'$,
and 2 RHON bases
$(\hatf_j)_{j=1,2,3}$ and
$(\hatf'_{j})_{j=1,2,3}$.
Call these the principal  parameters
of $\calg^{(2)}_2$
(see Fig.\ref{fig-principal-2cnot}).
More explicitly,

\beq
\calg^{(2)}_2=
\lam_{2r}  +
\Lam_{2r} + i\Lam_{2i}
\;,
\label{eq-diag-invariant-2bit}
\eeq
where

\beq
\lam_{2r}= c_{\alpha'} c_\alpha
\;,
\eeq

\beq
\Lam_{2r}=
-(s_{\alpha'}s_\alpha) \hatf_2'\hatf_2^T
\;,
\eeq

\beqa
\Lam_{2i}
&=&
(s_{\alpha'}c_\alpha) \hatf_3'\hatf_1^T
+
(c_{\alpha'}s_\alpha) \hatf_1'\hatf_3^T
\label{eq-mat-single-line-style}
\\
&=&
\begin{array}{l|ll}
        & \hatf_1^T & \hatf_3^T\\
\hline
\hatf_3'& s_{\alpha'}c_\alpha & 0\\
\hatf_1'& 0                   &c_{\alpha'}s_\alpha
\end{array}
\label{eq-mat-table-style}
\\
&=&
\left[
\begin{array}{cc}
\hatf'_3 & \hatf'_1
\end{array}
\right]
\left[
\begin{array}{cc}
s_{\alpha'}c_\alpha & 0 \\
0 & c_{\alpha'}s_\alpha
\end{array}
\right]
\left[
\begin{array}{cc}
\hatf_1 & \hatf_3
\end{array}
\right]^T
\;.
\label{eq-mat-mat-style}
\eeqa
(Eqs.(\ref{eq-mat-single-line-style}),
(\ref{eq-mat-table-style}), and
(\ref{eq-mat-mat-style})
are 3 different styles of representing the same
thing.)
\end{theo}
\proof

Define $\alpha'\in[0,\pi)$ to be the angle between
$\hata'$ and $\hatb'$. Thus

\beq
c_{\alpha'} = \hata'\cdot\hatb'
\;,\;\;
s_{\alpha'} = |\hata'\times\hatb'|
\;.
\eeq
If $s_{\alpha'}\neq 0$,
set

\beq
(f'_j)_{j=1,2,3}=
(\hatb',
\frac{\manyx{\hata'\hatb'\hatb'}}{s_{\alpha'}} ,
\frac{\manyx{\hata'\hatb'}}{s_{\alpha'}})
\;.
\eeq
If $s_{\alpha'}= 0$,
choose $(\hatf'_j)_{j=1,2,3}$
to be any
RHON basis
with $\hatf'_1=\hatb'$.

Use the previous paragraph with all
primes removed to define
$\alpha$ and $(\hatf_j)_{j=1,2,3}$.

\qed

Suppose we are given a matrix
which is known to be the LO-RHS invariant
$\calg_2^{(2)}$
of a quantum circuit with 2-qubits
and 2 DC-NOTs.
Furthermore, we are asked to extract
from this matrix
values (non-unique ones)
for $\hata$,$\hatb$,$\hata'$ and $\hatb'$.
Next we will give an algorithm for
accomplishing this task.
We will call it our
``Algorithm for Diagonalizing
$\calg_2^{(2)}$".
The algorithm first expresses
$\calg_2^{(2)}$
in term of its principal  parameters. Then it
solves for
$\hata$,$\hatb$,$\hata'$ and $\hatb'$
in terms of these parameters.

\vspace{.2in}
\noindent{\bf Algorithm for
Diagonalizing} $\calg_2^{(2)}$:

\begin{enumerate}
\item
Set $\lam_{2r}=\frac{1}{4}\tr(\calg_2^{(2)})$.
Set $\Delta = \calg_2^{(2)}- \lam_{2r}$,
$\Lam_{2r} = (\Delta + \Delta^\dagger)/2$
and $\Lam_{2i} = (\Delta - \Delta^\dagger)/(2i)$.
Hence, $\calg_2^{(2)}= \lam_{2r} + \Lam_{2r} +
i \Lam_{2i}$, where $\lam_{2r}$
is a real scalar, and
$\Lam_{2r},\Lam_{2i}$
are traceless Hermitian matrices.
\item
Calculate
$c_{\alpha'}c_\alpha$,
$s_{\alpha'}s_\alpha$,
$\hatf_2$ and
$\hatf'_2$
from $\lam_{2r}$ and $\Lam_{2r}$.
(If $\Lam_{2r}=0$, then
take $s_{\alpha'} s_\alpha=0$,
and choose
$\hatf_2$ and
$\hatf'_2$ to be
any 3d unit vectors.)
\item \label{item-diag-invar2-h-hprime}
Choose any
RHON basis $(\hath_j)_{j=1,2,3}$
such that $\hath_2=\hatf_2$, and
any RHON basis $(\hath'_j)_{j=1,2,3}$
such that $\hath'_2=\hatf'_2$.
\item \label{item-diag-invar2-m-matrix}
Find  a Singular Value
Decomposition (SVD) of the matrix

\beq
M =
\left[
\begin{array}{cc}
\hath^{'T}_3 \Lam_{2i}\hath_1&
\hath^{'T}_3 \Lam_{2i}\hath_3\\
\hath^{'T}_1 \Lam_{2i}\hath_1&
\hath^{'T}_1 \Lam_{2i}\hath_3
\end{array}
\right]
\;.
\eeq
In other words, find 2-dimensional orthogonal
matrices $U,V$ and
a non-negative 2-dimensional
diagonal matrix $D$ such that

\beq
M = U D V^T
\;.
\eeq

Now calculate $s_{\alpha'}c_\alpha$,
$c_{\alpha'}s_\alpha$,
$\hatf'_3$,
$\hatf'_1$,
$\hatf_3$,
$\hatf_1$ from

\beq
\left[
\begin{array}{cc}
s_{\alpha'}c_\alpha & 0\\
0 & c_{\alpha'}s_\alpha
\end{array}
\right]
= D
\;,
\eeq

\beq
[\hatf'_3, \hatf'_1]
=
[\hath'_3, \hath'_1] U
\;,
\label{eq-f-eq-hu}
\eeq
and

\beq
[\hatf_1, \hatf_3]
=
[\hath_1, \hath_3]V
\;.
\eeq
\item
By expressing $U$ on the RHS of
 Eq.(\ref{eq-f-eq-hu}) in component form,
it is easy to verify that

\beq
\hatf'_3 \times \hatf'_1
=
\det(U)\hath'_3 \times \hath'_1
\;.
\eeq
$\hath'_3\times \hath'_1\cdot \hath'_2= +1$
and $\hatf'_2=\hath'_2$ so

\beq
\hatf'_3\times \hatf'_1\cdot \hatf'_2
= \det(U)
\;.
\eeq
$\det(U)$ will
always equal either $+1$ or $-1$.
If $\det(U)=-1$, replace
$\hatf'_3\rarrow -\hatf'_3$ and
$s_{\alpha'}c_\alpha \rarrow -s_{\alpha'}c_\alpha$.
These replacements
make
$(\hatf'_1,\hatf'_2, \hatf'_3)$
a right handed basis.

If $\det(V)=-1$,
an analogous procedure can be
used  to convert
$(\hatf_1,\hatf_2, \hatf_3)$
into a right-handed basis.
\item
At this point, we know
the four quantities
$c_{\alpha'}c_\alpha$,
$s_{\alpha'}c_\alpha$,
$c_{\alpha'}s_\alpha$,
and
$s_{\alpha'}s_\alpha$.
Calculate $\alpha'\pm \alpha$
from

\begin{subequations}
\beq
\cos(\alpha'\pm\alpha)= c_{\alpha'}c_\alpha \mp
s_{\alpha'}s_\alpha
\;,
\eeq
and

\beq
\sin(\alpha'\pm\alpha)= s_{\alpha'}c_\alpha \pm
c_{\alpha'}s_\alpha
\;.
\eeq
\end{subequations}
Calculate $(\alpha', \alpha)$ from
$\alpha'\pm\alpha$.
\item
Calculate $\hata,\hatb,\hata',\hatb'$
from:

\beq
\left\{
\begin{array}{l}
\hatb = \hatf_1\\
\hata = c_{\alpha} \hatf_1 - s_{\alpha} \hatf_2
\end{array}
\right.
\;\;,\;\;\;
\left\{
\begin{array}{l}
\hatb' = \hatf'_1\\
\hata' = c_{\alpha'} \hatf'_1 - s_{\alpha'} \hatf'_2
\end{array}
\right.
\;.
\eeq

\end{enumerate}
\subsection{Invariant for Circuits with
3 DC-NOTs
\\{\footnotesize\tt[
ckt\_invar123.m,
ckt\_invar3.m,
diag\_ckt\_invar3.m,
test\_diag\_invar3.m
]}}

\label{sec-invariants-3cnots}

This part of our program is dedicated to the letters
$\calg^{(2)}_3$.

\begin{theo}\label{th-abc-invar3}
\beqa
\calg^{(2)}_3 &=&
\begin{array}{c}
\Qcircuit @C=1em @R=1em @!R{
&\ovalgate{\hatc}\qwx[1]
&\ovalgate{\hatb}\qwx[1]
&\ovalgate{\hata}\qwx[1]
&\ovalgate{-\hata}\qwx[1]
&\ovalgate{-\hatb}\qwx[1]
&\ovalgate{-\hatc}\qwx[1]
&\qw
\\
&\ovalgate{\hatc'}
&\ovalgate{\hatb'}
&\ovalgate{\hata'}
&\ovalgate{-\hata'}
&\ovalgate{-\hatb'}
&\ovalgate{-\hatc'}
&\qw
}
\end{array}
\label{eq-invariant-3bit-diag}
\\
&=&
\lam_{3r} + i\lam_{3i} + \Lam_{3r} + i\Lam_{3i}
\;,
\label{eq-invariant-3bit}
\eeqa
where

\beq
\lam_{3r} =
\manyx{\hata'\hatb' \hatb'}\cdot\hatc'
\;\;
\manyx{\hata\hatb\hatb}\cdot\hatc
\;,
\eeq

\beq
\lam_{3i} =
-(\hata\cdot\hatb)(\hatb\cdot\hatc)\calv'
-(\hata'\cdot\hatb')(\hatb'\cdot\hatc')\calv
\;,
\eeq

\beq
\Lam_{3r}=
\left\{
\begin{array}{l}
-(\hata'\cdot\hatb')(\hata\cdot\hatb)
\sigma_{\hatc',\hatc}
\\
+(\hata\cdot\hatb)(\hatb\cdot\hatc)
\sigma_{\manyx{\hata'\hatb'\hatc'},\hatc}
+(\hata'\cdot\hatb')(\hatb'\cdot\hatc')
\sigma_{\hatc',\manyx{\hata\hatb\hatc}}
\\
+(\hata'\cdot\hatb')\calv\sigma_{\manyx{\hatb'\hatc'},\hatc}
+(\hata\cdot\hatb)\calv'\sigma_{\hatc',\manyx{\hatb\hatc}}
\\
-\sigma_{
\manyx{\hata'\hatb'\hatb'\hatc'\hatc'},
\manyx{\hata\hatb\hatb\hatc\hatc}}
\;
\end{array}
\right.
\;,
\eeq

\beq
\Lam_{3i}=
\left\{
\begin{array}{l}
+(\hata\cdot\hatb)
\sigma_{\manyx{\hata'\hatb'\hatc'\hatc'},
\manyx{\hatb\hatc\hatc}}
+
(\hata'\cdot\hatb')
\sigma_{\manyx{\hatb'\hatc'\hatc'},
\manyx{\hata\hatb\hatc\hatc}}
\\
+\manyx{\hata\hatb\hatb}\cdot \hatc
\sigma_{\manyx{\hata'\hatb'\hatb'\hatc'},\hatc}
+
\manyx{\hata'\hatb'\hatb'}\cdot \hatc'
\sigma_{\hatc', \manyx{\hata\hatb\hatb\hatc}}
\end{array}
\right.
\;,
\eeq
where
$\calv = \hata\times\hatb\cdot\hatc$ and
$\calv' = \hata'\times\hatb'\cdot\hatc'$.
\end{theo}
\proof

\beqa
\calg^{(2)}_3 &=&
\begin{array}{c}
\Qcircuit @C=1em @R=1em @!R{
&\ovalgate{\hatc}\qwx[1]
&\qw
\\
&\ovalgate{\hatc'}
&\qw
}
\end{array}
\calg_2^{(2)}
\begin{array}{c}
\Qcircuit @C=1em @R=1em @!R{
&\ovalgate{-\hatc}\qwx[1]
&\qw
\\
&\ovalgate{-\hatc'}
&\qw
}
\end{array}\\
&=&
\begin{array}{c}
\Qcircuit @C=1em @R=1em @!R{
&\ovalgate{\hatc}\qwx[1]
&\qw
\\
&\ovalgate{\hatc'}
&\qw
}
\end{array}
\calg_2^{(2)}
\begin{array}{c}
\Qcircuit @C=1em @R=1em @!R{
&\ovalgate{\hatc}\qwx[1]
&\qw
\\
&\ovalgate{\hatc'}
&\qw
}
\end{array}
(-\sigma_{\hatc',\hatc})
\;.
\eeqa
An explicit expression
for $\calg^{(2)}_2$ was given
in Section \ref{sec-invariants-2cnots}.
Eq.(\ref{eq-sim-trans-of-sig-veca})
shows how to calculate the effect of DC-NOT
similarity transformations.

\qed

\begin{theo}\label{th-simple-orientation-invar3}
Suppose

\beq
\call=
\begin{array}{c}
\Qcircuit @C=1em @R=1em @!R{
&\ovalgate{\hatc}\qwx[1]
&\ovalgate{\hatb}\qwx[1]
&\ovalgate{\hata}\qwx[1]
&\qw
\\
&\ovalgate{\hatc'}
&\ovalgate{\hatb'}
&\ovalgate{\hata'}
&\qw
}
\end{array}
\;\;,\;\;
\calr=
\begin{array}{c}
\Qcircuit @C=1em @R=1em @!R{
&\ovalgate{\hatc_f}\qwx[1]
&\qw
&\ovalgate{\hatb_f}\qwx[1]
&\qw
&\ovalgate{\hata_f}\qwx[1]
&\qw
\\
&\ovalgate{\hatc'_f}
&\foil
&\ovalgate{\hatb'_f}
&\foil
&\ovalgate{\hata'_f}
&\qw
\gategroup{1}{2}{1}{6}{.7em}{--}
}
\end{array}
\;.
\eeq
For any $\call$,
it is possible to find an $\calr$ such that
$\call\sim_R \calr$, and such that
(a)$\hata_f\times\hatb_f\cdot\hatc_f=0$, and
(b)$\hatb'_f\perp span(\hatc'_f,\hata'_f)$.
\end{theo}
\proof

As pointed out in the introduction,
Ref.\cite{VD} shows how to express any
2-qubit unitary operation
 as a circuit with just 3-CNOTs.
It is easy to check that conditions (a) and (b)
are satisfied by the 3-CNOT
circuit given
in Ref.\cite{VD}. Hence, this theorem
has already been proven in Ref.\cite{VD},
although Ref.\cite{VD} does  not explicitly
point out this property of their
3-CNOT circuit. The
``Algorithm for Diagonalizing $\calg^{(2)}_3$",
that is presented later in this section,
also constitutes a (constructive) proof of this
theorem.

\qed

For $A, B \in \RR^{p\times q}$,
define the following two commutators:

\begin{subequations}\label{eq-left-right-commut}
\beq
[A,B]_L = A^T B - B^T A
\;,
\eeq

\beq
[A,B]_R = A B^T - B A^T
\;.
\eeq
\end{subequations}
(Here, the letters
$L$ and $R$ stand for left and
right. They indicate on which matrix
the transpose symbol acts,
either the left
or the right matrix in the matrix product.)
Ref.\cite{Tuc-KAK} presents a proof
(due to Eckart and Young) of the following
Theorem.
$A,B\in\RR^{p\times q}$ have
a simultaneous
Singular Value Decomposition (SVD)
  if and
only if $[A,B]_L$ and $[A,B]_R$
are both zero.
By a simultaneous SVD
we mean
orthogonal matrices $U,V$
and real diagonal matrices $D_A,D_B$
such that

\beq
A = UD_AV^T
\;\;,\;\;
B=U D_B V^T
\;.
\label{eq-sim-svd}
\eeq
When considering
the SVD of a single matrix $A$, one
usually insists in making
the entries of  $D_A$ non-negative,
and calling them
the singular values of $A$.
In the case of a simultaneous SVD,
one can't always make both
diagonal matrices non-negative,
but one can certainly make one of
them so.

Of course, the previous paragraph
applies almost intact if
$A$ and $B$ are elements of
$\CC^{p\times q}$ instead of
$\RR^{p\times q}$.
For $A,B$ complex, one must replace
the $T$ (transpose) symbol by the
$\dagger$ (Hermitian conjugate) symbol in
Eqs.(\ref{eq-left-right-commut}) and
(\ref{eq-sim-svd}). Also, the matrices $U,V$
in Eq.(\ref{eq-sim-svd})
must be unitary instead of orthogonal.

Note that when $A$ and $B$ are Hermitian,
the
condition that
$[A,B]_L$ and $[A,B]_R$
both vanish becomes simply the
condition that $A$ and $B$ commute.
The Eckart, Young theorem then
becomes a theorem
very familiar to practitioners
of Quantum Mechanics:  two Hermitian operators
can be simultaneously diagonalized
iff they commute.

\begin{theo}

\beq
[\Lam_{3r}, \Lam_{3i}]=0
\;,
\eeq

\beq
[\Lam^\Gamma_{3r}, \Lam^\Gamma_{3i}]_L=0
\;,
\eeq

\beq
[\Lam^\Gamma_{3r}, \Lam^\Gamma_{3i}]_R=0
\;.
\eeq
\end{theo}
\proof

Let

\beq
\Delta = \calg^{(2)}_3 - \tr(\calg^{(2)}_3)
\;,
\eeq
so

\beq
\Lam_{3r}= \frac{\Delta + \Delta^\dagger}{2}
\;\;,\;\;
\Lam_{3i}= \frac{\Delta - \Delta^\dagger}{2i}
\;.
\eeq
Thus,

\beq
[\Lam_{3r},\Lam_{3i}]=
\frac{1}{4i}[\Delta + \Delta^\dagger,
\Delta - \Delta^\dagger]
=
\frac{1}{2i}[\Delta^\dagger,
\Delta]
=
\frac{1}{2i}[\calg^{(2)\dagger}_3,
\calg^{(2)}_3]
=
0
\;,
\eeq
where the last commutator is zero because
$\calg^{(2)}_3$
is unitary.

Note that for any
$\hata, \hata',\hatb,\hatb'\in\unitvecs$,

\beqa
[\sigma_{\hata',\hata},\sigma_{\hatb',\hatb}]&=&
\left\{
\begin{array}{l}
+(\hata'\cdot\hatb' +i\sigma_{\hata'\times\hatb'})
\otimes
(\hata\cdot\hatb +i\sigma_{\hata\times\hatb})
\\
-
(\hatb'\cdot\hata' +i\sigma_{\hatb'\times\hata'})
\otimes
(\hatb\cdot\hata +i\sigma_{\hatb\times\hata})
\end{array}
\right.
\\
&=&
i2
[(\hata\cdot\hatb)\sigma_{\hata'\times\hatb',1}
-
(\hata'\cdot\hatb')\sigma_{1,\hata\times\hatb}]
\;.
\eeqa

From Theorem \ref{th-abc-invar3},
we know that $\Lam_{3r}$
and $\Lam_{3r}$
can be expressed in the form

\beq
\Lam_{3r} =
\sum_j \alpha_j \sigma_{\hata'_j,\hata_j}
\;,
\;\;
\Lam_{3i} =
\sum_k \beta_k \sigma_{\hatb'_k,\hatb_k}
\;,
\eeq
for some $\alpha_j,\beta_j\in \RR$
and
$\hata_j, \hata'_j, \hatb_k,\hatb'_k\in\unitvecs$.
Therefore,

\beqa
0&=& [\Lam_{3r}, \Lam_{3i}]
\\
&=&
\sum_{j,k} \alpha_j\beta_k
[ \sigma_{\hata'_j,\hata_j},
\sigma_{\hatb'_k,\hatb_k}]
\\
&=&
i2\sum_{j,k} \alpha_j\beta_k
[(\hata_j\cdot\hatb_k)\sigma_{\hata'_j\times\hatb_k',1}
-
(\hata'_j\cdot\hatb'_k)\sigma_{1,\hata_j\times\hatb_k}]
\;.
\eeqa
This implies that

\beq
\sum_{j,k} \alpha_j\beta_k
(\hata_j\cdot\hatb_k)
\hata'_j\times\hatb_k'=0
\;,\;\;
\sum_{j,k} \alpha_j\beta_k
(\hata'_j\cdot\hatb'_k)
\hata_j\times\hatb_k=0
\;.
\label{eq-sum-cross-prods-is-zero}
\eeq

Now note that
\beqa
[\Lam_{3r}^\Gamma, \Lam_{3i}^\Gamma]_R
&=&
(\sum_j \alpha_j \hata'_j\hata^T_j)
(\sum_k \beta_k \hatb_k\hatb^{'T}_k)
-
(\sum_k \beta_k \hatb'_k\hatb^{T}_k)
(\sum_j \alpha_j \hata_j\hata^{'T}_j)
\\
&=&
\sum_{j,k}
\alpha_j\beta_k(\hata^T_j\hatb_k)
[
\hata'_j \hatb^{'T}_k
-
\hatb'_k \hata^{'T}_j ]
\\
&=& 0
\;,
\eeqa
where the last expression vanishes
due to Eq.(\ref{eq-sum-cross-prods-is-zero}).
An analogous argument shows that
$[\Lam_{3r}^\Gamma, \Lam_{3i}^\Gamma]_L$
also vanishes.

\qed

It is convenient to parameterize
the expression for $\calg^{(2)}_3$
given by Theorem \ref{th-abc-invar3},
using as few parameters as possible.

\begin{figure}[h]
    \begin{center}
    \epsfig{file=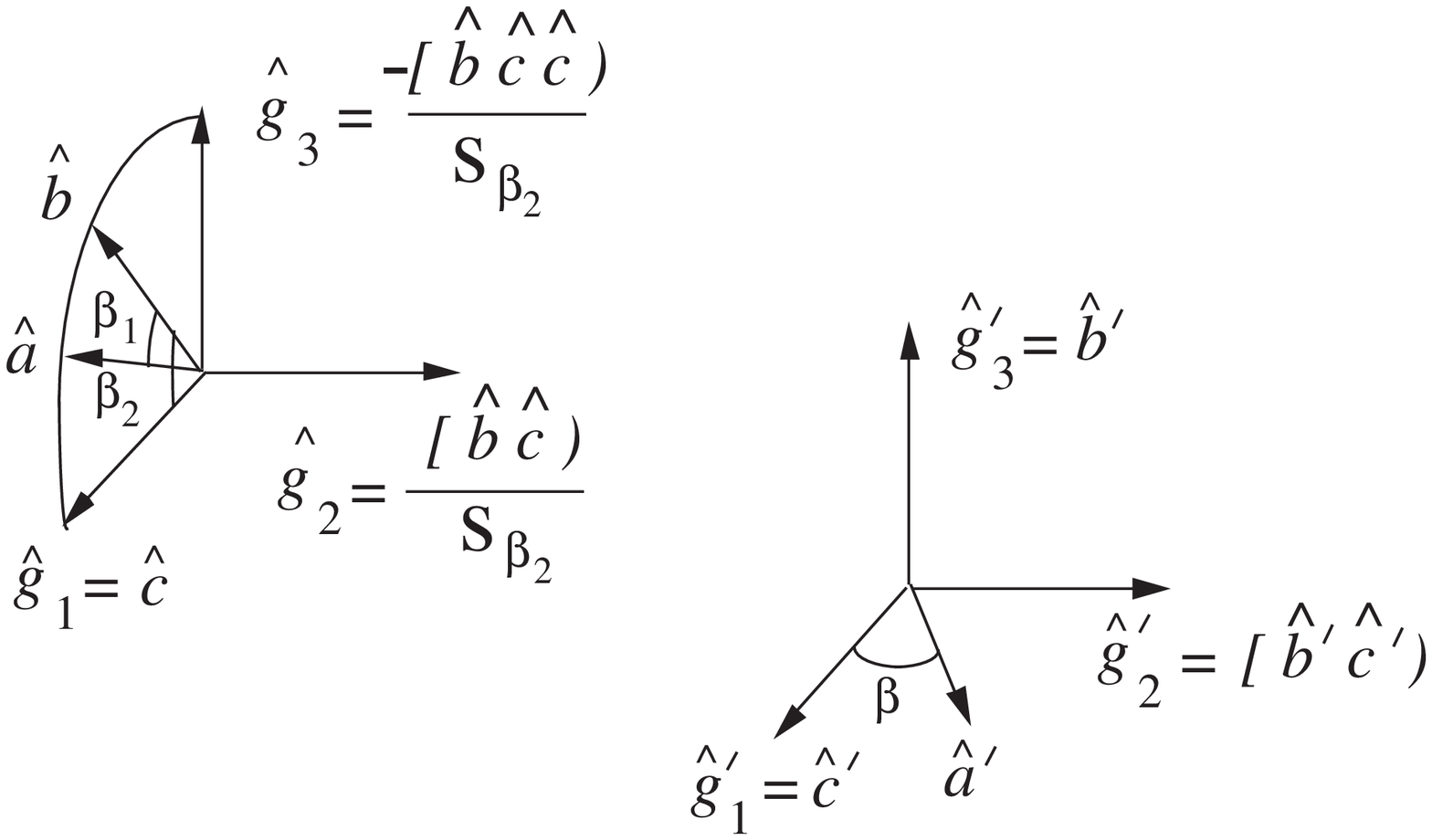, height=2.25in}
    \caption{Principal parameters of $\calg_3^{(2)}$.}
    \label{fig-principal-3cnot}
    \end{center}
\end{figure}

\begin{theo}
$\calg^{(2)}_3$ can be
parameterized with 3 real numbers
$\beta,\beta_1,\beta_2$,
and 2 RHON bases
$(\hatg_j)_{j=1,2,3}$ and
$(\hatg'_{j})_{j=1,2,3}$.
Call these the principal  parameters
of $\calg^{(2)}_3$
(see Fig.\ref{fig-principal-3cnot}).
More explicitly,

\beq
\calg^{(2)}_3=
\lam_{3r} + i\lam_{3i} +
\Lam_{3r} + i\Lam_{3i}
\;,
\label{eq-diag-invariant-3bit}
\eeq
where

\beq
\lam_{3r} = - X_o
\;,
\eeq

\beq
\lam_{3i} = - Y_o
\;,
\eeq

\beq
\Lam_{3r} =  \sum_{j=1}^3 \nu_j
\hatg_j'\hatg_{\pi(j)}^T
\;,
\label{eq-Lam3r-fin}
\eeq

\beq
\Lam_{3i} =  \sum_{j=1}^3 \mu_j
\hatg_j'\hatg_{\pi(j)}^T
\;,
\label{eq-Lam3i-fin}
\eeq
where

\beq
X_o = c_\beta \xi s_{\beta_1} s_{\beta_2}
\;,
\label{eq-xo-fin}
\eeq

\beq
Y_o = s_\beta c_{\beta_1} c_{\beta_2}
\;,
\label{eq-yo-fin}
\eeq

\beq
(\nu_j)_{j=1,2,3}
=
(
 s_\beta c_{\beta_1}s_{\beta_2},
 s_\beta s_{\beta_1}|c_{\beta_2}|,
 c_\beta c_{\beta_1}c_{\beta_2}
)
\;,
\label{eq-nuj-fin}
\eeq

\beq
(\mu_j)_{j=1,2,3}
=
(
-c_\beta s_{\beta_1}|c_{\beta_2}|,
-c_\beta c_{\beta_1}s_{\beta_2},
 s_\beta \xi s_{\beta_1}s_{\beta_2}
)
\;,
\label{eq-muj-fin}
\eeq
where $\xi\in\{+1,-1\}$ and
$\pi()$ is the permutation
$\left(
\begin{array}{ccc}
1 & 2 & 3\\
2 & 3 & 1
\end{array}
\right)$.

\end{theo}
\proof

We will assume from the onset of this
proof that
(a)$\hata\times\hatb\cdot\hatc=0$, and
(b)$\hatb'\perp span(\hatc',\hata')$.
This can be assumed without loss of
generality because of Theorem
\ref{th-simple-orientation-invar3}.

Let

\beq
\xi = {\rm sign}(\manyx{ab}\cdot\manyx{bc})
\;,\;\;
\xi_2 = {\rm sign}(\hatb\cdot\hatc)
\;.
\eeq
Without loss of generality, we will assume
that $-\xi\xi_2=+1$.
If $-\xi\xi_2$ is initially negative, we can
make it positive by
replacing both $\hata$
and $\hata'$ by their negatives.
This replacement will not change
$\calg^{(2)}_3$.
Using the circuit shown in
Eq.(\ref{eq-invariant-3bit-diag}),
it is easy to
prove that $\calg_3^{(2)}$ is
odd in both $\hata$ and $\hata'$.

Define
\beq
s_{\beta_2} = |\manyx{\hatb\hatc}|
\;,\;\;
\eta = |\manyx{\hata\hatb\hatb\hatc}|
=|\manyx{\hata\hatb}\hatb\cdot\hatc|
\;.
\eeq
To begin, we will assume that
$s_{\beta_2}\neq 0$
and $\eta\neq 0$. Later on, before ending
the proof, we will remove these two constraints.

If we define
\beq
X_o = (\hata'\cdot\hatc')
\manyx{\hata\hatb\hatb}\cdot \hatc
\;,
\label{eq-xo-init}
\eeq

\beq
Y_o= (\hata\cdot\hatb)(\hatb\cdot\hatc)\calv'
\;,
\label{eq-yo-init}
\eeq

\beq
(\hatg'_j)_{j=1,2,3}=
(
\hatc',
\manyx{\hatb'\hatc'},
\hatb'
)
\;,
\eeq

\beq
(\hatg_j)_{j=1,2,3}=
(
\hatc,
\frac{\manyx{\hatb\hatc}}{s_{\beta_2}},
\frac{-\manyx{\hatb\hatc\hatc}}{s_{\beta_2}}
)
\;,
\label{eq-gj-init}
\eeq

\begin{subequations}
\beq
(\nu_j)_{j=1,2,3}=
\left(
\hata\cdot\hatb\calv' s_{\beta_2},
\calv'\eta,
(\hata\cdot\hatb)(\hatb\cdot\hatc)(\hata'\cdot\hatc')
\right)
\;,
\label{eq-nuj-init}
\eeq

\beq
(\hatv_j)_{j=1,2,3}=
(
\frac{\manyx{\hatb\hatc}}{s_{\beta_2}},
\frac{-\manyx{\hata\hatb\hatb\hatc\hatc}}{\eta},
\hatc
)
\;,
\label{eq-vj-init}
\eeq
\end{subequations}

\begin{subequations}
\beq
(\mu_j)_{j=1,2,3}=
\left(
-\hata'\cdot\hatc'\eta,
-(\hata\cdot\hatb)(\hata'\cdot\hatc')s_{\beta_2},
\manyx{\hata\hatb\hatb}\cdot\hatc\calv'
\right)
\;,
\label{eq-muj-init}
\eeq

\beq
(\hatu_j)_{j=1,2,3}=
(
\frac{\manyx{\hata\hatb\hatb\hatc}}{\eta},
\frac{-\manyx{\hatb\hatc\hatc}}{s_{\beta_2}},
\hatc
)
\;,
\label{eq-uj-init}
\eeq
\end{subequations}
then

\beq
\calg^{(2)}_3=
-X_o -iY_o
+\sum_{j=1}^3 \nu_j
\hatg_j'\hatv_{j}^T
+i\sum_{j=1}^3 \mu_j
\hatg_j'\hatu_{j}^T
\;.
\eeq

Define an angle $\beta$ by

\beq
\cos(\beta)= \hata'\cdot\hatc'
\;,\;\;
\sin(\beta) = \calv'
\;.
\eeq
Define angles $\beta_1, \beta_2\in [0,\pi)$
by

\beq
\cos(\beta_1) = \hata\cdot\hatb
\;,\;\
\sin(\beta_1) = |\hata\times\hatb|
\;,
\eeq
and

\beq
\cos(\beta_2) = \hatb\cdot\hatc
\;,\;\
\sin(\beta_2) = |\hatb\times\hatc|
\;.
\eeq
Hence, $\manyx{\hata\hatb}/s_{\beta_1} =
\xi \manyx{\hatb\hatc}/s_{\beta_2}$.
 One finds

\beq
\eta = s_{\beta_1}|c_{\beta_2}|
\;,
\eeq

\beq
\frac{\manyx{\hata\hatb\hatb\hatc}}{\eta}
\cdot\hatg_2
=-\xi \xi_2
\;,
\eeq

\beq
\frac{-\manyx{\hata\hatb\hatb\hatc\hatc}}{\eta}
\cdot\hatg_3
=-\xi \xi_2
\;,
\eeq
and

\beq
\manyx{\hata\hatb\hatb}\cdot \hatc
=\xi s_{\beta_1}s_{\beta_2}
\;.
\eeq

At this point, it is easy
re-express various quantities
in terms of the principal  parameters.
Eq.(\ref{eq-xo-init}) for $X_o$,
Eq.(\ref{eq-yo-init}) for $Y_o$,
Eq.(\ref{eq-nuj-init}) for the $\nu_j$,
and
Eq.(\ref{eq-muj-init}) for the $\mu_j$,
yield, respectively,
Eq.(\ref{eq-xo-fin}),
Eq.(\ref{eq-yo-fin}),
Eq.(\ref{eq-nuj-fin}),
and
Eq.(\ref{eq-muj-fin}).

We can also re-express
Eqs.(\ref{eq-vj-init}) and (\ref{eq-uj-init})
for the $\hatv_j$ and $\hatu_j$
in terms of the principal  parameters.
One finds

\beq
(\hatv_j)_{j=1,2,3}=
(\hatg_2, -\xi\xi_2\hatg_3, \hatg_1)
=
(\hatg_2, \hatg_3, \hatg_1)
\;,
\eeq
and

\beq
(\hatu_j)_{j=1,2,3}=
(-\xi\xi_2\hatg_2, \hatg_3, \hatg_1)
=
(\hatg_2, \hatg_3, \hatg_1)
\;.
\eeq
Hence,
for $j=1,2,3$,
\beq
\hatv_j = \hatu_j=\hatg_{\pi(j)}
\;.
\label{eq-v-u-gpi}
\eeq

When $s_{\beta_2}$ or $\eta$ vanish,
Eq.(\ref{eq-gj-init}) fails to define
two of the vectors $\hatg_j$,
Eq.(\ref{eq-vj-init}) fails to define
on or two of the vectors $\hatv_j$,
and
Eq.(\ref{eq-uj-init}) fails to define
on or two of the vectors $\hatu_j$.
If $s_{\beta_2}=0$, the proof survives if
we define
$(\hatg_j)_{j=1,2,3}$ to be any
RHON basis such that $\hatg_1=\hatc$
and $\hatg_2\perp span(\hata,\hatb,\hatc)$.
Then define
the $\hatu_j$ and $\hatv_j$ vectors
in accordance with Eq.(\ref{eq-v-u-gpi}).
If $\eta=0$ but
$s_{\beta_2}\neq 0$,
define
the $\hatu_j$ and $\hatv_j$ vectors
in accordance with Eq.(\ref{eq-v-u-gpi}).

\qed

Suppose we are given a matrix
which is known to be the LO-RHS invariant
$\calg_3^{(2)}$
of a quantum circuit with 2-qubits
and 3 DC-NOTs.
Furthermore, we are asked to extract from
this matrix
values (non-unique ones)
for $\hata$, $\hatb$, $\hatc$,
$\hata'$, $\hatb'$ and $\hatc'$.
Next we will give an algorithm for
accomplishing this task.
We will call it our
``Algorithm for Diagonalizing
$\calg_3^{(2)}$".
The algorithm first expresses
$\calg_3^{(2)}$
in term of its principal parameters. Then it
solves for
$\hata$, $\hatb$, $\hatc$,
$\hata'$, $\hatb'$ and $\hatc'$
in terms of these parameters.

\vspace{.2in}
\noindent{\bf Algorithm for
Diagonalizing} $\calg_3^{(2)}:$

\begin{figure}[h]
    \begin{center}
    \epsfig{file=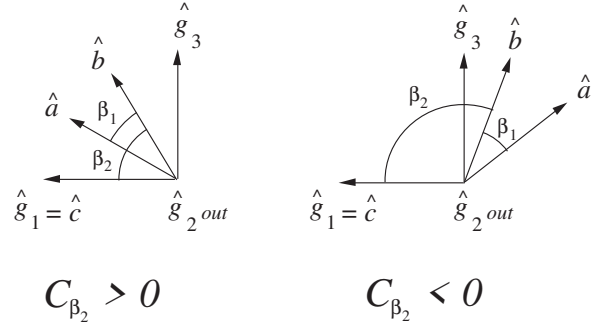, height=1.75in}
    \caption{Sign of $\beta_1$ is defined
    differently depending on whether $c_{\beta_2}$
    is positive or negative.}
    \label{fig-signs-principal-3cnot}
    \end{center}
\end{figure}

\begin{enumerate}
\item
Set $\lam_{3r}=\frac{1}{4}{\rm Re}[\tr(\calg_3^{(2)})]$
and $\lam_{3i}=\frac{1}{4}{ \rm Im}[\tr(\calg_3^{(2)})]$.
Set $\Delta = \calg_3^{(2)}- tr(\calg_3^{(2)})$,
$\Lam_{3r} = (\Delta + \Delta^\dagger)/2$
and $\Lam_{3i} = (\Delta - \Delta^\dagger)/(2i)$.
Hence, $\calg_3^{(2)}= \lam_{3r} + i\lam_{3i}+ \Lam_{3r} +
i \Lam_{3i}$, where $\lam_{3r}, \lam_{3i}$
are real scalars, and
$\Lam_{3r},\Lam_{3i}$
are traceless Hermitian matrices.

\item
Set $X_o=-\lam_{3r}$
and
$Y_o=-\lam_{3i}$.

\item
Do a simultaneous SVD of
$\Lam^\Gamma_{3r}$
and
$\Lam^\Gamma_{3i}$.
This decomposition is possible
since we have shown previously that
$[\Lam^\Gamma_{3r},\Lam^\Gamma_{3i}]_L$
and
$[\Lam^\Gamma_{3r},\Lam^\Gamma_{3i}]_R$
are both zero. The decomposition yields
orthogonal matrices $U,V$
and real diagonal matrices $D_{3r},D_{3i}$
such that

\beq
\Lam^\Gamma_{3r} = U D_{3r} V^T
\;,\;\;
\Lam^\Gamma_{3i} = U D_{3i} V^T
\;.
\eeq

For $j=1,2,3$, set

\beq
\nu_j = (D_{3r})_{jj}
\;\;,\;\;
\mu_j = (D_{3i})_{jj}
\;.
\eeq
Set

\beq
[\hatg'_1,\hatg'_2,\hatg'_3] = U
\;\;,\;\;
[\hatg_1,\hatg_2,\hatg_3] = V
\;.
\eeq
\item
Set $\xi = sign(\mu_3\nu_2)$.
Set $\xi_2 = -\xi$.
Calculate $\beta$ from

\beq
c_\beta
=
\xi \frac{X_o}
{\sqrt{\mu_3 ^2 + X_o^2}}
\;\;,\;\;
s_\beta=
\xi\frac{\mu_3}
{\sqrt{\mu_3 ^2 + X_o^2}}
\;.
\eeq
If $|c_\beta|\geq |s_\beta|$,
set

\beq
\left[
\begin{array}{cc}
c_{\beta_1} c_{\beta_2}& c_{\beta_1} s_{\beta_2}
\\
s_{\beta_1} c_{\beta_2}& s_{\beta_1} s_{\beta_2}
\end{array}
\right]
=
\frac{1}{c_\beta}
\left[
\begin{array}{cc}
\nu_3        & -\mu_2 \\
-\xi_2 \mu_1 & \xi X_o
\end{array}
\right]
\;.
\eeq
On the other hand, if $|s_\beta|\geq |c_\beta|$,
set

\beq
\left[
\begin{array}{cc}
c_{\beta_1} c_{\beta_2}& c_{\beta_1} s_{\beta_2}
\\
s_{\beta_1} c_{\beta_2}& s_{\beta_1} s_{\beta_2}
\end{array}
\right]
=
\frac{1}{s_\beta}
\left[
\begin{array}{cc}
Y_o        & \nu_1 \\
\xi_2 \nu_2 & \xi \mu_3
\end{array}
\right]
\;.
\eeq
\item
At this point, we know
the four quantities
$c_{\beta_1}c_{\beta_2}$,
$s_{\beta_1}c_{\beta_2}$,
$c_{\beta_1}s_{\beta_2}$,
and
$s_{\beta_1}s_{\beta_2}$.
Calculate $\beta_1\pm \beta_2$
from

\begin{subequations}
\beq
\cos(\beta_1\pm\beta_2)= c_{\beta_1}c_{\beta_2} \mp
s_{\beta_1}s_{\beta_2}
\;,
\eeq
and

\beq
\sin(\beta_1\pm\beta_2)= s_{\beta_1}c_{\beta_2} \pm
c_{\beta_1}s_{\beta_2}
\;.
\eeq
\end{subequations}
Calculate $(\beta_1, \beta_2)$ from
$\beta_1\pm\beta_2$.
\item
At this point,
$s_{\beta_1}s_{\beta_2}$
is guaranteed to be positive, but there
is not guarantee that $s_{\beta_1}$
and $s_{\beta_2}$ are individually positive (they
may both be negative). Furthermore,
at this point there is no guarantee
that $\xi_2 = sign(c_{\beta_2})$.
These disagreements with the
assumptions of our parameterization
can be fixed as follows.
If
$s_{\beta_1}<0$,
replace
$\beta_1$ and $\beta_2$ by their negatives,
and replace
$(\hatg'_1, \hatg'_2, \nu_1, \nu_2, \mu_1, \mu_2)$
each by its negative.
If
$\xi_2 c_{\beta_2}<0$,
replace
$\beta_1\rarrow \pi-\beta_1$
and
$\beta_2\rarrow \pi-\beta_2$,
and replace
$(\hatg'_1, \hatg'_2, \nu_1, \nu_2, \mu_1, \mu_2)$
each by its negative.

\item
Calculate $\hata,\hatb,\hatc,\hata',\hatb',\hatc'$
from:

\beq
\left\{
\begin{array}{l}
\hatc = \hatg_1\\
\hatb = c_{\beta_2}\hatg_1 + s_{\beta_2}\hatg_3\\
\hata =
\cos(\beta_2-\xi_2\beta_1) \hatg_1 +
\sin(\beta_2-\xi_2\beta_1) \hatg_3
\end{array}
\right.
\;\;,\;\;\;
\left\{
\begin{array}{l}
\hatc' = \hatg'_1\\
\hatb' = \hatg'_3\\
\hata' = c_{\beta} \hatg'_1 + s_{\beta} \hatg'_2
\end{array}
\right.
\;.
\eeq
Note the $\xi_2$'s in the
expression for $\hata$. The
reason for these $\xi_2$'s is
that in order to obey $-\xi\xi_2=+1$,
one must define
the sign of the angle $\beta_1$
differently depending on
whether $c_{\beta_2}$ is
positive or negative.
(See Fig.\ref{fig-signs-principal-3cnot})

\end{enumerate}

\begin{theo}\label{th-mu-nu-bilinear}
For any $j\in\{1,2,3\}$,

\beq
\mu_j\nu_j = X_o Y_o
\;.
\eeq
If $i,j,k$ are 3 distinct
element of $\{1,2,3\}$, then

\beq
\mu_i \mu_j = -X_o \nu_k
\;,
\eeq
and

\beq
\nu_i \nu_j = -Y_o \mu_k
\;.
\eeq
\end{theo}
\proof

Follows from the definitions
Eq.(\ref{eq-xo-fin}) for $X_o$,
Eq.(\ref{eq-yo-fin}) for $Y_o$,
Eq.(\ref{eq-nuj-fin}) for the $\nu_j$,
and
Eq.(\ref{eq-muj-fin}) for the $\mu_j$.

\qed

Define $\Pi$ to be
the permutation matrix that
corresponds to the permutation map $\pi()$
used above. Thus,

\beq
\Pi =
\left[
\begin{array}{ccc}
0 & 1 & 0\\
0 & 0 & 1\\
1 & 0 & 0
\end{array}
\right]
\;.
\eeq
If $(s_j)_{j=1,2,3}$ denotes
the standard basis,
define matrices $M_\mu$ and $M_\nu$ by

\beqa
M_\mu &=&
\sum_{j=1}^3 \mu_j \hats_j \hats_{\pi(j)}\\
&=&
diag(\mu_1, \mu_2, \mu_3)\Pi
\;,
\eeqa
and

\beqa
M_\nu &=&
\sum_{j=1}^3 \nu_j \hats_j \hats_{\pi(j)}\\
&=&
diag(\nu_1, \nu_2, \nu_3)\Pi
\;.
\eeqa
Note that $\Lam^\Gamma_{2r}$
(given by Eq.(\ref{eq-Lam3r-fin}) ) becomes
$M_\nu$
and
$\Lam^\Gamma_{2i}$
(given by Eq.(\ref{eq-Lam3i-fin}) ) becomes
$M_\mu$
when the bases
$(\hatg_j)_{j=1,2,3}$
and
$(\hatg'_j)_{j=1,2,3}$
are both rotated into
the standard basis.

\begin{theo}

\beq
M_\mu  M_\nu^T = M_\nu  M_\mu^T = X_o Y_o
\;,
\eeq

\beq
M_\mu^T  M_\nu = M^T_\nu  M_\mu = X_o Y_o
\;,
\eeq
and

\beq
(M_\mu^T)^2 = \tr(M_\nu) - M_\nu
\;.
\eeq
\end{theo}
\proof

Follows from
Theorem \ref{th-mu-nu-bilinear}.

\qed
\subsection{Invariant for Circuits with
4 DC-NOTs
\\{\footnotesize\tt[
ckt\_invar4.m
]}}

\label{sec-invariants-4cnots}

This part of our program is dedicated to the letters
$\calg^{(2)}_4$.

\begin{theo}\label{eq-invar4}
\beqa
\calg^{(2)}_4 &=&
\begin{array}{c}
\Qcircuit @C=1em @R=1em @!R{
&\ovalgate{\hatd}\qwx[1]
&\ovalgate{\hatc}\qwx[1]
&\ovalgate{\hatb}\qwx[1]
&\ovalgate{\hata}\qwx[1]
&\ovalgate{-\hata}\qwx[1]
&\ovalgate{-\hatb}\qwx[1]
&\ovalgate{-\hatc}\qwx[1]
&\ovalgate{-\hatd}\qwx[1]
&\qw
\\
&\ovalgate{\hatd'}
&\ovalgate{\hatc'}
&\ovalgate{\hatb'}
&\ovalgate{\hata'}
&\ovalgate{-\hata'}
&\ovalgate{-\hatb'}
&\ovalgate{-\hatc'}
&\ovalgate{-\hatd'}
&\qw
}
\end{array}\\
&=&
\lam_{4r} + i\lam_{4i} + \Lam_{4r} + i\Lam_{4i}
\;,
\label{eq-invariant-4bit}
\eeqa
where

\beq
\lam_{4r} =
-\sum_j (\hatg'_j\cdot\hatd')\nu_j
\hatg_{\pi(j)}\cdot\hatd
\;,\;\;
\lam_{4i} =
(\lam_{4r})_{\nu\rarrow\mu}
\;,
\eeq

\beq
\Lam_{4r}=
X_o \sigma_{\hatd',\hatd}
+ \sigma_{\vec{x'},\hatd}
+ \sigma_{\hatd',\vecx}
+ \Delta X
\;,
\eeq

\beq
\Lam_{4i}=
Y_o \sigma_{\hatd',\hatd}
- \sigma_{\vec{y'},\hatd}
- \sigma_{\hatd',\vecy}
+ \Delta Y
\;,
\eeq
where

\beq
\vecx = \sum_j \mu_j
(\hatg'_j\cdot\hatd')
\manyx{\hatg_{\pi(j)}\hatd}
\;,\;\;
\vec{y} = (\vecx)_{\mu\rarrow\nu}
\;,
\eeq

\beq
\vec{x'} = \sum_j \mu_j
(\hatg_{\pi(j)}\cdot\hatd)
\manyx{\hatg'_j\hatd'}
\;,\;\;
\vec{y'} = (\vec{x'})_{\mu\rarrow\nu}
\;,
\eeq

\beq
\Delta X =
\sum_j \nu_j
\sigma_{
\manyx{\hatg'_j\hatd'\hatd'},
\manyx{\hatg_{\pi(j)}\hatd\hatd}
}
\;,\;\;
\Delta Y =
(\Delta X)_{\nu\rarrow\mu}
\;,
\eeq
where any variables not already
defined
in the statement of this theorem
are defined in
Section \ref{sec-invariants-3cnots}.

\end{theo}
\proof

\beqa
\calg^{(2)}_4 &=&
\begin{array}{c}
\Qcircuit @C=1em @R=1em @!R{
&\ovalgate{\hatd}\qwx[1]
&\qw
\\
&\ovalgate{\hatd'}
&\qw
}
\end{array}
\calg_3^{(2)}
\begin{array}{c}
\Qcircuit @C=1em @R=1em @!R{
&\ovalgate{-\hatd}\qwx[1]
&\qw
\\
&\ovalgate{-\hatd'}
&\qw
}
\end{array}
\\
&=&
\begin{array}{c}
\Qcircuit @C=1em @R=1em @!R{
&\ovalgate{\hatd}\qwx[1]
&\qw
\\
&\ovalgate{\hatd'}
&\qw
}
\end{array}
\calg_3^{(2)}
\begin{array}{c}
\Qcircuit @C=1em @R=1em @!R{
&\ovalgate{\hatd}\qwx[1]
&\qw
\\
&\ovalgate{\hatd'}
&\qw
}
\end{array}
(-\sigma_{\hatd',\hatd})
\;.
\eeqa
An explicit expression
for $\calg^{(2)}_3$ was given
in Section \ref{sec-invariants-3cnots}.
Eq.(\ref{eq-sim-trans-of-sig-veca})
shows how to calculate the effect of DC-NOT
similarity transformations.

\qed

\begin{theo}
When the bases
$(\hatg_j)_{j=1,2,3}$
and
$(\hatg'_j)_{j=1,2,3}$
are both taken to be
the standard basis,
then the quantities
$\lam_{4r}$, $\lam_{4i}$
$\vecx$, $\vecy$,
$\vec{x'}$, $\vec{y'}$,
$\Delta X$ and $\Delta Y$
(all defined in
Theorem \ref{eq-invar4})
can be expressed in terms
of the matrices $M_\mu, M_\nu$
and the vectors $\hatd, \hatd'$ as follows:

\beq
\lam_{4r} =
-\hatd^{\;'T}M_\nu\hatd
\;,\;\;
\lam_{4i} =
(\lam_{4r})_{\nu\rarrow\mu}
\;,
\eeq

\beq
\vecx = \manyx{M_\mu^T\hatd',\hatd}
\;,\;\;
\vecy = (\vecx)_{\mu\rarrow\nu}
\;,
\eeq

\beq
\vec{x'} = \manyx{M_\mu\hatd,\hatd'}
\;,\;\;
\vec{y'} = (\vec{x'})_{\mu\rarrow\nu}
\;,
\eeq

\beq
\Delta X=
\hatd'\hatd^T (\hatd^{\;'T}M_\nu\hatd)
-M_\nu \hatd \hatd^T
-\hatd'\hatd^{\;'T} M_\nu
+M_\nu
\;,\;\;
\Delta Y =
(\Delta X)_{\nu\rarrow\mu}
\;.
\eeq
\end{theo}
\proof

Just algebra.

\qed

\begin{figure}[h]
    \begin{center}
    \epsfig{file=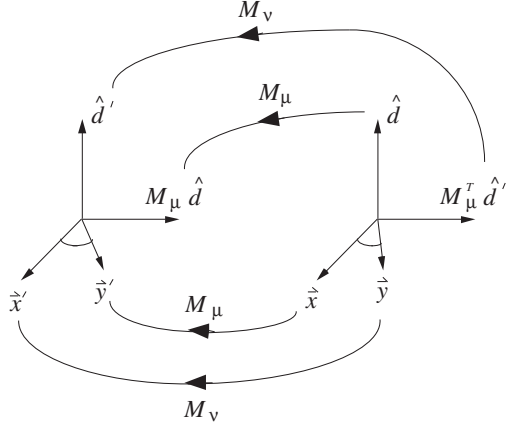, height=2.3in}
    \caption{Various vectors and
    what they are mapped into (up to a scalar
    factor) by $M_\mu$ and $M_\nu$.
    Since $M^T_\nu M_\mu$ and
    $M^T_\mu M_\nu$ are both
    proportional to the identity matrix,
    one can replace $M_\mu$ by $M_\nu^T$
    and $M_\nu$ by $M_\mu^T$ in this figure
    if one also reverses the direction of
    the mapping arrows.
    }
    \label{fig-m-mu-nu}
    \end{center}
\end{figure}

\begin{theo}
See Fig.\ref{fig-m-mu-nu}.

\begin{subequations}
\beq
M_\nu^T\vec{y'} = Y_o \vecx
\;,\;\;
M_\mu \vecx  = X_o \vec{y'}\;,
\eeq

\beq
M_\mu^T\vec{x'} = X_o \vecy
\;,\;\;
M_\nu \vecy  = Y_o \vec{x'}
\;.
\eeq
\end{subequations}
\end{theo}
\proof

Just algebra.

\qed

\section{Identities for Circuits with 2 Qubits}
\label{sec-two-qubit-ids}

This section deals with 2-qubit circuits, whereas
Section \ref{sec-3-qubit-ids}
deals with 3-qubit ones.
In this section, with its numerous
subsections, we start to reap the benefits
of all our preceding hard work.
The combination of dressed CNOTs and the LO-RHS
invariant proves to be very useful.
We find simple-to-check necessary
and sufficient conditions for the reduction of
a quantum circuit with $j$ CNOTs to fewer CNOTs, where
$j=2,3$. Plus we show how to express circuits with 1 or
2 controlled-U's as circuits with 2 or fewer CNOTs.
Plus we show how to
open and close a breach, a
procedure that can
reduce any 4-CNOT circuit to a 3-CNOT one.
\subsection{Reducing 2 DC-NOTs}
\label{sec-2-cnots}
\subsubsection{2 to 2 DC-NOTs (Angle Swapping)
\\{\footnotesize\tt[
swap\_angles.m,
test\_swang.m
]}}
\label{sec-2to2-cnots}

In this section we consider
a circuit with 2 DC-NOTs
acting on 2 qubits, and
show that a symmetry in $\calg^{(2)}_2$
allows one to swap
certain angles
without changing the effect of the circuit
(up to LO-RHS).

As motivation for
the main theorem of this section
(the Angle Swapping
Theorem), we present
the next theorem.
The next theorem shows
 that
the target and control qubits of
a controlled-U can be exchanged.

\begin{theo}\label{th-contr-u-flip}
For any $\theta\in\RR$,

\beq
\begin{array}{c}
\Qcircuit @C=1em @R=1em @!R{
&\ovalgate{\hata}\qwx[1]
&\qw
\\
&\gate{e^{i\theta\sigma_{\hatb'}}}
&\qw
}
\end{array}
=
\begin{array}{c}
\Qcircuit @C=1em @R=1em @!R{
&\gate{e^{i\theta\sigma_{\hata}}}\qwx[1]
&\gate{e^{-i\frac{\theta}{2}\sigma_\hata}}
&\qw
\\
&\ovalgate{\hatb'}
&\gate{e^{+i\frac{\theta}{2}\sigma_{\hatb'}}}
&\qw
}
\end{array}
\;.
\eeq
\end{theo}
\proof

\beqa
[e^{i\theta\sigma_{\hatb'}(1)}]^{n_\hata(0)}
&=&
e^{i\theta\sigma_{\hatb'}(1)
\left(
\frac{1-\sigma_\hata(0)}{2}
\right)
}\\
&=&
e^{i
\theta\sigma_{\hata}(0)
\left(
\frac{1-\sigma_{\hatb'}(1)}{2}
\right)
}
e^{-i\frac{\theta}{2}\sigma_\hata(0)}
e^{+i\frac{\theta}{2}\sigma_{\hatb'}(1)}
\\
&=&
[e^{i
\theta\sigma_{\hata}(0)}]
^{n_{\hatb'}(1)}
e^{-i\frac{\theta}{2}\sigma_\hata(0)}
e^{+i\frac{\theta}{2}\sigma_{\hatb'}(1)}
\;.
\eeqa

\qed

The previous theorem immediately implies the
next one, which states that
we can ``swap a breach" between two qubits.

\begin{theo}\label{th-swap-breach}
(Swapping a breach)
Suppose
\beq
\call=
\begin{array}{c}
\Qcircuit @C=1em @R=1em @!R{
&\ovalgate{\hata}\qwx[1]
&\breach
&\ovalgate{\hata}\qwx[1]
&\qw
\\
&\ovalgate{\hatp'}
&\qw
&\ovalgate{\hatq'}
&\qw
}
\end{array}
\;\;,\;\;\;
\calr=
\begin{array}{c}
\Qcircuit @C=1em @R=1em @!R{
&\ovalgate{\hatp}\qwx[1]
&\qw
&\ovalgate{\hatq}\qwx[1]
&\qw
\\
&\ovalgate{\hatb'}
&\breach
&\ovalgate{\hatb'}
&\qw
}
\end{array}
\;.
\eeq
For any $\call$, it is possible to
find an $\calr$ such that $\call\sim_R \calr$.
\end{theo}
\proof

Define $\theta$ to be the angle
between $\hatp'$ and $\hatq'$,
and $\hatb$ the direction of $\hatp'\times\hatq'$.
Then
$\hatp'\cdot\hatq'=\cos(\theta)$
and
$\hatp'\times\hatq'=\sin(\theta)\hatb'$ so
$\sigma_{\hatp'}\sigma_{\hatq'}=
e^{i\theta\sigma_{\hatb'}}$.
Thus,

\beq
\begin{array}{c}
\Qcircuit @C=1em @R=1em @!R{
&\ovalgate{\hata}\qwx[1]
&\breach
&\ovalgate{\hata}\qwx[1]
&\qw
\\
&\ovalgate{\hatp'}
&\qw
&\ovalgate{\hatq'}
&\qw
}
\end{array}
=
\begin{array}{c}
\Qcircuit @C=1em @R=1em @!R{
&\ovalgate{\hata}\qwx[1]
&\qw
\\
&\gate{e^{i\theta\sigma_{\hatb'}}}
&\qw
}
\end{array}
\;.
\label{eq-swap-breach-left}
\eeq

Given a unit vector $\hata$
and an angle $\theta$,
we can always find (non-unique)
unit vectors
$\hatp$ and $\hatq$ such that
$angle(\hatp,\hatq)=\theta$, and
$\hatp\times \hatq$ points along $\hata$. Then
$\hatp\cdot\hatq=\cos(\theta)$
and
$\hatp\times\hatq=\sin(\theta)\hata$ so
$\sigma_{\hatp}\sigma_{\hatq}=
e^{i\theta\sigma_{\hata}}$.
It follows that

\beq
\begin{array}{c}
\Qcircuit @C=1em @R=1em @!R{
&\gate{e^{i\theta\sigma_{\hata}}}\qwx[1]
&\qw
\\
&\ovalgate{\hatb'}
&\qw
}
\end{array}
=
\begin{array}{c}
\Qcircuit @C=1em @R=1em @!R{
&\ovalgate{\hatp}\qwx[1]
&\qw
&\ovalgate{\hatq}\qwx[1]
&\qw
\\
&\ovalgate{\hatb'}
&\breach
&\ovalgate{\hatb'}
&\qw
}
\end{array}
\;.
\label{eq-swap-breach-right}
\eeq
Now apply Theorem \ref{th-contr-u-flip}
to Eqs.(\ref{eq-swap-breach-left})
and (\ref{eq-swap-breach-right}).

\qed

Is it possible to swap a foil
instead of a breach?  Yes it is.
In fact, one can swap any angle,
 as the following
theorem shows.

\begin{theo}(Angle Swapping)
Let

\beq
\call =
\begin{array}{c}
\Qcircuit @C=1em @R=1em @!R{
&\ovalgate{\hatb}\qwx[1]
&\ovalgate{\hata}\qwx[1]
&\qw
\\
&\ovalgate{\hatb'}
&\ovalgate{\hata'}
&\qw
}
\end{array}
\;\;,\;\;
\calr =
\begin{array}{c}
\Qcircuit @C=1em @R=1em @!R{
&\ovalgate{\hatb_f}\qwx[1]
&\ovalgate{\hata_f}\qwx[1]
&\qw
\\
&\ovalgate{\hatb'_f}
&\ovalgate{\hata'_f}
&\qw
}
\end{array}
\;.
\eeq
For any $\call$, it is possible to
find an $\calr$ such that $\call \sim_R \calr$
and such that
angle$(\hatb, \hata)$ = angle$(\hatb'_f, \hata'_f)$
and
angle$(\hatb', \hata')$ = angle$(\hatb_f, \hata_f)$.
\end{theo}
\proof

As proven in Section \ref{sec-invariants-2cnots},
$\call^{(2)}$
can be parameterized as follows:

\beq
\call^{(2)}=
c_{\alpha'} c_\alpha
-(s_{\alpha'}s_\alpha) \hatf_2'\hatf_2^T
+i
\begin{array}{l|ll}
        & \hatf_1^T & \hatf_3^T\\
\hline
\hatf_3'& s_{\alpha'}c_\alpha & 0\\
\hatf_1'& 0                   &c_{\alpha'}s_\alpha
\end{array}
\;,
\eeq
where $\alpha, \alpha\in \RR$
and where $(\hatf_j)_{j=1,2,3}$ and
$(\hatf'_j)_{j=1,2,3}$ are two
RHON bases
such that

\beq
\hatb = \hatf_1
\;,\;\;
\hata = c_\alpha \hatf_1 - s_\alpha \hatf_2
\;,
\eeq
and

\beq
\hatb' = \hatf'_1
\;,\;\;
\hata' = c_{\alpha'} \hatf'_1 - s_{\alpha'} \hatf'_2
\;.
\eeq

$\calr^{(2)}$ can be
parameterized in
the same way as $\call^{(2)}$, but with
the replacements
$\alpha\rarrow\alpha_f$,
$\alpha'\rarrow\alpha'_f$,
$\hatf_j\rarrow (\hatf_j)_f$,
and
$\hatf'_j\rarrow (\hatf'_j)_f$.

Our goal is to construct an $\calr$ such that
$\call\sim_R \calr$. Such an $\calr$,
if it exists, must satisfy
$\hat{\call}^{(2)}= \pm \hat{\calr}^{(2)}$.
We will use the positive sign.
In light of
Eq.(\ref{eq-invar-of-hat-graph}), this gives

\beq
i^2 \call^{(2)} =  i^2 \calr^{(2)}
\;.
\eeq

From the symmetrical form of
the parameterized expressions
for $\call^{(2)}$ and
$\calr^{(2)}$,
it is clear
that these two invariants are
equal if their principal  parameters are
related in the following way:

\beq
\alpha_f = \alpha'
\;,\;\;
\alpha'_f = \alpha
\;,
\eeq

\beq
\hatf_{3f}=\hatf_1
\;,\;\;
\hatf_{1f}=\hatf_3
\;,\;\,
\hatf_{2f}=-\hatf_2
\;,
\eeq
and

\beq
\hatf'_{3f}=\hatf'_1
\;,\;\;
\hatf'_{1f}=\hatf'_3
\;,\;\,
\hatf'_{2f}=-\hatf'_2
\;.
\eeq
These relations between
the principal parameters of
$\call^{(2)}$ and
$\calr^{(2)}$ imply that

\beqa
\call &=&
\begin{array}{c}
\Qcircuit @C=1em @R=1em @!R{
&\ovalgate{\hatf_1}\qwx[1]
&\ovalgate{c_\alpha\hatf_1 - s_\alpha\hatf_2}\qwx[1]
&\qw
\\
&\ovalgate{\hatf'_1}
&\ovalgate{c_{\alpha'}\hatf'_1 - s_{\alpha'}\hatf'_2}
&\qw
}
\end{array}
\label{eq-2to2-left}
\\
&=&
\begin{array}{c}
\Qcircuit @C=1em @R=1em @!R{
&\ovalgate{\hatb}\qwx[1]
&\ovalgate{\hata}\qwx[1]
&\qw
\\
&\ovalgate{\hatb'}
&\ovalgate{\hata'}
&\qw
}
\end{array}
\;,
\eeqa
and

\beqa
\calr
&=&
\begin{array}{c}
\Qcircuit @C=1em @R=1em @!R{
&\ovalgate{\hatf_3}\qwx[1]
&\ovalgate{c_{\alpha'}\hatf_3 + s_{\alpha'}\hatf_2}\qwx[1]
&\qw
\\
&\ovalgate{\hatf'_3}
&\ovalgate{c_\alpha\hatf'_3 + s_\alpha\hatf'_2}
&\qw
}
\end{array}
\label{eq-2to2-right}
\\
&=&
\begin{array}{c}
\Qcircuit @C=1em @R=1em @!R{
&\ovalgate{\frac{\manyx{\hata\hatb}}{s_\alpha}}\qwx[1]
&\ovalgate{
\frac{
c_{\alpha'}\manyx{\hata\hatb}
+ s_{\alpha'}\manyx{\hata\hatb\hatb}
}{s_\alpha}
}\qwx[1]
&\qw
\\
&\ovalgate{\frac{\manyx{\hata'\hatb'}}{s_{\alpha'}}}
&\ovalgate{
\frac{
c_{\alpha}\manyx{\hata'\hatb'}
+ s_{\alpha}\manyx{\hata'\hatb'\hatb'}
}{s_{\alpha'}}
}
&\qw
}
\end{array}
\label{eq-2to2-singular}
\;.
\eeqa
(Eq.(\ref{eq-2to2-singular}) is valid only if
$s_\alpha$ and $s_{\alpha'}$
are both non-zero, whereas
Eq.(\ref{eq-2to2-right}) is always valid.
Theorem \ref{th-swap-breach}
corresponds to the case
$s_\alpha=0$.)

We are done proving the theorem,
but we will go one step further,
and give the value of the
local operations $U',U\in SU(2)$
such that

\beq
\call = \calr (U^{'\dagger}\otimes U^\dagger)
\;.
\eeq
When $\hatf_1=\hatf'_1=\hatx$
and $\hatf_3=\hatf'_3=\hatz$,
the right-hand sides
of Eqs.(\ref{eq-2to2-left})
and (\ref{eq-2to2-right})
appear in  Theorem \ref{th-2thirds-split}.
It follows from Theorem \ref{th-2thirds-split}
and Eq.(\ref{eq-2thirds-u-def})
that

\beq
U = e^{i \frac{\alpha}{2} \sigma_{\hatf_3}}
e^{-i \frac{\alpha'}{2} \sigma_{\hatf_1}}
\;,\;\;
U' = (U)_{\alpha\darrow \alpha', \hatf\rarrow \hatf'}
\;.
\eeq

\qed
\subsubsection{2 to 1 DC-NOTs}
\label{sec-2to1-cnots}

In this section, we give necessary
and sufficient conditions for
a circuit with 2 DC-NOTs acting
on 2 qubits to reduce to
1 DC-NOT.

\begin{figure}[h]
    \begin{center}
    \epsfig{file=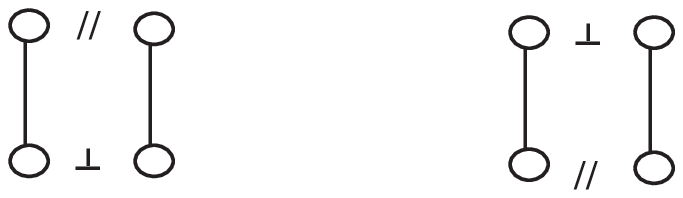, height=.75in}
    \caption{All circuits with 2 DC-NOTs
    that reduce to 1 DC-NOT.}
    \label{fig-2to1}
    \end{center}
\end{figure}

\begin{theo}
Suppose

\beq
\call =
\begin{array}{c}
\Qcircuit @C=1em @R=1em @!R{
&\ovalgate{\hatb}\qwx[1]
&\ovalgate{\hata}\qwx[1]
&\qw
\\
&\ovalgate{\hatb'}
&\ovalgate{\hata'}
&\qw
}
\end{array}
\;\;,\;\;
\calr =
\begin{array}{c}
\Qcircuit @C=1em @R=1em @!R{
&\ovalgate{\hata_f}\qwx[1]
&\qw
\\
&\ovalgate{\hata'_f}
&\qw
}
\end{array}
\;.
\eeq
For any $\call$, it is possible to
find an $\calr$ such that $\call\sim_R \calr$
 if and only if
($\hatb \parallel \hata$ and $\hatb'\perp \hata'$)
or
($\hatb \perp \hata$ and $\hatb'\parallel \hata'$).
See Fig.\ref{fig-2to1}.
\end{theo}
\proof

\lproof

Suppose $\hatb \perp \hata$ and
$\hatb'\parallel \hata'$
(the other case is analogous).
When $\hatb \perp \hata$,

\beq
\sigma_{\hatb}(0)^{n_{\hata'}(1)}
\sigma_{\hata}(0)^{n_{\hata'}(1)}
=
[i\sigma_{\hatb\times\hata}(0)]^{n_{\hata'}(1)}
\;.
\eeq
The last equation
can be expressed diagrammatically as

\beq
\begin{array}{c}
\Qcircuit @C=1em @R=1em @!R{
&\ovalgate{\hatb}\qwx[1]
&\foil
&\ovalgate{\hata}\qwx[1]
&\qw
\\
&\ovalgate{\hata'}
&\breach
&\ovalgate{\hata'}
&\qw
}
\end{array}
=
\begin{array}{c}
\Qcircuit @C=1em @R=1em @!R{
&\ovalgate{\hatb\times\hata}\qwx[1]
&\qw
&\qw
\\
&\ovalgate{\hata'}
&\gate{ i^{n_{\hata'}}}
&\qw
}
\end{array}
\;.
\eeq
Thus, when $\hatb\perp\hata$
and $\hatb'=\hata'$,
$\call$ reduces to a single DC-NOT.
More generally,
$\hata'=\pm \hatb'$.
Let $\call_{new}$ be a new circuit
obtained by replacing in $\call$:
$\hata'$ by its negative if $\hata'=-\hatb'$.
By virtue of Eq.(\ref{eq-dcnot-with-neg}),
$\call = \call_{new}(I_2\otimes U)$,
where $U\in U(2)$.
If $\call_{new}\sim_R \calr_{new}$, then
$\call\sim_R \calr_{new}$.

\rproof

$\call\sim_R \calr$ so
$\hat{\call}^{(2)}= \pm \hat{\calr}^{(2)}$.
In light of
Eq.(\ref{eq-invar-of-hat-graph}), this gives

\beq
i^2\call^{(2)} = \pm i \calr^{(2)}
\;.
\eeq
It follows that

\beq
\lam_{2r} + \Lam_{2r} + i\Lam_{2i}
=
\pm i\sigma_{\hata'_f, \hata_f}
\;,
\eeq
where

\beq
\lam_{2r}= (\hata\cdot\hatb)(\hata'\cdot\hatb')
\;,
\eeq

\beq
\Lam_{2r}=
-\sigma_{\manyx{\hata'\hatb'\hatb'},\manyx{\hata\hatb\hatb}}
\;,
\eeq

\beq
\Lam_{2i}=
\hata\cdot\hatb\sigma_{\hata'\times\hatb',\hatb}
+
\hata'\cdot\hatb'\sigma_{\hatb',\hata\times\hatb}
\;.
\eeq
$\lam_{2r}=0$ so
$\hata\cdot\hatb=0$ or $\hata'\cdot\hatb'=0$.
Assume the former (the other case is analogous).
Then $\hata\perp\hatb$.
$\Lam_{2r}=0$ and $\hata\cdot\hatb=0$
so $\manyx{\hata'\hatb'\hatb'}=0$,
which in turn implies that
$\hata'\parallel\hatb'$.

\qed
\subsubsection{2 to 0 DC-NOTs}
\label{sec-2to0-cnots}

In this section, we give necessary
and sufficient conditions for
a circuit with 2 DC-NOTs acting
on 2 qubits to reduce to
zero DC-NOTs (i.e., to merely local operations).

\begin{figure}[h]
    \begin{center}
    \epsfig{file=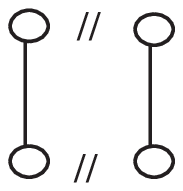, height=.75in}
    \caption{All circuits with 2 DC-NOTs
    that reduce to 0 DC-NOTs.}
    \label{fig-2to0}
    \end{center}
\end{figure}

\begin{theo}
Suppose

\beq
\call=
\begin{array}{c}
\Qcircuit @C=1em @R=1em @!R{
&\ovalgate{\hatb}\qwx[1]
&\ovalgate{\hata}\qwx[1]
&\qw
\\
&\ovalgate{\hatb'}
&\ovalgate{\hata'}
&\qw
}
\end{array}
\;.
\eeq
For any $\call$,
$\call\sim_R 1$ if and only if
$\hata\parallel\hatb$ and
$\hata'\parallel\hatb'$.
See Fig.\ref{fig-2to0}.
\end{theo}
\proof

\lproof

When $\hata=\hatb$ and $\hata'=\hatb'$,
$\call$ equals 1.
More generally,
$\hata=\pm\hatb$
and
$\hata'=\pm\hatb'$.
Let $\call_{new}$ be a new circuit
obtained by replacing in $\call$:
(1)$\hata$ by its negative if $\hata=-\hatb$,
(2)$\hata'$ by its negative if $\hata'=-\hatb'$.
By virtue of Eq.(\ref{eq-dcnot-with-neg}),
$\call = \call_{new}(U'\otimes U)$,
where $U',U\in U(2)$.
If $\call_{new}\sim_R 1$, then
$\call\sim_R 1$.

\rproof

$\call\sim_R 1$ so
$\hat{\call}^{(2)}= \pm 1$.
In light of
Eq.(\ref{eq-invar-of-hat-graph}), this gives

\beq
i^2 \call^{(2)}= \pm 1
\;.
\eeq
It follows that

\beq
\lam_{2r} + \Lam_{2r} + i\Lam_{2i}
=\pm 1
\;.
\eeq
Thus
$\lam_{2r}=(\hata\cdot\hatb)(\hata'\cdot\hatb')=\pm 1$,
which implies
$\hata\parallel\hatb$ and
$\hata'\parallel\hatb'$.

\qed
\subsection{Reducing 3 DC-NOTs}
\label{sec-3-cnots}
\subsubsection{3 to 2 DC-NOTs
\\{\footnotesize\tt[
dr\_3to2.m,
test\_dr\_3to2.m
]}}
\label{sec-3to2-cnots}

In this section, we give necessary
and sufficient conditions for
a circuit with 3 DC-NOTs acting
on 2 qubits to reduce to
2 DC-NOTs.

\begin{figure}[h]
    \begin{center}
    \epsfig{file=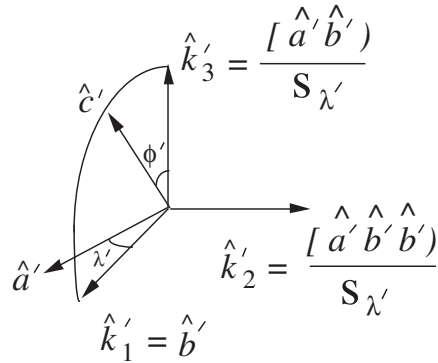, height=2in}
    \caption{Vectors and angles associated with
    bit-1 space spanned by $\hata',\hatb',\hatc'$. }
    \label{fig-phi-lam-prime}
    \end{center}
\end{figure}

The constraint
$\manyx{\hata\hatb\hatb}\cdot\hatc=0$
shows up below. The field of Spherical Geometry
sheds some light on this constraint.
If we connect the points
$\hata, \hatb, \hatc$ by mayor-circle arcs
on the unit sphere, then we get what is called a
spherical triangle.
$\manyx{\hata\hatb\hatb}\cdot\hatc=0$
if and only if this spherical triangle
has a right angle at
vertex $\hatb$.(See Fig.\ref{fig-phi-lam-prime}
for an example of
$\manyx{\hata'\hatb'\hatb'}\cdot\hatc'=0$.)

\begin{figure}[h]
    \begin{center}
    \epsfig{file=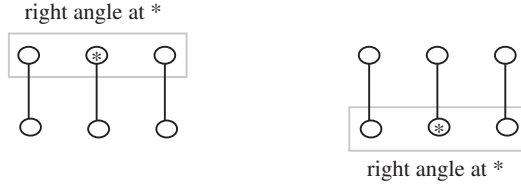, height=1in}
    \caption{All circuits with 3 DC-NOTs
    that reduce to 2 DC-NOTs.}
    \label{fig-3to2}
    \end{center}
\end{figure}

\begin{theo}
Suppose

\beq
\call=
\begin{array}{c}
\Qcircuit @C=1em @R=1em @!R{
&\ovalgate{\hatc}\qwx[1]
&\ovalgate{\hatb}\qwx[1]
&\ovalgate{\hata}\qwx[1]
&\qw
\\
&\ovalgate{\hatc'}
&\ovalgate{\hatb'}
&\ovalgate{\hata'}
&\qw
}
\end{array}
\;\;,\;\;
\calr=
\begin{array}{c}
\Qcircuit @C=1em @R=1em @!R{
&\ovalgate{\hatb_f}\qwx[1]
&\ovalgate{\hata_f}\qwx[1]
&\qw
\\
&\ovalgate{\hatb'_f}
&\ovalgate{\hata'_f}
&\qw
}
\end{array}
\;.
\eeq
For any $\call$,
it is possible to find an $\calr$ such that
$\call\sim_R \calr$
if and only if
either
$\manyx{\hata\hatb\hatb}\cdot\hatc=0$
\;\;or\;\;
$\manyx{\hata'\hatb'\hatb'}\cdot\hatc'=0$.
See Fig.\ref{fig-3to2}.
\end{theo}
\proof

Before we start the proof in earnest,
let us restate some pertinent formulas
taken from previous sections.

From Section \ref{sec-invariants-2cnots}, we know that

\beqa
\calr^{(2)}&=&
\lam_{2r} + \Lam_{2r} + i \Lam_{2i}\\
&=&
c_{\alpha'} c_\alpha
-(s_{\alpha'}s_\alpha) \hatf_2'\hatf_2^T
+i\;
\begin{array}{l|ll}
        & \hatf_1^T & \hatf_3^T\\
\hline
\hatf_3'& s_{\alpha'}c_\alpha & 0\\
\hatf_1'& 0                   &c_{\alpha'}s_\alpha
\end{array}
\;.
\label{eq-r-invars-3to2}
\eeqa

From Section \ref{sec-invariants-3cnots}, we know that
\beq
\call^{(2)}=
\lam_{3r} + i\lam_{3i} + \Lam_{3r} + i \Lam_{3i}
\;,
\eeq
where

\beq
\lam_{3r} =
\manyx{\hata'\hatb' \hatb'}\cdot\hatc'
\;\;
\manyx{\hata\hatb\hatb}\cdot\hatc
\;,
\eeq

\beq
\lam_{3i} =
-(\hata\cdot\hatb)(\hatb\cdot\hatc)\calv'
-(\hata'\cdot\hatb')(\hatb'\cdot\hatc')\calv
\;,
\label{eq-lam3i-3to2}
\eeq

\beq
\Lam_{3r}=
\left\{
\begin{array}{l}
-(\hata'\cdot\hatb')(\hata\cdot\hatb)
\hatc'\hatc^T
\\
+(\hata\cdot\hatb)(\hatb\cdot\hatc)
\manyx{\hata'\hatb'\hatc'}\hatc^T
+(\hata'\cdot\hatb')(\hatb'\cdot\hatc')
\hatc'\manyx{\hata\hatb\hatc}^T
\\
+(\hata'\cdot\hatb')\calv \manyx{\hatb'\hatc'}\hatc^T
+(\hata\cdot\hatb)\calv'\hatc'\manyx{\hatb\hatc}^T
\\
-
\manyx{\hata'\hatb'\hatb'\hatc'\hatc'}
\manyx{\hata\hatb\hatb\hatc\hatc}^T
\;,
\end{array}
\right.
\eeq
and

\beq
\Lam_{3i}=
\left\{
\begin{array}{l}
+(\hata\cdot\hatb)
\manyx{\hata'\hatb'\hatc'\hatc'}
\manyx{\hatb\hatc\hatc}^T
+
(\hata'\cdot\hatb')
\manyx{\hatb'\hatc'\hatc'}
\manyx{\hata\hatb\hatc\hatc}^T
\\
+\manyx{\hata\hatb\hatb}\cdot \hatc
\manyx{\hata'\hatb'\hatb'\hatc'}\hatc^T
+
\manyx{\hata'\hatb'\hatb'}\cdot \hatc'
\hatc'\manyx{\hata\hatb\hatb\hatc}^T
\end{array}
\right.
\;.
\eeq

Now we begin the proof in earnest.

\rproof

$\call\sim_R \calr$ so
$\hat{\call}^{(2)}= \pm \hat{\calr}^{(2)}$.
In light of
Eq.(\ref{eq-invar-of-hat-graph}), this gives

\beq
i^3 \call^{(2)}=
\pm i^2 \calr^{(2)}
\;.
\eeq
It follows that

\beq
0=\lam_{3r}=
\manyx{\hata'\hatb'\hatb'}\cdot\hatc'
\;\;
\manyx{\hata\hatb\hatb}\cdot\hatc
\;.
\eeq
Thus, either
$\manyx{\hata'\hatb'\hatb'}\cdot\hatc'$
or
$\manyx{\hata\hatb\hatb}\cdot\hatc$.

\lproof

Assume
$\manyx{\hata'\hatb'\hatb'}\cdot\hatc'=0$.
(The other case,
$\manyx{\hata\hatb\hatb}\cdot\hatc=0$,
is analogous).

It is convenient at this point to define
a RHON basis $(\hatk'_j)_{j=1,2,3}$
for the 3d real space spanned by
$\hata',\hatb', \hatc'$. Let
$s_{\lam'} = |\manyx{\hata'\hatb'}|$. If
$s_{\lam'}\neq 0$, let

\beq
(\hatk'_j)_{j=1,2,3}=
(\hatb',
\frac{\manyx{\hata'\hatb'\hatb'}}{s_{\lam'}},
\frac{\manyx{\hata'\hatb'}}{s_{\lam'}}
)
\;.
\label{eq-hprime-basis-3to2}
\eeq
If $s_{\lam'} = 0$,
define $(\hatk'_j)_{j=1,2,3}$
to be any RHON basis such that
$\hatk'_1=\hatb'$ and
$\hatk'_2$ is perpendicular to
$span(\hatb',\hatc')$.
Let
$\phi'=angle(\hatc', \hatk'_3)$.
Since
$\manyx{\hata'\hatb'\hatb'}\cdot\hatc'=0$,

\beq
\hata' = c_{\lam'}\hatk'_1 - s_{\lam'} \hatk'_2,
\;,\;\;
\hatb' = \hatk'_1
\;,\;\;
\hatc' = s_{\phi'} \hatk'_1 + c_{\phi'} \hatk'_3
\;.
\label{eq-abcprime-3to2}
\eeq
Eqs.(\ref{eq-hprime-basis-3to2})
and (\ref{eq-abcprime-3to2})
are illustrated in Fig.\ref{fig-phi-lam-prime}.

Our goal is to construct an $\calr$ such that
$\call\sim_R \calr$. Such an $\calr$ must satisfy
$\hat{\call}^{(2)}= \pm \hat{\calr}^{(2)}$.
We will use the positive sign.
In light of
Eq.(\ref{eq-invar-of-hat-graph}), this gives

\beq
i^3 \call^{(2)} =  i^2 \calr^{(2)}
\;.
\eeq
It follows that:

\begin{subequations}
\beq
\lam_{2r} = -\lam_{3i}
\;,
\label{eq-cond1-3to2}
\eeq

\beq
0=\lam_{3r}
\;,
\label{eq-cond2-3to2}
\eeq

\beq
\Lam_{2r} = -\Lam_{3i}
\;,
\label{eq-cond3-3to2}
\eeq

\beq
\Lam_{2i} = \Lam_{3r}
\;.
\label{eq-cond4-3to2}
\eeq
\end{subequations}

By evaluating Eq.(\ref{eq-cond1-3to2}), we get

\beqa
c_{\alpha'}c_\alpha
&=&
(\hata\cdot\hatb)(\hatb\cdot\hatc)\calv'
+
(\hata'\cdot\hatb')(\hatb'\cdot\hatc')\calv
\\
&=&
(\hata\cdot\hatb)(\hatb\cdot\hatc)
s_{\lam'}c_{\phi'}
+
c_{\lam'}s_{\phi'}\calv
\;.
\eeqa

Eq.(\ref{eq-cond2-3to2})
is satisfied since
$\manyx{\hata'\hatb'\hatb'}\cdot\hatc'=0$
by assumption.

By evaluating Eq.(\ref{eq-cond3-3to2}), we get

\beq
-s_{\alpha'}s_\alpha \hatf'_2 \hatf_2^T
=
\left\{
\begin{array}{l}
-(\hata\cdot\hatb)\manyx{\hata'\hatb'\hatc'\hatc'}
\manyx{\hatb\hatc\hatc}^T
\\
-(\hata'\cdot\hatb')\manyx{\hatb'\hatc'\hatc'}
\manyx{\hata\hatb\hatc\hatc}^T
\\
-\manyx{\hata\hatb\hatb}\cdot\hatc
\manyx{\hata'\hatb'\hatb'\hatc'}\hatc^T
\end{array}
\right.
\;.
\eeq
Define $\vech$ by

\beq
\vech =
\left\{
\begin{array}{l}
+s_{\lam'}s_{\phi'}
(\hata\cdot\hatb)
\manyx{\hatb\hatc\hatc}^T
\\
-c_{\lam'}c_{\phi'}
\manyx{\hata\hatb\hatc\hatc}^T
\\
+s_{\lam'}\manyx{\hata\hatb\hatb}\cdot\hatc\hatc^T
\end{array}
\right.
\;.
\eeq
If $s_{\lam'}\neq 0$
and $|\vech|\neq 0$, let

\beq
s_{\alpha'}s_\alpha = |\vech|
\;,\;\;
\hatf'_2 =
\frac{
\manyx{\hata'\hatb'\hatb'\hatc'}
}{s_{\lam'}}
\;,\;\;
\hatf_2 = \frac{\vech}{|\vech|}
\;.
\label{eq-sin-sin-3to2}
\eeq
If $|\vech|=0$, set
$s_{\alpha'}s_\alpha=0$ and
choose
any unit vectors for $\hatf_2$
and $\hatf'_2$.
If $|\vech|\neq 0$
but $s_{\lam'}=0$,
keep  Eq.(\ref{eq-sin-sin-3to2})
for $s_{\alpha'}s_\alpha$
and $\hatf_2$ but
use $\hatf'_2 = \hatk'_2\times\hatc'$.

By evaluating Eq.(\ref{eq-cond4-3to2}), we get

\beq
\Lam_{2i} = \hatc'\vecv_1^T + \hatk'_2\vecv_2^T
\;,
\label{eq-Lam2i-svd}
\eeq
where

\begin{subequations} \label{eq-defs-v1-v2}
\beq
\vecv_1 =
-c_{\lam'}(\hata\cdot\hatb)\hatc
+c_{\lam'}s_{\phi'}\manyx{\hata\hatb\hatc}
+s_{\lam'}c_{\phi'}(\hata\cdot\hatb)\manyx{\hatb\hatc}
\;,
\eeq
and

\beq
\vecv_2 =
s_{\lam'}s_{\phi'}(\hata\cdot\hatb)(\hatb\cdot\hatc)\hatc
-c_{\lam'}c_{\phi'}\calv\hatc
+s_{\lam'}\manyx{\hata\hatb\hatb\hatc\hatc}
\;.
\eeq
\end{subequations}

At this point, we can follow from
step \ref{item-diag-invar2-h-hprime}
to the end of the
Algorithm for Diagonalizing $\calg^{(2)}_2$
that was given in
Section \ref{sec-invariants-2cnots}.
This will yield values for
$\hata_f$,
$\hata'_f$,
$\hatb_f$, and
$\hatb'_f$.

\qed

Compared with the previous
Theorem,
the next theorem
imposes more constraints on $\call$,
and obtains a
more constrained $\calr$.

\begin{theo}\label{th-persistence}
Suppose

\beq
\call=
\begin{array}{c}
\Qcircuit @C=1em @R=1em @!R{
&\ovalgate{\hatc}\qwx[1]
&\ovalgate{\hatb}\qwx[1]
&\ovalgate{\hata}\qwx[1]
&\qw
\\
&\ovalgate{\hatc'}
&\ovalgate{\hatb'}
&\ovalgate{\hata'}
&\qw
}
\end{array}
\;\;,\;\;
\calr=
\begin{array}{c}
\Qcircuit @C=1em @R=1em @!R{
&\ovalgate{\hatb_f}\qwx[1]
&\ovalgate{\hata_f}\qwx[1]
&\qw
\\
&\ovalgate{\hatb'_f}
&\ovalgate{\hata'_f}
&\qw
}
\end{array}
\;.
\eeq
Let $\lam'=angle(\hata',\hatb')$
and $\phi'=angle(\hatc',\hata'\times\hatb')$.
For any $\call$, if

\begin{subequations}
\beq
\manyx{\hata'\hatb'\hatb'}\cdot\hatc'=0
\;,
\eeq
and

\beq
\left[c_{\phi'}(\hata\cdot\hatb)
\manyx{\hata\hatb} -
s_{\lam'}c_{\lam'}s_{\phi'}\hatb\right]\cdot\hatc=0
\;,
\eeq
\end{subequations}
then
it is possible to find an $\calr$ such that
$\call\sim_R \calr$
and such that $\hatb'_f=\hatc'$.
(Hence,
$\hatc'$
``persists", from initial
circuit $\call$ to  final circuit $\calr$,
as the bottom defining vector
of the leftmost DC-NOT for both circuits.
)

\end{theo}
\proof

The $(\Larrow)$ part of the
proof of the previous theorem
still applies.

Using the definitions
of $\vecv_1$ and $\vecv_2$
given by Eqs.(\ref{eq-defs-v1-v2}), it is not hard
to show that

\beq
\vecv_1^T\vecv_2=0
\;\;\iff\;\;
\left[c_{\phi'}(\hata\cdot\hatb)
\manyx{\hata\hatb} -
s_{\lam'}c_{\lam'}s_{\phi'}\hatb\right]\cdot\hatc=0
\;.
\eeq

Since
$\vecv_1$ and $\vecv_2$ are
orthogonal,
the singular
values and singular vectors
of $\Lam_{2i}$
can be obtained simply by inspection
of Eq.(\ref{eq-Lam2i-svd}).
If $|\hatv_1|\neq 0$ and
$|\hatv_2|\neq 0$, then
one can immediately set

\beq
\hatf'_3 = \hatk'_2
\;,\;\;
\hatf_1 = \frac{\vecv_2}{|\vecv_2|}
\;,\;\;
s_{\alpha'}c_\alpha = |\vecv_2|
\;,
\eeq
and

\beq
\hatf'_1 = \hatc'
\;,\;\;
\hatf_3 = \frac{\vecv_1}{|\vecv_1|}
\;,\;\;
c_{\alpha'}s_\alpha = |\vecv_1|
\;.
\eeq
If $|\vecv_1|=0$ but
$|\vecv_2|\neq 0$,
choose $\hatf_3=\hatf_1\times\hatf_2$.
If $|\vecv_1|\neq 0$ but
$|\vecv_2|= 0$,
choose $\hatf_1=\hatf_2\times\hatf_3$.
If $|\vecv_1|=0$ and $|\vecv_2|=0$,
choose $\hatf_1$ and $\hatf_3$
to be any vectors that make
$(\hatf_j)_{j=1,2,3}$ a RHON basis.

\qed
\subsubsection{3 to 1 DC-NOTs}
\label{sec-3to1-cnots}

In this section, we give necessary
and sufficient conditions for
a circuit with 3 DC-NOTs acting
on 2 qubits to reduce to
1 DC-NOT.

\begin{figure}[h]
    \begin{center}
    \epsfig{file=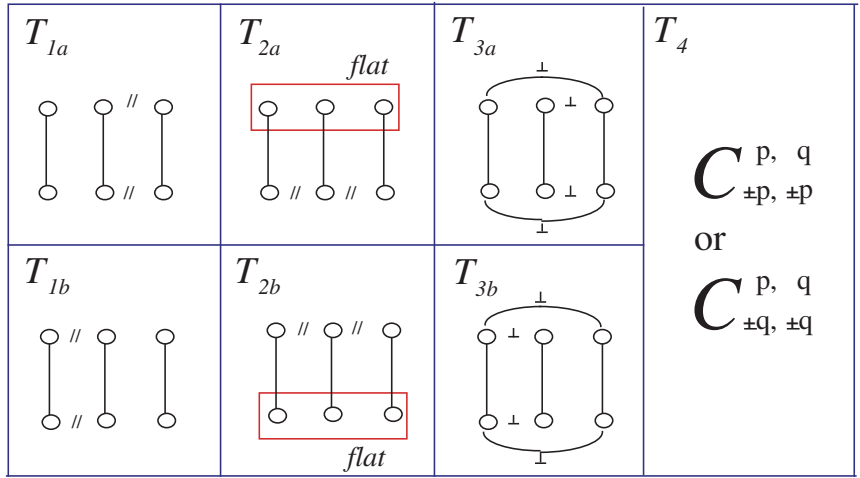, height=2.5in}
    \caption{All circuits with 3 DC-NOTs
    that reduce to 1 DC-NOTs. The 8 circuits
    $\cpq_{\pm p, \pm p}$ and $\cpq_{\pm q, \pm q}$
    are defined by Eq.(\ref{eq-all-t4}).}
    \label{fig-3to1}
    \end{center}
\end{figure}

\begin{theo}
Suppose

\beq
\call=
\begin{array}{c}
\Qcircuit @C=1em @R=1em @!R{
&\ovalgate{\hatc}\qwx[1]
&\ovalgate{\hatb}\qwx[1]
&\ovalgate{\hata}\qwx[1]
&\qw
\\
&\ovalgate{\hatc'}
&\ovalgate{\hatb'}
&\ovalgate{\hata'}
&\qw
}
\end{array}
\;\;,\;\;
\calr=
\begin{array}{c}
\Qcircuit @C=1em @R=1em @!R{
&\ovalgate{\hata_f}\qwx[1]
&\qw
\\
&\ovalgate{\hata'_f}
&\qw
}
\end{array}
\;.
\eeq
Let
$\calv = \hata\times\hatb\cdot\hatc$,
and $\calv' = \hata'\times\hatb'\cdot\hatc'$.
For any $\call$,
it is possible to find an $\calr$ such that
$\call\sim_R \calr$
if and only if
one or more of the following
 are true:
(See Fig.\ref{fig-3to1})
\begin{enumerate}

\item[$T_{1a}$]:
$(\hatb \parallel \hata)$ and
$(\hatb' \parallel \hata')$

\item[$T_{1b}$]:
$(\hatc \parallel \hatb)$ and
$(\hatc' \parallel \hatb')$

\item[$T_{2a}$]:
$(\hatc' \parallel\hatb' \parallel \hata')$
and
$\calv=0$

\item[$T_{2b}$]:
$(\hatc \parallel \hatb \parallel \hata)$
and
$\calv'=0$

\item[$T_{3a}$]:
$\hata \perp span(\hatb,\hatc)$
and
$\hata' \perp span(\hatb',\hatc')$

\item[$T_{3b}$]:
$\hatc \perp span(\hata,\hatb)$
and
$\hatc' \perp span(\hata',\hatb')$

\item[$T_4$]:
\beq
\left\{
\begin{array}{l}
\manyx{\hata\hatb\hatb}\cdot\hatc=0
\;\;and\;\;
\manyx{\hata'\hatb'\hatb'}\cdot\hatc'=0
\\
\frac{|\hata\times\hatb|}{|\hata\cdot\hatb|}=
\frac{|\hata'\times\hatb'|}{|\hata'\cdot\hatb'|}
\;\;and\;\;
\frac{|\hatb\times\hatc|}{|\hatb\cdot\hatc|}=
\frac{|\hatb'\times\hatc'|}{|\hatb'\cdot\hatc'|}
\\
\sign
\left(
\frac{\calv}{(\hata\cdot\hatb)(\hatb\cdot\hatc)}
\right)
=
-\sign
\left(
\frac{\calv'}{(\hata'\cdot\hatb')(\hatb'\cdot\hatc')}
\right)
\end{array}
\right.
\;.
\label{eq-t4-def}
\eeq
\end{enumerate}

\end{theo}
\proof

\lproof

Consider a circuit of type $T_{1b}$
($T_{1a}$ case is analogous).
When $\hatc=\hatb$ and $\hatc'=\hatb'$,
it is obvious that a $T_{1b}$ circuit
reduces to a single DC-NOT.
More generally,
$\hatc=\pm\hatb$
and
$\hatc'=\pm\hatb'$.
Let $\call_{new}$ be a new circuit
obtained by replacing in $\call$:
(1)$\hatc$ by its negative if $\hatc=-\hatb$,
(2)$\hatc'$ by its negative if $\hatc'=-\hatb'$.
By virtue of Eq.(\ref{eq-dcnot-with-neg}),
$\call = (U'\otimes U)\call_{new}$,
where $U',U\in U(2)$.
If $\call_{new}\sim_R \calr_{new}$, then
$\call\sim_R (U'\otimes U)\calr_{new}
(U^{'\dagger}\otimes U^\dagger)$.

Now consider a circuit of type $T_{2a}$
($T_{2b}$ case is analogous).
Note that when $\calv=0$,

\beq
\sigma_\hatc\sigma_\hatb\sigma_\hata=
\sigma_{
(\hata\cdot\hatb)\hatc-\manyx{\hata\hatb\hatc}
}
=\sigma_{\hata_f}
\;,
\eeq
so

\beq
\begin{array}{c}
\Qcircuit @C=1em @R=1em @!R{
&\ovalgate{\hatc}\qwx[1]
&\ovalgate{\hatb}\qwx[1]
&\ovalgate{\hata}\qwx[1]
&\qw
\\
&\ovalgate{\hata'}
&\ovalgate{\hata'}
&\ovalgate{\hata'}
&\qw
}
\end{array}
=
\begin{array}{c}
\Qcircuit @C=1em @R=1em @!R{
&\gate{\sigma_\hatc\sigma_\hatb\sigma_\hata}\qwx[1]
&\qw
\\
&\ovalgate{\hata'}
&\qw
}
\end{array}
=
\begin{array}{c}
\Qcircuit @C=1em @R=1em @!R{
&\ovalgate{\hata_f}\qwx[1]
&\qw
\\
&\ovalgate{\hata'}
&\qw
}
\end{array}
\;.
\eeq
Thus, when
$\hata'=\hatb'=\hatc'$,
a $T_{2a}$ circuit reduces to a
single DC-NOT.
More generally, $\hata'=\pm \hatb'$
and  $\hatc'=\pm \hatb'$.
Let $\call_{new}$ be a new circuit
obtained by replacing in $\call$:
(1)$\hata'$ by its negative if $\hata'=-\hatb'$,
(2)$\hatc'$ by its negative if $\hatc'=-\hatb'$.
By virtue of Eq.(\ref{eq-dcnot-with-neg}),
$\call = (I_2\otimes U)\call_{new}(I_2\otimes V)$,
where $U,V\in U(2)$.
If $\call_{new}\sim_R \calr_{new}$, then
$\call\sim_R (I_2\otimes U)\calr_{new}
(I_2\otimes U^\dagger)$.

Circuits of type $T_{3a}$
($T_{3b}$ case is analogous)
reduce to a single DC-NOT by
virtue of Theorem \ref{th-2thirds-ab}.

\begin{figure}[h]
    \begin{center}
    \epsfig{file=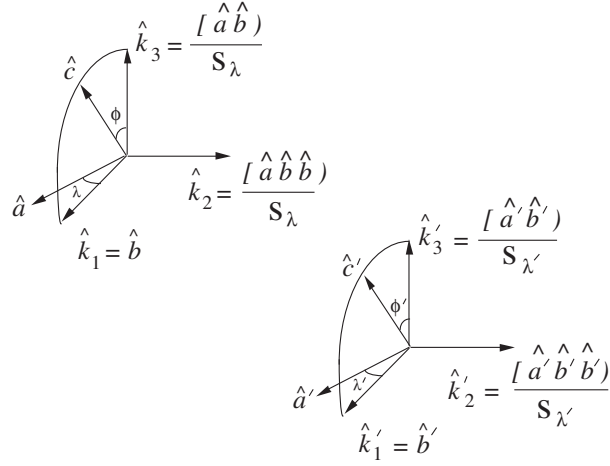, height=2.5in}
    \caption{Vectors and angles associated with
    bit-0 space spanned by $\hata,\hatb,\hatc$.
    Vectors and angles associated with
    bit-1 space spanned by $\hata',\hatb',\hatc'$. }
    \label{fig-double-phi-lam}
    \end{center}
\end{figure}

Now consider a circuit of type $T_4$.
For any $w_1,w_2\in\{x,y,z\}$ and
$\xi\in \RR$,
let $\hatp_{w_1,w_2}^\xi$ and
$\hatq_{w_1,w_2}^\xi$ be defined as in
Eq.(\ref{eq-general-p-q-def}).
Because of the first line
of Eq.(\ref{eq-t4-def}),
one can choose a special coordinate system
for bit 0 such that
$\hatc\rarrow \hatpzx^\phi$,
$\hatb\rarrow \hatx$,
$\hata\rarrow \hatqxy^\lam$,
and a special coordinate system for bit 1 such that
$\hatc'\rarrow \hatpzx^{\phi'}$,
$\hatb'\rarrow \hatx$,
$\hata'\rarrow \hatqxy^{\lam'}$.
See Fig.\ref{fig-double-phi-lam}.
$\hatc,\hatb,\hata$ and
$\hatc',\hatb',\hata'$ are portrayed in
Fig.\ref{fig-double-phi-lam},
when $(\hatk_j)_{j=1,2,3}$ and
$(\hatk'_j)_{j=1,2,3}$ are
the standard basis.
In the special coordinate systems, the first line of
Eq.(\ref{eq-t4-def})
is satisfied by construction.
The second line of Eq.(\ref{eq-t4-def}) becomes

\beq
|\tan \lam| = |\tan \lam'|
\;\;{\rm and}\;\;
|\tan \phi| = |\tan \phi'|
\;,
\label{eq-tan-lam-phi-abs}
\eeq
and the third line

\beq
\sign\left(
\frac{\tan\lam}{\tan\phi}
\right)
=
-\sign\left(
\frac{\tan\lam'}{\tan\phi'}
\right)
\;.
\label{eq-tan-lam-phi-sig}
\eeq

In general, Eq.(\ref{eq-tan-lam-phi-abs})
 is satisfied iff
$\lam'\in\{\pm\lam, \pi\pm \lam\} + 2\pi \ZZ$
and
$\phi'\in\{\pm\phi, \pi\pm \phi\} + 2\pi \ZZ$.
This gives
16 sign possibilities, but
only 8 of them satisfy
Eq.(\ref{eq-tan-lam-phi-sig}).
Indeed,
let
$\cpq_{\pm p, \pm p}$ and
$\cpq_{\pm q, \pm q}$
denote the following 8 circuits:

\beq
\cpq_{(-1)^{m}r, (-1)^{n}r}
=
\begin{array}{c}
\Qcircuit @C=1em @R=1em @!R{
&\ovalgate{\hatpzx^\phi}\qwx[1]
&\ovalgate{\hatx}\qwx[1]
&\ovalgate{\hatqxy^\lam}\qwx[1]
&\qw
\\
&\ovalgate{(-1)^{m}r^\phi_{zx}}
&\ovalgate{\hatx}
&\ovalgate{(-1)^{n}r^\lam_{xy}}
&\qw
}
\end{array}
\;,
\label{eq-all-t4}
\eeq
where $r\in \{p,q\}$
and $m,n\in Bool$.
The following $4\times 4$
matrix has rows
labeled by
the 4 possible values of $\phi'$,
and columns labeled by the
4 possible values of $\lam'$.
As its $(\phi',\lam')$ entry,
the matrix has: the  $T_4$
circuit implied by
 that
value of $(\phi',\lam')$,
if such a circuit exists, or an
$\times$ if none exists.

\beq
\begin{array}{l||c|c|c|c}
 \phi'=\downarrow,\lam'=\rarrow & \lam   &\pi-\lam & \pi+\lam & -\lam \\
         \hline\hline
\phi     & \times & \cpq_{p,-p}  & \times   &\cpq_{p,p}\\
\hline
\pi-\phi & \cpq_{-q,q} & \times  & \cpq_{-q,-q}  & \times\\
\hline
\pi+\phi & \times & \cpq_{-p,-p} & \times   &\cpq_{-p,p}\\
\hline
-\phi    & \cpq_{q,q}  & \times  & \cpq_{q,-q}   & \times
\end{array}
\;.
\label{eq-table-t4-ckts}
\eeq
In conclusion,
the 3 lines of Eq.(\ref{eq-t4-def})
imply, in the special coordinate systems,
a circuit of
type Eq.(\ref{eq-all-t4}).

For $\cpq_{q,q}$
(ditto, for $\cpq_{p,p}$), there
exists an $\calr$ such that
$\call\sim_R\calr$ by
virtue of Eq.(\ref{eq-split-sim-trans-id})
(ditto, Eq.(\ref{eq-split-sim-trans-id2})).
The other 6 circuits of table
Eq.(\ref{eq-table-t4-ckts}) can be handled as follows.
Let $\call_{new}$ be a new circuit
obtained by replacing in $\call$:
(1)$\lam'$ by $\lam'-\pi$
 if $\lam'=\pi\pm\lam \mod(2\pi)$,
(2)$\phi'$ by $\phi'-\pi$
if $\phi'=\pi\pm\phi\mod(2\pi)$.
By virtue of Eq.(\ref{eq-dcnot-with-neg}),
$\call = (U'\otimes U)\call_{new}(V'\otimes V)$
where $U',U,V',V\in U(2)$, and where
$\call_{new}$
is of type
$\cpq_{q,q}$ or $\cpq_{p,p}$.
If $\call_{new}\sim_R \calr_{new}$, then
$\call\sim_R (U'\otimes U)\calr_{new}
(U^{'\dagger}\otimes U^\dagger)$.

\rproof

$\call\sim_R \calr$ so
$\hat{\call}^{(2)}= \pm \hat{\calr}^{(2)}$.
In light of
Eq.(\ref{eq-invar-of-hat-graph}), this gives

\beq
i^3 \call^{(2)}
= \pm i \calr^{(2)}
\;.
\eeq
It follows that

\beq
\lam_{3r} + i\lam_{3i} + \Lam_{3r} + i\Lam_{3i} = \pm \sigma_{\hata'_f,\hata_f}
\;,
\eeq
where

\beq
\lam_{3r} =
\manyx{\hata'\hatb' \hatb'}\cdot\hatc'
\;\;
\manyx{\hata\hatb\hatb}\cdot\hatc
\;,
\eeq

\beq
\lam_{3i} =
-(\hata\cdot\hatb)(\hatb\cdot\hatc)\calv'
-(\hata'\cdot\hatb')(\hatb'\cdot\hatc')\calv
\;,
\label{eq-lam3i-3to1}
\eeq

\beq
\Lam_{3r}=
\left\{
\begin{array}{l}
-(\hata'\cdot\hatb')(\hata\cdot\hatb)
\hatc'\hatc^T
\\
+(\hata\cdot\hatb)(\hatb\cdot\hatc)
\manyx{\hata'\hatb'\hatc'}\hatc^T
+(\hata'\cdot\hatb')(\hatb'\cdot\hatc')
\hatc'\manyx{\hata\hatb\hatc}^T
\\
+(\hata'\cdot\hatb')\calv \manyx{\hatb'\hatc'}\hatc^T
+(\hata\cdot\hatb)\calv'\hatc'\manyx{\hatb\hatc}^T
\\
-
\manyx{\hata'\hatb'\hatb'\hatc'\hatc'}
\manyx{\hata\hatb\hatb\hatc\hatc}^T
\;,
\end{array}
\right.
\eeq
and

\beq
\Lam_{3i}=
\left\{
\begin{array}{l}
+(\hata\cdot\hatb)
\manyx{\hata'\hatb'\hatc'\hatc'}
\manyx{\hatb\hatc\hatc}^T
+
(\hata'\cdot\hatb')
\manyx{\hatb'\hatc'\hatc'}
\manyx{\hata\hatb\hatc\hatc}^T
\\
+\manyx{\hata\hatb\hatb}\cdot \hatc
\manyx{\hata'\hatb'\hatb'\hatc'}\hatc^T
+
\manyx{\hata'\hatb'\hatb'}\cdot \hatc'
\hatc'\manyx{\hata\hatb\hatb\hatc}^T
\end{array}
\right.
\;.
\eeq

{\bf First assume that there
are no breaches in} $\call$ (i.e.,
$\hata \not\parallel \hatb$,
$\hatb \not\parallel \hatc$,
$\hata' \not\parallel \hatb'$,
$\hatb' \not\parallel \hatc'$).

 Note that

\beq
\manyx{\hata\hatb\hatb}\cdot\hatc=0
\;\;{\rm and}\;\;
\manyx{\hata'\hatb'\hatb'}\cdot\hatc'=0
\;.
\label{eq-2right-triangles}
\eeq
This is why.
From $\lam_{3r}=0$, we must
have either
$\manyx{\hata\hatb\hatb}\cdot\hatc=0$
or
$\manyx{\hata'\hatb'\hatb'}\cdot\hatc'=0$.
But if one of these holds,
then the other one follows. Indeed,
assume
$\manyx{\hata\hatb\hatb}\cdot\hatc=0$.
Since also
$\manyx{\hata\hatb}\neq 0$,
it follows that
$\manyx{\hata\hatb\hatb\hatc}\neq 0$.
From $\Lam_{3i}=0$,
$\hatc'^T\Lam_{3i}=
\manyx{\hata'\hatb'\hatb'}\cdot\hatc'
\manyx{\hata\hatb\hatb\hatc}^T=0$
so
$\manyx{\hata'\hatb'\hatb'}\cdot\hatc'=0$.
By an analogous argument,
assuming
$\manyx{\hata'\hatb'\hatb'}\cdot\hatc'=0$
leads to
$\manyx{\hata\hatb\hatb}\cdot\hatc=0$.

Next note that

\begin{subequations}
\label{eq-perp-spreading}
\beq
\hata\cdot\hatb=\hata'\cdot\hatb'=0
\;\;\Rarrow\;\;
\hata\cdot\hatc=\hata'\cdot\hatc'=0
\;,
\label{eq-perp-spreading-from-a}
\eeq
and

\beq
\hatc\cdot\hatb=\hatc'\cdot\hatb'=0
\;\;\Rarrow\;\;
\hatc\cdot\hata=\hatc'\cdot\hata'=0
\;.
\label{eq-perp-spreading-from-c}
\eeq
\end{subequations}
Eqs.(\ref{eq-perp-spreading})
become obvious if one
uses the BAC minus CAB identity
to expand Eqs.(\ref{eq-2right-triangles}).

It is convenient at this point to define
a RHON basis $(\hatk'_j)_{j=1,2,3}$
for the 3d real space spanned by
$\hata',\hatb', \hatc'$. Let
$s_{\lam'} = |\manyx{\hata'\hatb'}|$.
If $s_{\lam'}\neq 0$, let

\beq
(\hatk'_j)_{j=1,2,3}=
(\hatb',
\frac{\manyx{\hata'\hatb'\hatb'}}{s_{\lam'}},
\frac{\manyx{\hata'\hatb'}}{s_{\lam'}}
)
\;.
\label{eq-gprime-basis-3to1}
\eeq
If $s_{\lam'}=0$,
let $(\hatk_j)_{j=1,2,3}$ be any
RHON basis such that $\hatk'_1=\hatb'$
and $\hatk'_2$ is perpendicular
to $span(\hatb',\hatc')$.
Let
$\phi'=angle(\hatc', \hatk'_3)$.
Since
$\manyx{\hata'\hatb'\hatb'}\cdot\hatc'=0$,

\beq
\hata' = c_{\lam'}\hatk'_1 - s_{\lam'} \hatk'_2,
\;,\;\;
\hatb' = \hatk'_1
\;,\;\;
\hatc' = s_{\phi'} \hatk'_1 + c_{\phi'} \hatk'_3
\;.
\label{eq-abcprime-3to1}
\eeq
Eqs.(\ref{eq-gprime-basis-3to1})
and (\ref{eq-abcprime-3to1})
are illustrated in Fig.\ref{fig-double-phi-lam}.

Use the previous paragraph with all
the primes removed
to define angles $\lam,\phi$ and
a RHON basis
$(\hatk_j)_{j=1,2,3}$
for the 3d real space spanned by
$\hata,\hatb,\hatc$.

When expressed in terms of
$\lam,\lam',\phi$ and $\phi'$,
the constraint
$\lam_{3i}=0$
reduces to

\beq
-[
 s_{\lam'}c_\lam c_{\phi'}s_\phi
+c_{\lam'}s_\lam s_{\phi'}c_\phi] = 0
\;.
\label{eq-eq1-of-system}
\eeq
Likewise, the constraint
$\Lam_{3i}=0$ reduces to

\beq
-[
 s_{\lam'}c_\lam s_{\phi'}c_\phi
+c_{\lam'}s_\lam c_{\phi'}s_\phi]
\manyx{\hatk'_2\hatc'}\manyx{\hatk_2\hatc}^T = 0
\;.
\label{eq-eq2-of-system}
\eeq

Eqs.(\ref{eq-eq1-of-system}) and
(\ref{eq-eq2-of-system}) imply the following
system of 2 equations:

\beq
\left[
\begin{array}{cc}
c_{\phi'}s_\phi & s_{\phi'}c_\phi\\
s_{\phi'}c_\phi & c_{\phi'}s_\phi
\end{array}
\right]
\left[
\begin{array}{c}
s_{\lam'}c_\lam \\
c_{\lam'}s_\lam
\end{array}
\right]
=0
\;.
\label{eq-system-lam-sol}
\eeq
This system of equations can
also be rewritten in the form:

\beq
\left[
\begin{array}{cc}
c_{\lam'}s_\lam & s_{\lam'}c_\lam\\
s_{\lam'}c_\lam & c_{\lam'}s_\lam
\end{array}
\right]
\left[
\begin{array}{c}
s_{\phi'}c_\phi \\
c_{\phi'}s_\phi
\end{array}
\right]
=0
\;.
\label{eq-system-phi-sol}
\eeq

For Eq.(\ref{eq-system-lam-sol}),
either (i)the solution is the zero vector,
or (ii)the determinant of
the coefficient matrix vanishes. (i)If the
solution is zero, then
$s_{\lam'}c_\lam=c_{\lam'} s_\lam=0$.
Since we are assuming no breaches,
$s_{\lam'}\neq 0$ and $s_{\lam}\neq 0$,
so we must have $c_\lam= c_{\lam'}=0$.
By virtue of Eq.(\ref{eq-perp-spreading-from-a}),
 this means that the circuit
must be of type $T_{3a}$. (ii) If the
determinant is zero, then

\beq
|\tan \phi\;| = |\tan \phi'\;|
\;.
\label{eq-tan-phi}
\eeq
We will pursue this possibility later on.

Likewise, for Eq.(\ref{eq-system-phi-sol}),
either (i)the solution is the zero vector,
or (ii)the determinant of
the coefficient matrix vanishes. (i)If the
solution is zero, then
$s_{\phi'}c_\phi=c_{\phi'} s_\phi=0$.
Since we are assuming no breaches,
$c_{\phi}\neq 0$ and $c_{\phi'}\neq 0$,
so we must have $s_{\phi'}= s_{\phi}=0$.
By virtue of Eq.(\ref{eq-perp-spreading-from-c}),
 this means that the circuit
must be of type $T_{3b}$. (ii) If the
determinant is zero, then

\beq
|\tan \lam\;| = |\tan \lam'\;|
\;.
\label{eq-tan-lam}
\eeq
We will pursue this possibility later on.

 Suppose we assume that the circuit $\call$
is not of type $T_3$. Then, we have shown
that it must satisfy
Eqs.(\ref{eq-tan-phi}) and (\ref{eq-tan-lam}). But
these two equations are
the second line of Eq.(\ref{eq-t4-def}).
To prove that the circuit must be
of type $T_4$,
it remains for us to prove that the third
line of Eq.(\ref{eq-t4-def}) also holds.
This third line
clearly follows from $\lam_{3i}=0$,
where $\lam_{3i}$
is given by Eq.(\ref{eq-lam3i-3to1}).

{\bf Next , assume that there is at least one
breach in} $\call$. Without loss of generality,
assume there is a breach between $\hata$ and
$\hatb$ (i.e., $\hata \parallel \hatb$).

$\hata \parallel \hatb$ implies
that $\calv=0$.

The constraint $\lam_{3i}=0$ reduces to

\beq
(\hatb\cdot\hatc)\calv'=0
\;,
\label{eq-b-c-volp}
\eeq
which implies that either $\hatb\cdot\hatc=0$
or $\calv'=0$.
The constraint $\Lam_{3i}=0$ reduces
to

\beq
\manyx{\hata'\hatb'\hatc'\hatc'}
\manyx{\hatb\hatc\hatc}^T =0
\;,
\eeq
which implies that either
(i)$\hatb\parallel \hatc$
or (ii)$\hata'\parallel \hatb'$
or (iii)$\manyx{\hata'\hatb'}\parallel \hatc'$.
(i)If $\hatb\parallel \hatc$,
then, by Eq.(\ref{eq-b-c-volp}), $\calv'=0$.
$(\hata\parallel \hatb \parallel \hatc)$
and $\calv'=0$ so $\call$ is of
type $T_{2b}$.
(ii)If $\hata'\parallel \hatb'$, then, since
also $\hata\parallel \hatb$,
$\call$ is of type $T_{1a}$.
(iii)Suppose $\manyx{\hata'\hatb'}\parallel \hatc'$.
Assume that $\hata'\not \parallel \hatb'$
as the case when these two vectors are parallel
has already been considered. It follows that
the conditions $T_{3b}$ are satisfied.
This is why.
$\manyx{\hata'\hatb'}\parallel \hatc'$
and $\hata'\not \parallel \hatb'$ imply that
$\calv'\neq 0$, and, therefore, by virtue
of Eq.(\ref{eq-b-c-volp}), $\hatc\perp\hatb$.
Now $\hatc\perp\hatb$ and $\hata\parallel\hatb$
imply that $\hatc\perp\hata$.
Thus, $\hatc\perp span(\hatb,\hata)$.
Also, $\manyx{\hata'\hatb'}\parallel \hatc'$
implies that $\hatc'\perp span(\hatb',\hata')$.

\qed
\subsubsection{3 to 0 DC-NOTs}
\label{sec-3to0-cnots}

In this section, we give necessary
and sufficient conditions for
a circuit with 3 DC-NOTs acting
on 2 qubits to reduce to
zero DC-NOTs (i.e., to merely local operations).

\begin{figure}[h]
    \begin{center}
    \epsfig{file=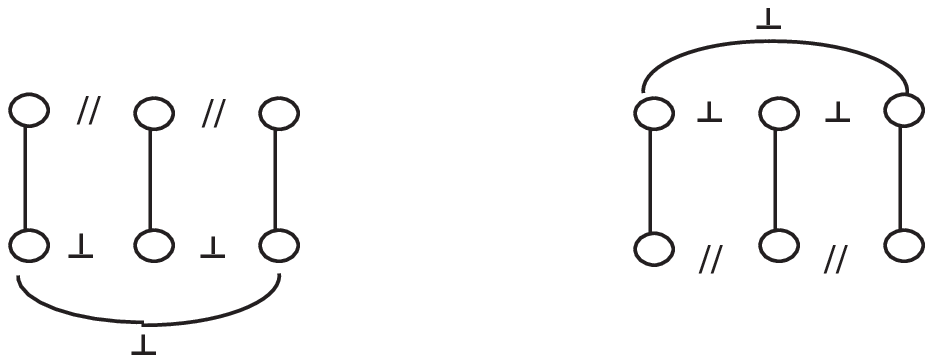, height=1in}
    \caption{All circuits with 3 DC-NOTs
    that reduce to 0 DC-NOTs.}
    \label{fig-3to0}
    \end{center}
\end{figure}

\begin{theo}
Suppose

\beq
\call=
\begin{array}{c}
\Qcircuit @C=1em @R=1em @!R{
&\ovalgate{\hatc}\qwx[1]
&\ovalgate{\hatb}\qwx[1]
&\ovalgate{\hata}\qwx[1]
&\qw
\\
&\ovalgate{\hatc'}
&\ovalgate{\hatb'}
&\ovalgate{\hata'}
&\qw
}
\end{array}
\;.
\eeq
For any $\call$,
$\call\sim_R 1$ if and only if
one of the following is true
(see Fig.\ref{fig-3to0})
\begin{enumerate}
\item[$T_a$]:
$(\hatc',\hatb',\hata')$ are mutually orthogonal,
and
$(\hatc\parallel\hatb\parallel\hata)$

\item[$T_b$]:
$(\hatc,\hatb,\hata)$ are mutually orthogonal,
and
$(\hatc'\parallel\hatb'\parallel\hata')$

\end{enumerate}
\end{theo}
\proof

\lproof

Consider a circuit of type $T_b$
($T_a$ case is analogous).
Note that when $(\hatc,\hatb,\hata)$
are mutually orthogonal,

\beq
\sigma_\hatc\sigma_\hatb\sigma_\hata=
i\hatc\cdot\manyx{\hatb\hata} = \pm i
\;.
\eeq
Hence,

\beq
\begin{array}{c}
\Qcircuit @C=1em @R=1em @!R{
&\ovalgate{\hatc}\qwx[1]
&\ovalgate{\hatb}\qwx[1]
&\ovalgate{\hata}\qwx[1]
&\qw
\\
&\ovalgate{\hata'}
&\ovalgate{\hata'}
&\ovalgate{\hata'}
&\qw
}
\end{array}
=
\begin{array}{c}
\Qcircuit @C=1em @R=1em @!R{
&\gate{\sigma_\hatc\sigma_\hatb\sigma_\hata}\qwx[1]
&\qw
\\
&\ovalgate{\hata'}
&\qw
}
\end{array}
=
\begin{array}{c}
\Qcircuit @C=1em @R=1em @!R{
&\qw
&\qw
\\
&\gate{(\pm i)^{n_{\hata'}}}
&\qw
}
\end{array}
\;.
\eeq
Thus, when
 $\hata'=\hatb'=\hatc'$,
 a $T_b$ circuit reduces to zero DC-NOTs.
More generally, $\hata'=\pm \hatb'$
and  $\hatc'=\pm \hatb'$.
Let $\call_{new}$ be a new circuit
obtained by replacing in $\call$:
(1)$\hata'$ by its negative if $\hata'=-\hatb'$,
(2)$\hatc'$ by its negative if $\hatc'=-\hatb'$.
By virtue of Eq.(\ref{eq-dcnot-with-neg}),
$\call = (I_2\otimes U)\call_{new}(I_2\otimes V)$
where $U,V\in U(2)$.
If $\call_{new}\sim_R 1$, then
$\call\sim_R 1$.

\rproof

$\call\sim_R 1$ so
$\hat{\call}^{(2)}= \pm 1$.
In light of
Eq.(\ref{eq-invar-of-hat-graph}), this gives

\beq
i^3 \call^{(2)}
= \pm 1
\;.
\eeq
It follows that

\beq
\lam_{3r} + i\lam_{3i}
+ \Lam_{3r} + i\Lam_{3i} =
\pm i
\;,
\eeq
where

\beq
\lam_{3r} =
\manyx{\hata'\hatb' \hatb'}\cdot\hatc'
\;\;
\manyx{\hata\hatb\hatb}\cdot\hatc
\;,
\eeq

\beq
\lam_{3i} =
-(\hata\cdot\hatb)(\hatb\cdot\hatc)\calv'
-(\hata'\cdot\hatb')(\hatb'\cdot\hatc')\calv
\;,
\eeq

\beq
\Lam_{3r}=
\left\{
\begin{array}{l}
-(\hata'\cdot\hatb')(\hata\cdot\hatb)
\hatc'\hatc^T
\\
+(\hata\cdot\hatb)(\hatb\cdot\hatc)
\manyx{\hata'\hatb'\hatc'}\hatc^T
+(\hata'\cdot\hatb')(\hatb'\cdot\hatc')
\hatc'\manyx{\hata\hatb\hatc}^T
\\
+(\hata'\cdot\hatb')\calv \manyx{\hatb'\hatc'}\hatc^T
+(\hata\cdot\hatb)\calv'\hatc'\manyx{\hatb\hatc}^T
\\
-
\manyx{\hata'\hatb'\hatb'\hatc'\hatc'}
\manyx{\hata\hatb\hatb\hatc\hatc}^T
\;,
\end{array}
\right.
\eeq
and

\beq
\Lam_{3i}=
\left\{
\begin{array}{l}
+(\hata\cdot\hatb)
\manyx{\hata'\hatb'\hatc'\hatc'}
\manyx{\hatb\hatc\hatc}^T
+
(\hata'\cdot\hatb')
\manyx{\hatb'\hatc'\hatc'}
\manyx{\hata\hatb\hatc\hatc}^T
\\
+\manyx{\hata\hatb\hatb}\cdot \hatc
\manyx{\hata'\hatb'\hatb'\hatc'}\hatc^T
+
\manyx{\hata'\hatb'\hatb'}\cdot \hatc'
\hatc'\manyx{\hata\hatb\hatb\hatc}^T
\end{array}
\right.
\;.
\eeq

$\hatc^{\;'T}\Lam_{3r}\hatc = 0$ so
$\hata'\cdot\hatb'=0$ or $\hata\cdot\hatb=0$.
Both can't be true at once or else
we would have $\lam_{3i}=0$,
which is false. Assume henceforth that
$\hata'\cdot\hatb'\neq 0$ and $\hata\cdot\hatb=0$
(the case
$\hata'\cdot\hatb'= 0$ and $\hata\cdot\hatb\neq 0$
is analogous).
When $\hata\cdot\hatb=0$, $|\lam_{3i}|=1$
reduces to
$|(\hata'\cdot\hatb')(\hatb'\cdot\hatc')\calv|=1$,
which immediately implies that
$(\hatc'\parallel\hatb'\parallel\hata')$,
and
$(\hatc,\hatb,\hata)$ are mutually orthogonal.
Thus, circuit $\call$ must be of type $T_b$.

\qed
\subsection{Reducing Controlled-$U$'s}
\label{sec-contr-us}
\subsubsection{One Controlled-$U$}
\label{sec-1-contr-u}

In this section, we show that any controlled-U can
be expressed with just two CNOTs. This result
was first proven by
Barenco et al. in Ref.\cite{Bar95}. Their method of
proof is long and opaque compared
with the ultra simple proof given
below. This attests to the benefits
of using dressed CNOTs.

\begin{theo}
Let $\theta\in \RR$.
Suppose

\beq
\call =
\begin{array}{c}
\Qcircuit @C=1em @R=1em @!R{
&\gate{e^{i\theta\sigma_{\hatw}}}\qwx[1]
&\qw
\\
&\dotgate
&\qw
}
\end{array}
\;\;,\;\;
\calr=
\begin{array}{c}
\Qcircuit @C=1em @R=1em @!R{
&\ovalgate{\hatb}\qwx[1]
&\ovalgate{\hata}\qwx[1]
&\qw
\\
&\ovalgate{\hatz}
&\ovalgate{\hatz}
&\qw
}
\end{array}
\;.
\eeq
For any $\call$, it is possible to
find an $\calr$ such that $\call= \calr$.
\end{theo}
\proof

Given a unit vector $\hatw$
and an angle $\theta$,
we can always find (non-unique)
unit vectors
$\hatb$ and $\hata$ such that
$angle(\hatb,\hata)=\theta$, and
$\hatb\times\hata$ points along $\hatw$. Then
$\hatb\cdot\hata=\cos(\theta)$
and
$\hatb\times\hata=\sin(\theta)\hatw$ so
$\sigma_\hatb \sigma_\hata =e^{i\theta\sigma_\hatw}$.

\beq
[e^{i\theta\sigma_\hatw(0)}]^{n(1)}=
[\sigma_\hatb(0)\sigma_\hata(0)]^{n(1)}
=
\sigma_\hatb(0)^{n(1)}
\sigma_\hata(0)^{n(1)}
\;.
\eeq
\qed
\subsubsection{Two Controlled-$U$'s
(The Deflation Identity)
\\{\footnotesize\tt[
deflate\_dcnots.m,
test\_deflate\_dcnots.m
]}}
\label{sec-2-contr-u}

In this section, we show that
a product of two controlled-Us can
be expressed with just two CNOTs. This
``Deflation Identity"
was first proven in Ref.\cite{Tuc-deflation}.
Unlike the
proof of Ref.\cite{Tuc-deflation}, the one below
uses dressed CNOTs.

\begin{figure}[h]
    \begin{center}
    \epsfig{file=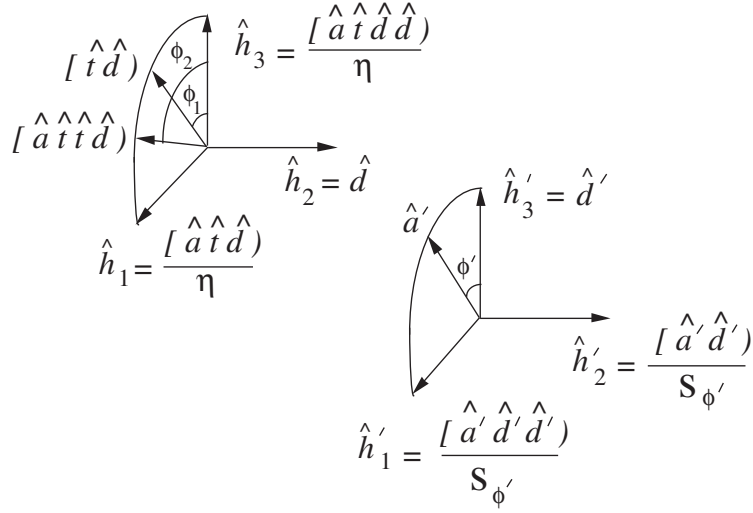, height=2.75in}
    \caption{Variables used in
    Theorem \ref{th-deflation}.}
    \label{fig-deflation}
    \end{center}
\end{figure}

\begin{theo}\label{th-deflation}
Let $A\in SU(2)$ and $\theta_L, \theta_R\in \RR$.
Suppose

\beq
\call =
\begin{array}{c}
\Qcircuit @C=1em @R=1em @!R{
&\gate{e^{i\theta_L\sigma_{\hatw_L}}}\qwx[1]
&\qw
&\gate{e^{i\theta_R\sigma_{\hatw_R}}}\qwx[1]
&\qw
\\
&\dotgate
&\gate{A}
&\dotgate
&\qw
}
\end{array}
\;\;,\;\;
\calr=
\begin{array}{c}
\Qcircuit @C=1em @R=1em @!R{
&\ovalgate{\hatb_f}\qwx[1]
&\ovalgate{\hata_f}\qwx[1]
&\qw
\\
&\ovalgate{\hatb'_f}
&\ovalgate{\hata'_f}
&\qw
}
\end{array}
\;.
\eeq
For any $\call$, it is possible to
find an $\calr$ such that $\call\sim_R \calr$.

\end{theo}
\proof

Given a unit vector $\hatw_L$
and an angle $\theta_L$,
we can always find (non-unique)
unit vectors
$\hatd$ and $\hatc$
such that
$angle(\hatd,\hatc)=\theta_L$, and
$\hatd\times \hatc$ points along $\hatw_L$. Then
$\hatd\cdot\hatc=\cos(\theta_L)$
and
$\hatd\times\hatc=\sin(\theta_L)\hatw_L$ so
$\sigma_\hatd \sigma_\hatc =e^{i\theta_L\sigma_{\hatw_L}}$.
Likewise, given a unit vector $\hatw_L$
and an angle $\theta_L$,
we can always find (non-unique)
unit vectors
$\hatb$ and $\hata$
such that
$\sigma_\hatb \sigma_\hata =e^{i\theta_R\sigma_{\hatw_R}}$.
We are free to rotate the vectors
$\hatd$ and $\hatc$ (ditto,
$\hatb$ and $\hata$) within the plane
they initially span, as long as we don't change the
angle between them.
In particular, we can choose both $\hatc$
and $\hatb$ to lie along
the line of intersection
between the planes
$span(\hatd,\hatc)$ and
$span(\hatb,\hata)$.
In other words, we can always choose
$\hatc=\hatb$. Call $\hatt$ their common value .
It is now clear that, without loss of generality,
we can replace
$\call$ by

\beq
\call =
\begin{array}{c}
\Qcircuit @C=1em @R=1em @!R{
&\ovalgate{\hatd}\qwx[1]
&\qw
&\ovalgate{\hatt}\qwx[1]
&\breach
&\ovalgate{\hatt}\qwx[1]
&\qw
&\ovalgate{\hata}\qwx[1]
&\qw
\\
&\ovalgate{\hatd'}
&\breach
&\ovalgate{\hatd'}
&\qw
&\ovalgate{\hata'}
&\breach
&\ovalgate{\hata'}
&\qw
}
\end{array}
\;.
\eeq

Our goal is to construct an $\calr$ such that
$\call\sim_R \calr$. Such an $\calr$,
if it exists, must satisfy
$\hat{\call}^{(2)}= \pm \hat{\calr}^{(2)}$.
We will use the positive sign.
In light of
Eq.(\ref{eq-invar-of-hat-graph}), this gives

\beq
i^4 \call^{(2)} =  i^2 \calr^{(2)}
\;.
\eeq
Using the same calculational techniques that
were used in
Section \ref{sec-two-bit-dcnot-rhs-invariants},
one finds

\beq
\call^{(2)}=
\left\{
\begin{array}{l}
(\hata\cdot\hatt)(\hatt\cdot\hatd)
-(\hata'\cdot\hatd')
\manyx{\hata\hatt\hatt}\cdot\hatd
\\
i\left[
(\hata\cdot\hatt)\hatd'\manyx{\hatt\hatd}^T
-(\hata'\cdot\hatd')
\hatd'\manyx{\hata\hatt\hatt\hatd}^T
+\manyx{\hata'\hatd'\hatd'}
\manyx{\hata\hatt\hatd\hatd}^T
\right]
\\
+\manyx{\hata\hatt}\cdot\hatd
\manyx{\hata'\hatd'}\hatd^T
\end{array}
\right.
\;.
\eeq
From Section \ref{sec-invariants-2cnots},
we know that $\calr^{(2)}$ can be expressed as

\beqa
\calr^{(2)}&=&
\lam_{2r} + \Lam_{2r} + i \Lam_{2i}\\
&=&
c_{\alpha'} c_\alpha
-(s_{\alpha'}s_\alpha) \hatf_2'\hatf_2^T
+i\;
\begin{array}{l|ll}
        & \hatf_1^T & \hatf_3^T\\
\hline
\hatf_3'& s_{\alpha'}c_\alpha & 0\\
\hatf_1'& 0                   &c_{\alpha'}s_\alpha
\end{array}
\;.
\label{eq-r-invar-2contr-u}
\eeqa

We must have

\beq
\lam_{2r}= c_{\alpha'}c_\alpha =
-(\hata\cdot\hatt)(\hatt\cdot\hatd)
+(\hata'\cdot\hatd')
\manyx{\hata\hatt\hatt}\cdot\hatd
\;,
\eeq
and

\beq
\Lam_{2r}=
-s_{\alpha'}s_\alpha
\hatf_2'\hatf_2^T =
-\manyx{\hata\hatt}\cdot\hatd
\manyx{\hata'\hatd'}\hatd^T
\;.
\label{eq-ssff}
\eeq
Define

\beq
s_{\phi'} = |\manyx{\hata'\hatd'}|
\;,\;\;
\eta = |\manyx{\hata\hatt\hatd}|
\;.
\eeq
If $s_{\phi'}\neq 0$,
Eq.(\ref{eq-ssff}) is satisfied by

\beq
s_{\alpha'}s_\alpha =
\manyx{\hata\hatt}\cdot\hatd
\; s_{\phi'}
\;,\;\;
\hatf'_2 =
\frac{\manyx{\hata'\hatd'}}
{s_{\phi'}}
\;,\;\;
\hatf_2 = \hatd
\;.
\label{eq-fprime2-def-2-contr-u}
\eeq
If $s_{\phi'}=0$,
choose $s_\alpha s_{\alpha'}$
and $\hatf_2$ the
same way, but choose $\hatf'_2$ to
be any vector perpendicular to $\hatd'$.

If $s_{\phi'}\neq 0$ and $\eta\neq 0$,
define the following
two  RHON bases
(illustrated in Fig.\ref{fig-deflation}):

\beq
(\hath'_j)_{j=1,2,3}=
(
\frac{\manyx{\hata'\hatd'\hatd'}}
{ s_{\phi'}},
\frac{\manyx{\hata'\hatd'}}
{ s_{\phi'}},
\hatd'
)
\;,
\label{eq-hprimej-def-2-contr-u}
\eeq
and

\beq
(\hath_j)_{j=1,2,3}=
(
\frac{\manyx{\hata\hatt\hatd}}
{ \eta},
\hatd,
\frac{\manyx{\hata\hatt\hatd\hatd}}
{\eta}
)
\;.
\label{eq-hj-def-2-contr-u}
\eeq
If $s_{\phi'}=0$, pick $(\hath'_j)_{j=1,2,3}$
to be any RHON basis such that $\hath'_3=\hatd'$.
If $\eta=0$, pick $(\hath_j)_{j=1,2,3}$
to be any RHON basis such that $\hath_2=\hatd$.
Define the following  two angles
(illustrated in Fig.\ref{fig-deflation}):

\beq
\phi_2 = angle(\manyx{\hata\hatt\hatt\hatd},
\hath_3)
\;,\;\;
\phi_1 = angle(\manyx{\hatt\hatd},
\hath_3)
\;.
\eeq
We must  have

\beqa
\Lam_{2i}&=&
-(\hata\cdot\hatt)\hatd'\manyx{\hatt\hatd}^T
+(\hata'\cdot\hatd')
\hatd'\manyx{\hata\hatt\hatt\hatd}^T
-\manyx{\hata'\hatd'\hatd'}
\manyx{\hata\hatt\hatd\hatd}^T
\\
&=&
\begin{array}{l|ll}
        & \hath_1^T & \hath_3^T\\
\hline
\hath_3'& -\hata\cdot\hatt s_{\phi_1} + c_{\phi'}s_{\phi_2} &
 -\hata\cdot\hatt c_{\phi_1} + c_{\phi'}c_{\phi_2}\\
\hath_1'& 0                   &-s_{\phi'}\eta
\end{array}
\;.
\eeqa

At this point, we can follow from
step \ref{item-diag-invar2-m-matrix}
to the end of the
Algorithm for Diagonalizing $\calg^{(2)}_2$
that was given in
Section \ref{sec-invariants-2cnots}.
This will yield values for
$\hata_f$,
$\hata'_f$,
$\hatb_f$, and
$\hatb'_f$.

\qed
\subsection{Opening and Closing a Breach
\\{\footnotesize\tt[
breach.m,
test\_breach.m
]}}
\label{sec-opening-breach}

\begin{quote}
{\it
Once more unto the breach, dear friends, once more;
    Or close the wall up with our English dead!}
(from ``King Henry V"
by W. Shakespeare)
\end{quote}

In this section, we show how to ``open and
close a breach" in 2-qubit circuits. This
is a procedure whereby one can reduce any
2-qubit circuit with 4 CNOTs into
a circuit with 3 CNOTs.
Applying this procedure repeatedly, one
can reduce any 2-qubit circuit with more than
3 CNOTs into a circuit with only 3 CNOTs. The fact that
all 2-qubit circuits can be expressed
with 3 (or fewer) CNOTs was first proven in Ref.\cite{VD}.
Unlike the proof below, their proof was based on
Cartan's KAK decomposition\cite{Tuc-KAK}.

\begin{theo}(Opening a Breach)
Suppose

\beq
\call=
\begin{array}{c}
\Qcircuit @C=1em @R=1em @!R{
&\ovalgate{\hatp_L}\qwx[1]
&\ovalgate{\hatq_L}\qwx[1]
&\ovalgate{\hatq_R}\qwx[1]
&\ovalgate{\hatp_R}\qwx[1]
&\qw
\\
&\ovalgate{\hatp'_L}
&\ovalgate{\hatq'_L}
&\ovalgate{\hatq'_R}
&\ovalgate{\hatp'_R}
&\qw
}
\end{array}
\;,
\label{eq-op-cl-breach-call}
\eeq

\beq
\calr=
\begin{array}{c}
\Qcircuit @C=1em @R=1em @!R{
&\ovalgate{\hatp_{Lf}}\qwx[1]
&\ovalgate{\hatq_{Lf}}\qwx[1]
&\qw
&\ovalgate{\hatq_{Rf}}\qwx[1]
&\ovalgate{\hatp_{Rf}}\qwx[1]
&\qw
\\
&\ovalgate{\hatp'_{Lf}}
&\ovalgate{\hatt'}
&\breach
&\ovalgate{\hatt'}
&\ovalgate{\hatp'_{Rf}}
&\qw
}
\end{array}
\;.
\label{eq-op-cl-breach-calr}
\eeq
For any $\call$, it is possible to
find an $\calr$ such that $\call\sim_R \calr$.
\end{theo}
\proof

We begin by inserting a ``unit wedge" into $\call$:

\beq
\call=
\begin{array}{c}
\Qcircuit @C=1em @R=1em @!R{
&\ovalgate{\hatp_L}\qwx[1]
&\ovalgate{\hatq_L}\qwx[1]
&\ovalgate{\hatt}\qwx[1]
&\ovalgate{\hatt}\qwx[1]
&\ovalgate{\hatq_R}\qwx[1]
&\ovalgate{\hatp_R}\qwx[1]
&\qw
\\
&\ovalgate{\hatp'_L}
&\ovalgate{\hatq'_L}
&\ovalgate{\hatt'}
&\ovalgate{\hatt'}
&\ovalgate{\hatq'_R}
&\ovalgate{\hatp'_R}
&\qw
}
\end{array}
\;.
\label{eq-op-cl-breach-call-wedge}
\eeq
In Eq.(\ref{eq-op-cl-breach-call-wedge}),
$\hatt$ and $\hatt'$
are auxiliary parameters
whose values are still to be defined.

Consider separately
each half of the circuit in
Eq.(\ref{eq-op-cl-breach-call-wedge}).
Our goal is to re-express each
half as follows:

\beq
\begin{array}{c}
\Qcircuit @C=1em @R=1em @!R{
&\ovalgate{\hatt}\qwx[1]
&\ovalgate{\hatq_R}\qwx[1]
&\ovalgate{\hatp_R}\qwx[1]
&\qw
\\
&\ovalgate{\hatt'}
&\ovalgate{\hatq'_R}
&\ovalgate{\hatp'_R}
&\qw
}
\end{array}
=
\begin{array}{c}
\Qcircuit @C=1em @R=1em @!R{
&\ovalgate{\hatq_{Rf}}\qwx[1]
&\ovalgate{\hatp_{Rf}}\qwx[1]
&\gate{U_{Rf}}
&\qw
\\
&\ovalgate{\hatt'}
&\ovalgate{\hatp'_{Rf}}
&\gate{U'_{Rf}}
&\qw
}
\end{array}
\;,
\label{eq-op-cl-breach-rhs}
\eeq
and

\beq
\begin{array}{c}
\Qcircuit @C=1em @R=1em @!R{
&\ovalgate{\hatp_L}\qwx[1]
&\ovalgate{\hatq_L}\qwx[1]
&\ovalgate{\hatt}\qwx[1]
&\qw
\\
&\ovalgate{\hatp'_L}
&\ovalgate{\hatq'_L}
&\ovalgate{\hatt'}
&\qw
}
\end{array}
=
\begin{array}{c}
\Qcircuit @C=1em @R=1em @!R{
&\gate{U_{Lf}}
&\ovalgate{\hatp_{Lf}}\qwx[1]
&\ovalgate{\hatq_{Lf}}\qwx[1]
&\qw
\\
&\gate{U'_{Lf}}
&\ovalgate{\hatp'_{Lf}}
&\ovalgate{\hatt'}
&\qw
}
\end{array}
\;.
\label{eq-op-cl-breach-lhs}
\eeq
From Theorem \ref{th-persistence}, we
know that Eq.(\ref{eq-op-cl-breach-rhs})
will be achieved if we constrain our
auxiliary parameters by:

\begin{subequations}
\label{eq-op-cl-breach-contraints}
\beq
\manyx{\hatp'_R\hatq'_R\hatq'_R}\cdot\hatt'=0
\;,
\eeq
and

\beq
\left[c_{\phi'_R}(\hatp_R\cdot\hatq_R)
\hatp_R\times\hatq_R
-s_{\lam'_R}c_{\lam'_R}s_{\phi'_R}\hatq_R
\right]\cdot \hatt
=0
\;.
\eeq
\end{subequations}
Likewise,
Eq.(\ref{eq-op-cl-breach-lhs})
will be achieved if we constrain our
auxiliary parameters by the
same pair of equations as
Eqs.(\ref{eq-op-cl-breach-contraints}),
but with $R$ subscripts replaced by
$L$ subscripts.
These 4 constraint equations
can be used to solve for the
4 degrees of freedom contained in the
auxiliary parameters
$\hatt$ and $\hatt'$.

\qed

By a ``unit wedge" we mean
a circuit element which
equals one. An analogous
concept is a ``partition of unity".
If it equals one, why use it?
Because it depends on new,
auxiliary parameters, and, by merging
the unit wedge with its surroundings,
we get a new expression
which contains the auxiliary parameters,
but is functionally independent of them.
We can then choose convenient values for
the auxiliary parameters.
The net result is that we
can transform the original circuit to
a new one that performs exactly as the old one
but appears different.

Note that in Eq.(\ref{eq-op-cl-breach-call-wedge})
we used a unit wedge consisting of a
single DC-NOT
times itself. There was no a priori obvious reason
why this unit wedge would lead us to a proof
of the theorem. We could have
chosen
a unit wedge
that provided more auxiliary parameters.
For instance, we could have chosen
a product of 3 DC-NOTs (times the inverse
of the product).
After all, 1 DC-NOT
can express only a limited subset of
all possible 2-qubit transformations
whereas 3 DC-NOTs can be used to express any
of them. For proving the above
theorem, using a
unit wedge with only
1 DC-NOT turned out to be sufficient.
But one can envisage this theorem proving
technique being
used elsewhere
with more complicated unit wedges.

Suppose one starts with a circuit
which,
like $\call$ in  Eq.(\ref{eq-op-cl-breach-call}),
possesses 4 DC-NOTs.
By the last theorem,
one can ``open a breach" in it; that is,
transform it
into a circuit which, like $\calr$ in
Eq.(\ref{eq-op-cl-breach-calr}),
possesses two
adjacent oval nodes both carrying a $\hatt'$.
Then one can combine the two
adjacent DC-NOTs
with a $\hatt'$ node and obtain
a controlled-U.
Finally, one can  use the
 Deflation
Identity of Sec.\ref{sec-2-contr-u}
to express the just created controlled-U
and an adjacent DC-NOT
as a circuit with two CNOTs.
The net effect of this
procedure is to reduce any
2-qubit circuit with 4 CNOTs into one with 3 CNOTs.
\section{Identities for Circuits with 3 Qubits}
\label{sec-3-qubit-ids}
\subsection{Pass-Through Identities}
\label{sec-pass-thru}

In the following 3 subsections, we consider
the following 3 ``identities" (one subsection
per identity):

\begin{subequations}
\label{eq-bare-pass-thru}
\beq
\begin{array}{c}
\Qcircuit @C=1em @R=.5em @!R{
&\ovalgate{\;}\qwx[2]
&\ovalgate{\;}\qwx[1]
&\qw
\\
&\qw
&\ovalgate{\;}
&\qw
\\
&\ovalgate{\;}
&\qw
&\qw
}
\end{array}
\sim_R
\begin{array}{c}
\Qcircuit @C=1em @R=.5em @!R{
&\ovalgate{\;}\qwx[1]
&\ovalgate{\;}\qwx[2]
&\qw
\\
&\ovalgate{\;}
&\qw
&\qw
\\
&\qw
&\ovalgate{\;}
&\qw
}
\end{array}
\;,
\label{eq-bare-pass-1}
\eeq

\beq
\begin{array}{c}
\Qcircuit @C=1em @R=.5em @!R{
&\ovalgate{\;}\qwx[2]
&\ovalgate{\;}\qwx[1]
&\ovalgate{\;}\qwx[1]
&\qw
\\
&\qw
&\ovalgate{\;}
&\ovalgate{\;}
&\qw
\\
&\ovalgate{\;}
&\qw
&\qw
&\qw
}
\end{array}
\sim_R
\begin{array}{c}
\Qcircuit @C=1em @R=.5em @!R{
&\ovalgate{\;}\qwx[1]
&\ovalgate{\;}\qwx[2]
&\ovalgate{\;}\qwx[1]
&\qw
\\
&\ovalgate{\;}
&\qw
&\ovalgate{\;}
&\qw
\\
&\qw
&\ovalgate{\;}
&\qw
&\qw
}
\end{array}
\;,
\label{eq-bare-pass-2}
\eeq

\beq
\begin{array}{c}
\Qcircuit @C=1em @R=.5em @!R{
&\ovalgate{\;}\qwx[2]
&\ovalgate{\;}\qwx[1]
&\ovalgate{\;}\qwx[1]
&\ovalgate{\;}\qwx[1]
&\qw
\\
&\qw
&\ovalgate{\;}
&\ovalgate{\;}
&\ovalgate{\;}
&\qw
\\
&\ovalgate{\;}
&\qw
&\qw
&\qw
&\qw
}
\end{array}
\sim_R
\begin{array}{c}
\Qcircuit @C=1em @R=.5em @!R{
&\ovalgate{\;}\qwx[1]
&\ovalgate{\;}\qwx[2]
&\ovalgate{\;}\qwx[1]
&\ovalgate{\;}\qwx[1]
&\qw
\\
&\ovalgate{\;}
&\qw
&\ovalgate{\;}
&\ovalgate{\;}
&\qw
\\
&\qw
&\ovalgate{\;}
&\qw
&\qw
&\qw
}
\end{array}
\;.
\label{eq-bare-pass-3}
\eeq
\end{subequations}
Note that in all 3 identities, the
initial and final circuits
both have
the same number of
DC-NOTs, acting on the same 3 qubits.
In all 3 cases, we pass
a DC-NOT (the mobile one) acting on qubits
0 and 1 through another
DC-NOT (the static one) acting on qubits 0 and 2.
Thus, the mobile and static DC-NOTs
both act on qubit 0, but
the second qubit on which they act
differs.
We will refer to
Eq.(\ref{eq-bare-pass-1}),
Eq.(\ref{eq-bare-pass-2}),
and
Eq.(\ref{eq-bare-pass-3})
as the Pass-Through Identities 1,2, and 3, respectively.
In the initial circuit of Pass-Through Identity $n$,
the mobile DC-NOT
is part of a group of $n$ adjacent DC-NOTs
acting on qubits 0 and 1.

The Pass-Through Identities
Eqs.(\ref{eq-bare-pass-thru})
do not, per se, change the number of DC-NOTs.
However, in some situations,
they can be used to reduce the
number of DC-NOTs. For example,

\beqa
\label{eq-eg-pass-thru}
\begin{array}{c}
\Qcircuit @C=1em @R=.5em @!R{
&\ovalgate{\;}\qwx[1]
&\ovalgate{\;}\qwx[1]
&\ovalgate{\;}\qwx[1]
&\ovalgate{\;}\qwx[2]
&\ovalgate{\;}\qwx[1]
&\qw
\\
&\ovalgate{\;}
&\ovalgate{\;}
&\ovalgate{\;}
&\qw
&\ovalgate{\;}
&\qw
\\
&\qw
&\qw
&\qw
&\ovalgate{\;}
&\qw
&\qw
}
\end{array}
&\sim_R &
\begin{array}{c}
\Qcircuit @C=1em @R=.5em @!R{
&\ovalgate{\;}\qwx[1]
&\ovalgate{\;}\qwx[1]
&\ovalgate{\;}\qwx[1]
&\ovalgate{\;}\qwx[1]
&\ovalgate{\;}\qwx[2]
&\qw
\\
&\ovalgate{\;}
&\ovalgate{\;}
&\ovalgate{\;}
&\ovalgate{\;}
&\qw
&\qw
&\qw
\\
&\qw
&\qw
&\qw
&\qw
&\ovalgate{\;}
&\qw
&\qw
}
\end{array}
\\
&\sim_R &
\begin{array}{c}
\Qcircuit @C=1em @R=.5em @!R{
&\ovalgate{\;}\qwx[1]
&\ovalgate{\;}\qwx[1]
&\ovalgate{\;}\qwx[1]
&\ovalgate{\;}\qwx[2]
&\qw
\\
&\ovalgate{\;}
&\ovalgate{\;}
&\ovalgate{\;}
&\qw
&\qw
&\qw
\\
&\qw
&\qw
&\qw
&\ovalgate{\;}
&\qw
&\qw
}
\end{array}
\;.
\eeqa
In Eq.(\ref{eq-eg-pass-thru}),  there are
initially 3 adjacent DC-NOTs
on the LHS of the static
DC-NOT.
Using Pass-Through Identity 1,
we produce 4 adjacent
DC-NOTs on the LHS of the static
DC-NOT. As shown
in Section \ref{sec-opening-breach},
these 4 adjacent
DC-NOTs
can always be reduced to 3 DC-NOTs.
\subsubsection{Pass-Through Identity 1}
\label{sec-pass1}

\begin{theo}
Suppose

\beq
\call =
\begin{array}{c}
\Qcircuit @C=1em @R=.5em @!R{
&\ovalgate{\hatb}\qwx[2]
&\ovalgate{\hata}\qwx[1]
&\qw
\\
&\qw
&\ovalgate{\hata'}
&\qw
\\
&\ovalgate{\hatb''}
&\qw
&\qw
}
\end{array}
\;\;,\;\;
\calr =
\begin{array}{c}
\Qcircuit @C=1em @R=.5em @!R{
&\ovalgate{\hata_f}\qwx[1]
&\ovalgate{\hatb_f}\qwx[2]
&\qw
\\
&\ovalgate{\hata'_f}
&\qw
&\qw
\\
&\qw
&\ovalgate{\hatb''_f}
&\qw
}
\end{array}
\;.
\eeq
For any $\call$, it is possible to
find an $\calr$ such that $\call\sim_R \calr$
 if and only if $\hata\parallel\hatb$.
\end{theo}
\proof

\lproof
Let $\hata'_f=\hata'$ and
$\hatb''_f = \hatb''$.
Clearly,
if $\hata=\hatb$, then $\call= \calr$.
More generally, $\hata=\pm\hatb$.
Let $\call_{new}$ be a new circuit
obtained by replacing in $\call$:
$\hata$ by its negative if $\hata=-\hatb$.
By virtue of Eq.(\ref{eq-dcnot-with-neg}),
$\call = \call_{new}(I_2\otimes U \otimes I_2)$
where $U\in U(2)$.
If $\call_{new}\sim_R \calr_{new}$, then
$\call\sim_R \calr_{new}$.

\rproof

Using the same calculational techniques that
were used in
Section \ref{sec-two-bit-dcnot-rhs-invariants},
one finds

\beq
\call^{(2)}=
\hata\cdot\hatb \sigma_{\hatb'',\hata',1}
+i \sigma_{1,\hata',\manyx{\hata\hatb}}
\;,
\eeq
and

\beq
\calr^{(2)}=
\hatb_f\cdot\hata_f \sigma_{\hatb''_f,\hata'_f,1}
+i \sigma_{\hatb''_f,1,\manyx{\hata_f\hatb_f}}
\;.
\eeq
$\call\sim_R \calr$ implies that
$\call^{(2)}$ is proportional to
$\calr^{(2)}$. Therefore,
$\sigma_{1,\hata',\manyx{\hata\hatb}}$ must vanish.
Hence, $\manyx{\hata\hatb}=0$, which is
implies $\hata\parallel \hatb$.

\qed
\subsubsection{Pass-Through Identity 2}
\label{sec-pass2}

\begin{theo}
Suppose

\beq
\call =
\begin{array}{c}
\Qcircuit @C=1em @R=.5em @!R{
&\ovalgate{\hate}\qwx[2]
&\ovalgate{\hatb}\qwx[1]
&\ovalgate{\hata}\qwx[1]
&\qw
\\
&\qw
&\ovalgate{\hatb'}
&\ovalgate{\hata'}
&\qw
\\
&\ovalgate{\hate''}
&\qw
&\qw
&\qw
}
\end{array}
\;\;,\;\;
\calr =
\begin{array}{c}
\Qcircuit @C=1em @R=.5em @!R{
&\ovalgate{\hatb_f}\qwx[1]
&\ovalgate{\hate_f}\qwx[2]
&\ovalgate{\hata_f}\qwx[1]
&\qw
\\
&\ovalgate{\hatb'_f}
&\qw
&\ovalgate{\hata'_f}
&\qw
\\
&\qw
&\ovalgate{\hate''_f}
&\qw
&\qw
}
\end{array}
\;.
\eeq
For any $\call$, if there exists $\hatt'$
such that

\beq
\begin{array}{c}
\Qcircuit @C=1em @R=1em @!R{
&\ovalgate{\hate}\qwx[1]
&\ovalgate{\hatb}\qwx[1]
&\ovalgate{\hata}\qwx[1]
&\qw
\\
&\ovalgate{\hatt'}
&\ovalgate{\hatb'}
&\ovalgate{\hata'}
&\qw
}
\end{array}
\sim_R
\begin{array}{c}
\Qcircuit @C=1em @R=1em @!R{
&\ovalgate{\hata_f}\qwx[1]
&\qw
\\
&\ovalgate{\hata'_f}
&\qw
}
\end{array}
\;,
\label{eq-3to1-for-some-t}
\eeq
then
it is possible to
find an $\calr$ such that $\call\sim_R \calr$.
\end{theo}
\proof

One has

\beqa
\call &=&
\begin{array}{c}
\Qcircuit @C=1em @R=.5em @!R{
&\ovalgate{\hate}\qwx[2]
&\ovalgate{\hate}\qwx[1]
&\ovalgate{\hate}\qwx[1]
&\ovalgate{\hatb}\qwx[1]
&\ovalgate{\hata}\qwx[1]
&\qw
\\
&\qw
&\ovalgate{\hatt'}
&\ovalgate{\hatt'}
&\ovalgate{\hatb'}
&\ovalgate{\hata'}
&\qw
\\
&\ovalgate{\hate''}
&\qw
&\qw
&\qw
&\qw
&\qw
}
\end{array}\label{eq-pass2-a}
\\
&=&
\begin{array}{c}
\Qcircuit @C=1em @R=.5em @!R{
&\ovalgate{\hate}\qwx[1]
&\ovalgate{\hate}\qwx[2]
&\ovalgate{\hate}\qwx[1]
&\ovalgate{\hatb}\qwx[1]
&\ovalgate{\hata}\qwx[1]
&\qw
\\
&\ovalgate{\hatt'}
&\qw
&\ovalgate{\hatt'}
&\ovalgate{\hatb'}
&\ovalgate{\hata'}
&\qw
\\
&\qw
&\ovalgate{\hate''}
&\qw
&\qw
&\qw
&\qw
}
\end{array}\label{eq-pass2-b}
\\
&=&
\begin{array}{c}
\Qcircuit @C=1em @R=.5em @!R{
&\ovalgate{\hate}\qwx[1]
&\ovalgate{\hate}\qwx[2]
&\ovalgate{\hata_f}\qwx[1]
&\qw
\\
&\ovalgate{\hatt'}
&\qw
&\ovalgate{\hata'_f}
&\qw
\\
&\qw
&\ovalgate{\hate''}
&\qw
&\qw
}
\end{array}\label{eq-pass2-c}
\;.
\eeqa
In (a), we
introduced a unit wedge.
To go from (a) to (b),
we passed half of that unit wedge
across the ``static" DC-NOT.
Finally, to go from (b) to (c),
we used Eq.(\ref{eq-3to1-for-some-t}).

\qed

Note that Section \ref{sec-3to1-cnots}
gives necessary and sufficient
conditions for a 2-qubit circuit with
3 DC-NOTs to reduce to an equivalent
circuit with 1 DC-NOT.
Using those necessary and sufficient
conditions, it is easy to check
in any particular instance
whether
there exists a $\hatt'$
such that Eq.(\ref{eq-3to1-for-some-t})
is satisfied.
\subsubsection{Pass-Through Identity 3
\\{\footnotesize\tt[
pass3.m,
test\_pass3.m
]}}
\label{sec-pass3}

\begin{theo}\label{th-pass-thru3}
Suppose

\beq
\call =
\begin{array}{c}
\Qcircuit @C=1em @R=.5em @!R{
&\ovalgate{\hate}\qwx[2]
&\ovalgate{\hatc}\qwx[1]
&\ovalgate{\hatb}\qwx[1]
&\ovalgate{\hata}\qwx[1]
&\qw
\\
&\qw
&\ovalgate{\hatc'}
&\ovalgate{\hatb'}
&\ovalgate{\hata'}
&\qw
\\
&\ovalgate{\hate''}
&\qw
&\qw
&\qw
&\qw
}
\end{array}
\;\;,\;\;
\calr =
\begin{array}{c}
\Qcircuit @C=1em @R=.5em @!R{
&\ovalgate{\hatc_f}\qwx[1]
&\ovalgate{\hate_f}\qwx[2]
&\ovalgate{\hatb_f}\qwx[1]
&\ovalgate{\hata_f}\qwx[1]
&\qw
\\
&\ovalgate{\hatc'_f}
&\qw
&\ovalgate{\hatb'_f}
&\ovalgate{\hata'_f}
&\qw
\\
&\qw
&\ovalgate{\hate''_f}
&\qw
&\qw
&\qw
}
\end{array}
\;.
\eeq
For any $\call$, it is possible to
find an $\calr$ such that $\call\sim_R \calr$.
\end{theo}
\proof

\beqa
\call &=&
\begin{array}{c}
\Qcircuit @C=1em @R=.5em @!R{
&\ovalgate{\hate}\qwx[2]
&\ovalgate{\hate}\qwx[1]
&\ovalgate{\hate}\qwx[1]
&\ovalgate{\hatc}\qwx[1]
&\ovalgate{\hatb}\qwx[1]
&\ovalgate{\hata}\qwx[1]
&\qw
\\
&\qw
&\ovalgate{\hatt'}
&\ovalgate{\hatt'}
&\ovalgate{\hatc'}
&\ovalgate{\hatb'}
&\ovalgate{\hata'}
&\qw
\\
&\ovalgate{\hate''}
&\qw
&\qw
&\qw
&\qw
&\qw
&\qw
}
\end{array}
\\
&=&
\begin{array}{c}
\Qcircuit @C=1em @R=.5em @!R{
&\ovalgate{\hate}\qwx[1]
&\ovalgate{\hate}\qwx[2]
&\ovalgate{\hate}\qwx[1]
&\ovalgate{\hatc}\qwx[1]
&\ovalgate{\hatb}\qwx[1]
&\ovalgate{\hata}\qwx[1]
&\qw
\\
&\ovalgate{\hatt'}
&\qw
&\ovalgate{\hatt'}
&\ovalgate{\hatc'}
&\ovalgate{\hatb'}
&\ovalgate{\hata'}
&\qw
\\
&\qw
&\ovalgate{\hate''}
&\qw
&\qw
&\qw
&\qw
&\qw
}
\end{array}
\\
&=&
\begin{array}{c}
\Qcircuit @C=1em @R=.5em @!R{
&\ovalgate{\hate}\qwx[1]
&\ovalgate{\hate}\qwx[2]
&\ovalgate{\hatb_f}\qwx[1]
&\ovalgate{\hata_f}\qwx[1]
&\qw
\\
&\ovalgate{\hatt'}
&\qw
&\ovalgate{\hatb'_f}
&\ovalgate{\hata'_f}
&\qw
\\
&\qw
&\ovalgate{\hate''}
&\qw
&\qw
&\qw
}
\end{array}
\;.
\eeqa
In (a), we
introduced a unit wedge.
To go from (a) to (b),
we passed half of that unit wedge
across the ``static" DC-NOT.
Finally, to go from (b) to (c),
we used Theorem \ref{eq-th-aux-pass3}.

\qed

The next theorem is used in the proof
of Theorem \ref{th-pass-thru3}.

\begin{theo}\label{eq-th-aux-pass3}
Suppose

\beq
\call(\hatd')=
\begin{array}{c}
\Qcircuit @C=1em @R=1em @!R{
&\ovalgate{\hatd}\qwx[1]
&\ovalgate{\hatc}\qwx[1]
&\ovalgate{\hatb}\qwx[1]
&\ovalgate{\hata}\qwx[1]
&\qw
\\
&\ovalgate{\hatd'}
&\ovalgate{\hatc'}
&\ovalgate{\hatb'}
&\ovalgate{\hata'}
&\qw
}
\end{array}
\;\;,\;\;
\calr=
\begin{array}{c}
\Qcircuit @C=1em @R=1em @!R{
&\ovalgate{\hatb_f}\qwx[1]
&\ovalgate{\hata_f}\qwx[1]
&\qw
\\
&\ovalgate{\hatb'_f}
&\ovalgate{\hata'_f}
&\qw
}
\end{array}
\;.
\eeq
For any $\call(\cdot)$, there exists a $\hatd'$
and an $\calr$ such that
$\call\sim_R \calr$.
\end{theo}
\proof

Our goal is to find a $\hatd'$ and
to construct an $\calr$ such that
$\call\sim_R \calr$. Such an $\calr$ must satisfy
$\hat{\call}^{(2)}= \pm \hat{\calr}^{(2)}$.
We will use the positive sign.
In light of
Eq.(\ref{eq-invar-of-hat-graph}),
the
following must be true:

\beq
i^4\call^{(2)} = i^2 \calr^{(2)}
\;.
\eeq

From Section \ref{sec-invariants-4cnots}, we know that
\beq
\call^{(2)}=
\lam_{4r} + i\lam_{4i} + \Lam_{4r} + i \Lam_{4i}
\;,
\eeq
where

\beq
\lam_{4r} =
-\hatd^{\;'T}M_\nu\hatd
\;,
\eeq

\beq
\lam_{4i} =
-\hatd^{\;'T}M_\mu\hatd
\;,
\eeq

\beq
\Lam_{4r}=
X_o \hatd'\hatd^T
+ \vec{x'}\hatd^T
+ \hatd'\vecx^T
+ \Delta X
\;,
\eeq

\beq
\Lam_{4i}=
Y_o \hatd'\hatd^T
- \vec{y'}\hatd^T
- \hatd'\vecy^T
+ \Delta Y
\;.
\eeq
The precise definitions of $(X_o,Y_o)$,
$(\vecx,\vec{x'},\vecy,\vec{y'})$,
$(\Delta X, \Delta Y)$, and $(M_\mu, M_\nu)$
in terms of $(\hata,\hata')$,
$(\hatb,\hatb')$, $(\hatc,\hatc')$, and
$(\hatd,\hatd')$
are given in
Section \ref{sec-invariants-4cnots}.

From Section \ref{sec-invariants-2cnots},
we know that

\beqa
\calr^{(2)}&=&
\lam_{2r} + \Lam_{2r} + i \Lam_{2i}\\
&=&
c_{\alpha'} c_\alpha
-(s_{\alpha'}s_\alpha) \hatf_2'\hatf_2^T
+i\;
\begin{array}{l|ll}
        & \hatf_1^T & \hatf_3^T\\
\hline
\hatf_3'& s_{\alpha'}c_\alpha & 0\\
\hatf_1'& 0                   &c_{\alpha'}s_\alpha
\end{array}
\;.
\eeqa

We must have

\begin{subequations}

\beq
\lam_{2r} = - \lam_{4r}
\;,
\label{eq-cond1-pass-w3}
\eeq

\beq
0 = \lam_{4i}
\;,
\label{eq-cond2-pass-w3}
\eeq

\beq
\Lam_{2r} = - \Lam_{4r}
\;,
\label{eq-cond3-pass-w3}
\eeq
and

\beq
\Lam_{2i} = - \Lam_{4i}
\;.
\label{eq-cond4-pass-w3}
\eeq
\end{subequations}

To begin, we will assume that $X_o\neq 0$.
Later on, before ending
the proof, we will remove this assumption.

By evaluating Eq.(\ref{eq-cond1-pass-w3}), we get

\beq
c_{\alpha'}c_\alpha =
\hatd^{\;'T}M_\nu\hatd
\;.
\eeq

By evaluating Eq.(\ref{eq-cond2-pass-w3}),
we get

\beq
0 = \hatd^{\;'T}M_\mu \hatd
\;.
\eeq
Let $\hatd'$ be
any unit vector that satisfies this equation.

By evaluating Eq.(\ref{eq-cond3-pass-w3}), we get

\beq
-s_{\alpha'}s_\alpha \hatf'_2\hatf_2^T
=-(
X_o \hatd'\hatd^T
+ \vec{x'}\hatd^T
+ \hatd'\vecx^T
+ \Delta X)
\;.
\label{eq-cond3-unfactored}
\eeq
For Eq.(\ref{eq-cond3-unfactored}) to be true,
the RHS of that equation must
factor into the product of a column vector
times a row vector:

\beq
-s_{\alpha'}s_\alpha \hatf'_2\hatf_2^T
=-X_o\left(
\hatd' + \frac{\vec{x'}}{X_o}\right)
\left(\hatd + \frac{\vec{x}}{X_o}\right)^T
\;.
\label{eq-rank1-rhs}
\eeq
Let

\beq
s_{\alpha'}s_\alpha =
X_o \eta'_2 \eta_2
\;,\;\;
\hatf'_2 =
\frac{\hatd' + \frac{\vec{x'}}{X_o}}{\eta'_2}
\;,\;\;
\hatf_2 =
\frac{\hatd + \frac{\vec{x}}{X_o}}{\eta_2}
\;,
\eeq
where

\beq
\eta'_2 = \left|\hatd' + \frac{\vec{x'}}{X_o}\right|
=\sqrt{ 1 + \frac{(\vec{x'})^2}{(X_o)^2}}
\;,\;\;
\eta_2 = (\eta'_2)_\unprime
\;.
\eeq

Note that since
Eqs.(\ref{eq-cond3-unfactored})
and (\ref{eq-rank1-rhs})
are both true, the following must be true:

\beq
\frac{\vec{x'}\vec{x}^T}{X_o} = \Delta X
\;.
\label{eq-Delta-x-id}
\eeq
Eq.(\ref{eq-Delta-x-id}) can also
be proven  by expressing it
 in terms of
$(\hata,\hata')$,
$(\hatb,\hatb')$, $(\hatc,\hatc')$, and
$(\hatd,\hatd')$.

By evaluating Eq.(\ref{eq-cond4-pass-w3}), we get

\beq
\Lam_{2i}
=
-(
Y_o \hatd'\hatd^T
- \vec{y'}\hatd^T
- \hatd'\vecy^T
+ \Delta Y)
\;.
\label{eq-diag-equal-not-diag}
\eeq

At this point, we can follow from
step \ref{item-diag-invar2-h-hprime}
to the end of the
Algorithm for Diagonalizing $\calg^{(2)}_2$
that was given in
Section \ref{sec-invariants-2cnots}.
This will yield values for
$\hata_f$,
$\hata'_f$,
$\hatb_f$, and
$\hatb'_f$.

Now assume $X_o=0$.
By Eq.(\ref{eq-Delta-x-id}), either
$\vec{x'}=0$ or $\vecx=0$. When
$\vec{x'}=0$ and $\vecx\neq 0$
(the case $\vec{x'}\neq 0$ and $\vecx= 0$
is analogous),
Eq.(\ref{eq-rank1-rhs}) becomes

\beq
-s_{\alpha'}s_\alpha \hatf'_2\hatf_2^T
=-\left(
\hatd' + \frac{\vec{x'}}{X_o}\right)
\vec{x}^T
\;,
\eeq
where
$\frac{\vec{x'}}{X_o}$ is defined
as the obvious limit. Thus, we can set

\beq
s_{\alpha'}s_\alpha =
\eta'_2 |\vecx|
\;,\;\;
\hatf'_2 =
\frac{\hatd' + \frac{\vec{x'}}{X_o}}{\eta'_2}
\;,\;\;
\hatf_2 =
\frac{\vecx}{|\vecx|}
\;.
\eeq
If $\vec{x'}=\vecx=0$, then
Eq.(\ref{eq-rank1-rhs}) becomes
$-s_{\alpha'}s_\alpha \hatf'_2\hatf_2^T=0$,
so we can set $s_{\alpha'}s_\alpha=0$
and define $\hatf_2$ and $\hatf'_2$ to
be arbitrary unit vectors.

Additional observations:

Note that
$\hatf^{\;'T}_2\Lam_{2i} = 0$ implies

\begin{subequations}\label{eq-f2-Lam2i}
\beq
\vec{x'}\cdot\vec{y'} = X_o Y_o
\;,
\eeq
and

\beq
\Delta Y^T \vec{x'} = X_o \vec{y}
\;.
\eeq
\end{subequations}
Likewise, note that
$\Lam_{2i}\hatf_2 = 0$ implies

\begin{subequations}\label{eq-Lam2i-f2}
\beq
\vec{x}\cdot\vec{y} = X_o Y_o
\;,
\eeq
and

\beq
\Delta Y\vec{x} = X_o \vec{y'}
\;.
\eeq
\end{subequations}
Eqs.(\ref{eq-f2-Lam2i}) and
(\ref{eq-Lam2i-f2}) can also
be proven  by expressing them
 in terms of
$(\hata,\hata')$,
$(\hatb,\hatb')$, $(\hatc,\hatc')$, and
$(\hatd,\hatd')$.

If $|\vecx|$ and $|\vec{x'}|$
are both non-zero, it is possible to
introduce 2 RHON bases
$(\hath'_j)_{j=1,2,3}$
and $(\hath_j)_{j=1,2,3}$,
defined as follows.
Define $\hath'_2$ and $\hath_2$ by
\beq
\hath'_2 = \hatf'_2
\;,\;\;
\hath_2 = \hatf_2
\;.
\eeq
Define $\hath'_3$ and $\hath_3$ by

\beq
\hath'_3 =
\frac{\hatd' - \frac{\vec{x'}X_o}{(\vec{x'})^2}}
{\eta'_3}
\;,\;\;
\hath_3 = (\hath'_3)_\unprime
\;,
\eeq
where

\beq
\eta'_3 = \left|\hatd' - \frac{\vec{x'}X_o}{(\vec{x'})^2}\right|
=\sqrt{ 1 + \frac{(X_o)^2}{(\vec{x'})^2}}=
\frac{X_o}{|\vec{x'}|}\eta_2
\;,\;\;
\eta_3 = (\eta'_3)_\unprime
\;.
\eeq
Define $\hath'_1$ and $\hath_1$ by

\beq
\hath'_1 =\frac{\manyx{\vec{x'}\hatd'} \sign(X_o)}
{\eta'_1}
\;,\;\;
\hath_1 = (\hath'_1)_\unprime
\;,
\eeq
where

\beq
\eta'_1=|\manyx{\vec{x'}\hatd'}| = |\vec{x'}|
\;,\;\;
\eta_1 = (\eta'_1)_\unprime
\;.
\eeq
After some algebra,
one can show that
Eq.(\ref{eq-diag-equal-not-diag})
becomes

\beq
\Lam_{2i}=
\begin{array}{l|ll}
        & \frac{\hath_1^T}{|\vec{x}|} & \hath_3^T\eta_3\\
\hline
\hath_3'\eta'_3&
\vec{y}^T\manyx{\vec{x}\hatd} \sign(X_o)& -Y_o
\\ \frac{\hath_1'}{|\vec{x'}|}&
 -\manyx{\vec{x'}\hatd'}^T\Delta Y\manyx{\vec{x}\hatd}
 &\vec{y}^{\;'T}\manyx{\vec{x'}\hatd'} \sign(X_o)
\end{array}
\;.
\eeq
The entries of the previous table
can be expressed solely
in terms of $(\hatd,\hatd')$
and $(M_\mu, M_\nu)$.
After some algebra, one finds
that

\beq
\vec{y}^T\manyx{\vec{x}\hatd}=
(M_\nu^T\hatd')\cdot\manyx{M^T_\mu\hatd',\hatd}
\;,
\eeq

\beq
\vec{y}^{\;'T}\manyx{\vec{x'}\hatd'}=
(M_\nu\hatd)\cdot\manyx{M_\mu\hatd,\hatd'}
\;,
\eeq
and

\beq
\manyx{\vec{x'}\hatd'}^T\Delta Y \manyx{\vec{x}\hatd}=
\hatd^T M_\mu^T M_\mu M_\mu^T \hatd'
\;.
\eeq

\qed
\vskip1cm
\noindent{\bf Acknowledgements}

\noindent I thank E. Rains for
kind and instructive email on the group
theoretic aspects of circuit invariants.

\end{document}